\documentclass{aa}  

\usepackage{natbib}
\usepackage{graphicx}
\usepackage{txfonts}
\begin{document}

\title{The Near-Infrared Spectrograph (NIRSpec) on the James Webb Space Telescope}

\subtitle{I. Overview of the instrument and its capabilities}

\author{ P. Jakobsen\thanks{NIRSpec Project Scientist 1997-2011}\inst{1}
\and P. Ferruit\thanks{NIRSpec Project Scientist 2012-present}\inst{2}
\and C.    Alves de Oliveira  \inst{2}
\and S.    Arribas            \inst{3}
\and G.    Bagnasco           \inst{4}
\and R.    Barho              \inst{5}
\and T. L. Beck               \inst{6}
\and S.    Birkmann           \inst{7}
\and T.    B\"{o}ker          \inst{7}
\and A. J. Bunker             \inst{8}
\and S.    Charlot            \inst{9}
\and P.    de Jong            \inst{4}
\and G.    de Marchi          \inst{4}
\and R.    Ehrenwinkler       \inst{10}
\and M.    Falcolini          \inst{4}
\and R.    Fels               \inst{4}
\and M.    Franx              \inst{11}
\and D.    Franz              \inst{12}
\and M.    Funke              \inst{5}
\and G.    Giardino           \inst{13}
\and X.    Gnata              \inst{14}
\and W.    Holota             \inst{15}
\and K.    Honnen             \inst{5}
\and P. L. Jensen             \inst{4}
\and M.    Jentsch            \inst{5}
\and T.    Johnson            \inst{12}
\and D.    Jollet             \inst{4}
\and H.    Karl               \inst{10}
\and G.    Kling              \inst{5}
\and J.    K\"{o}hler         \inst{10}
\and M.-G. Kolm               \inst{10}
\and N.    Kumari             \inst{16}
\and M. E. Lander             \inst{12}
\and R.    Lemke              \inst{10}
\and M.    L\'{o}pez-Caniego  \inst{17}
\and N.    L\"{u}tzgendorf    \inst{7}
\and R.    Maiolino           \inst{18}
\and E.    Manjavacas         \inst{16}
\and A.    Marston            \inst{2}
\and M.    Maschmann          \inst{5}
\and R.    Maurer             \inst{10}
\and B.    Messerschmidt      \inst{5}
\and S. H. Moseley            \inst{19}
\and P.    Mosner             \inst{10}
\and D. B. Mott               \inst{12}
\and J.    Muzerolle          \inst{6}
\and N.    Pirzkal            \inst{16}
\and J-F.  Pittet             \inst{10}
\and A.    Plitzke            \inst{4}
\and W.    Posselt            \inst{20}
\and B.    Rapp               \inst{12}
\and B. J. Rauscher           \inst{12}
\and T.    Rawle              \inst{7}
\and H.-W. Rix                \inst{21}
\and A.    R\"{o}del          \inst{10}
\and P.    Rumler             \inst{4}
\and E.    Sabbi              \inst{6}
\and J.-C. Salvignol          \inst{4}
\and T.    Schmid             \inst{5}
\and M.    Sirianni           \inst{7}
\and C.    Smith              \inst{21}
\and P.    Strada             \inst{4}
\and M.    te Plate           \inst{7}
\and J.    Valenti            \inst{6}
\and T.    Wettemann          \inst{10}
\and T.    Wiehe              \inst{5}
\and M.    Wiesmayer          \inst{10}
\and C. J. Willott            \inst{23}
\and R.    Wright             \inst{24}
\and P.    Zeidler            \inst{16} 
\and C.    Zincke             \inst{12}
}

\institute{ 
%1
Cosmic Dawn Center, Niels Bohr Institute, University of Copenhagen, Denmark \and
%2
European Space Agency, European Space Astronomy Centre, Madrid, Spain \and
%3
Centro de Astrobiologia (CSIC-INTA), Departamento de Astrofisica, Madrid, Spain \and
%4
European Space Agency, European Space Research and Technology Centre, Noordwijk, The Netherlands \and
%5*
Airbus Defence and Space GmbH, Friedrichshafen, Germany \and
%6
Space Telescope Science Institute, Baltimore, Maryland, USA \and
%7
European Space Agency, Space Telescope Science Institute, Baltimore, Maryland, USA \and
%8
Department of Physics, University of Oxford, United Kingdom \and
%9
Sorbonne Universit\'{e}, CNRS, UMR 7095, Institut d’Astrophysique de Paris, France \and
%10
Airbus Defence and Space GmbH, Ottobrunn, Germany \and
%11
Leiden Observatory, Leiden University, The Netherlands \and
%12
NASA Goddard Space Flight Center, Greenbelt, Maryland, USA \and
%13
ATG Europe for the European Space Agency, European Space Research and Technology Centre, Noordwijk, The Netherlands \and
%14
Airbus Defence and Space SAS, Toulouse, France \and
%15
Holota Optics, Neuhaus, Germany \and
%16
AURA for the European Space Agency, Space Telescope Science Institute, Baltimore, Maryland, USA \and
%17
Aurora Technology for the European Space Agency, European Space Astronomy Centre, Madrid, Spain \and
%18
Kavli Institute for Cosmology, University of Cambridge, United Kingdom \and
%19
Quantum Circuits, Inc., New Haven, Connecticut, USA \and
%20
OHB System AG, Bremen, Germany \and
%21
Max-Planck Institute for Astronomy, Heidelberg, Germany \and
%22
Northrup Grumman Innovation Systems, Beltsville, Maryland, USA \and
%23
NRC Herzberg, National Research Council, Victoria, British Columbia, Canada \and
%24
Ball Aerospace, Boulder, Colorado, USA
}

\date{Received 15 November 2021; accepted 26 January 2022}

% \abstract{}{}{}{}{} 
% 5 {} token are mandatory
 
\abstract {We provide an overview of the design and capabilities of the near-infrared spectrograph (NIRSpec) onboard the James Webb Space Telescope. NIRSpec is designed to be capable of carrying out low-resolution ({$R\!=30\!-330$}) prism spectroscopy over the wavelength range $0.6-5.3\!~\mu$m and higher resolution ({$R\!=500\!-1340$}  or {$R\!=1320\!-3600$}) grating spectroscopy over $0.7-5.2\!~\mu$m, both in single-object mode employing any one of five fixed slits, or a 3.1$\times$3.2 arcsec$^2$ integral field unit, or in multiobject mode  employing a novel programmable micro-shutter device covering a 3.6$\times$3.4~arcmin$^2$ field of view. The all-reflective optical chain of NIRSpec and the performance of its different components are described, and some of the trade-offs made in designing the instrument are touched upon. The faint-end spectrophotometric sensitivity expected of NIRSpec, as well as its dependency on the energetic particle environment that its two detector arrays are likely to be subjected to in orbit are also discussed.
}

               
\maketitle

\titlerunning{NIRSpec on the James Webb Space Telescope I}
\authorrunning{P. Jakobsen et al.}

%________________________________________________________________
\section{Introduction}
\label{sec:intro}

The James Webb Space Telescope (JWST) is a collaborative project between NASA, ESA, and the Canadian Space Agency (CSA). The history and scientific objectives driving the JWST mission, as well as the top level requirements of the observatory and its instrument suite have been described by \citet{gard06}. Although fundamentally different in design and emphasizing the near-infrared part of the electromagnetic spectrum, JWST can be regarded as the direct result of an unanticipated scientific finding of its predecessor, the Hubble Space Telescope (HST).

Prior to its launch in 1990, it was generally thought that HST would contribute relatively little to the study of galaxies at redshifts $z\ga2$, on the grounds that remote galaxies were presumed to be large, extended, and relatively featureless, such that the higher spatial resolution afforded by the diffraction-limited HST would offer little or no advantage for their detection and study \citep[cf.][]{bahcall90}. Nonetheless, the first deep exposures obtained with HST gradually revealed this presumption to be incorrect -- the star-forming regions of faint galaxies at high redshifts turned out to be small and physically compact, at least in the rest-frame ultraviolet wavelengths probed, with the faintest and most remote objects only just resolvable with HST. This realization soon led to a series of lengthy targeted observations of the high redshift universe, culminating in the Hubble Deep Field and subsequent follow-on programs that have pushed the HST observatory and its successive series of imaging cameras to their limits in sensitivity \citep{will96, will00, giav04, beck06, koek07, grog11, illi13}. That fact that the vast majority of the galaxies detected in these deep surveys are too faint for spectroscopic follow-up with current facilities, has led to the development of techniques for determining the approximate ``photometric'' redshift of the objects by fitting assumed template spectra to their broadband colors \citep[e.g.,][]{dahl13}. This ``poor man's spectroscopy'' approach is greatly aided at high redshift by the strong intervening intergalactic neutral hydrogen absorption, which causes near total absorption of the spectra shortward of the onset of the Lyman Forest at the rest frame Ly$\alpha$ line. Although many such high redshift ``drop out'' galaxy candidates at $z\!>\!6$ up to $z\!\sim\!12$ are presently known \citep[e.g.,][]{oesc16}, nearly all of these objects await proper spectroscopic confirmation and a detailed follow-up study.

The goal of picking up where HST comes up short in exploring the early universe with a larger aperture near-infrared-optimized space telescope has driven much of the initial design of the JWST observatory and its instruments. This is especially true of the Near-Infrared Spectrograph (NIRSpec) onboard JWST. The required capabilities of NIRSpec were the subject of considerable debate during the early design phase of the mission. The overarching scientific objective set for NIRSpec since its inception was to enable near-infrared 1-5\!~$\mu$m spectroscopic observations of the most distant galaxies imaged by HST and eventually obtain spectra of even more remote objects discovered by JWST itself, thereby not only allowing their precise redshifts to be determined, but bringing the battery of spectroscopic diagnostic tools of modern astrophysics to bear on their study \citep[e.g.,][]{chev19}.

Obtaining adequate signal-to-noise spectra of the most remote and faint galaxies inevitably requires considerably longer exposure times than broad-band imaging. It was therefore universally acknowledged from the outset that NIRSpec would need to have a significant multiobject capability and be able to capture the spectra of $\sim$100 objects simultaneously over a field of view covering several arcminutes matching that of the foreseen JWST Near-Infrared Camera (NIRCam). Faint-end limiting sensitivity was also a prime consideration, which ruled out high background slit-less devices such as Fourier Transform Spectrographs \citep{grah00}, and leaving a conventional dispersive slit-based approach. The remaining issue was the choice of slit selection mechanism, where ground-based customized aperture plate and fiber optical approaches proved to be impractical to transfer to a space application. An entirely image-slicer based approach \citep{bonn03} proved to be too large and expensive for the required wide field of view. Space limitations also meant that a mechanical slit-jaw device \citep{hene04} would at best only be capable of accommodating a few dozen or so movable slits. This left the option of employing a novel Micro-Electro-Mechanical System (MEMS) device, either in the form of an array of individually tiltable micro-mirrors \citep{mack04} or an equivalent array of individually addressable micro-shutters \citep{mose04}. Although micro-mirror arrays had already been developed for commercial use in projection systems, a custom built micro-shutter array (MSA) was in the end adopted on the grounds that it allowed a simpler optical design, and  was expected to provide higher contrast and display fewer adverse diffraction effects when exposed to the coherent light of the diffraction-limited point spread function (PSF) of JWST.

In 2001 agreement was reached between ESA and NASA that the former would provide the NIRSpec instrument as part of its contributions to the JWST mission, with NASA Goddard Space Flight Center providing the two critical micro-shutter-array (MSA) and near-IR detector subsystems. NIRSpec was subsequently designed and built by European industry under ESA leadership, with Airbus Defence and Space (formerly EADS Astrium) as prime contractor. Following a succesful test and calibration campaign in Europe, the instrument was formally delivered to NASA in late 2013. At the time of writing NIRSpec has successfully undergone integration and testing alongside the other three JWST instruments, and the integrated JWST observatory is being prepared for a launch from French Guiana on an ESA-supplied Ariane V in December 2021. 

Although, the NIRSpec instrument was primarily designed with multiobject observations of faint high redshift galaxies in mind, it was from the beginning acknowledged that the range of potential applications of such an instrument flown on JWST would be far broader. For this reason, every effort was made during the design phase to assure that NIRSpec would also be capable of addressing a wide range of other astrophysical topics.

The remainder of this paper aims to provide an overview of the design and scientific capabilities of NIRSpec, touching upon the trade-offs made along the way, and describing some of the perhaps lesser known aspects of the instrument that its future users should be cognizant of. The three accompanying papers respectively provide further details of NIRSpec's multiobject capabilities \citep[Paper II]{ferr20}, the specifics of the Integral Field Unit \citep[Paper III]{boek20}, and prospects for carrying out Fixed Slit spectroscopic observations of occulting exoplanets \citep[Paper IV]{birk20}.

%__________________________________________________________________
\section{Overview of the instrument}

NIRSpec is designed to be capable of performing both single- and multiobject spectroscopic observations over the 0.6-5.3\!~$\mu$m near-infrared wavelength range at three spectral resolutions. Briefly stated, the lowest resolution $R\simeq100$ mode is intended for obtaining exploratory continuum spectra and redshifts of remote galaxies; the intermediate resolution $R\simeq1000$ mode is designed for accurately measuring their nebular emission lines, and the higher resolution $R\simeq2700$ mode for performing kinematic studies using these emission lines.

\begin{figure}
  \resizebox{\hsize}{!}{\includegraphics{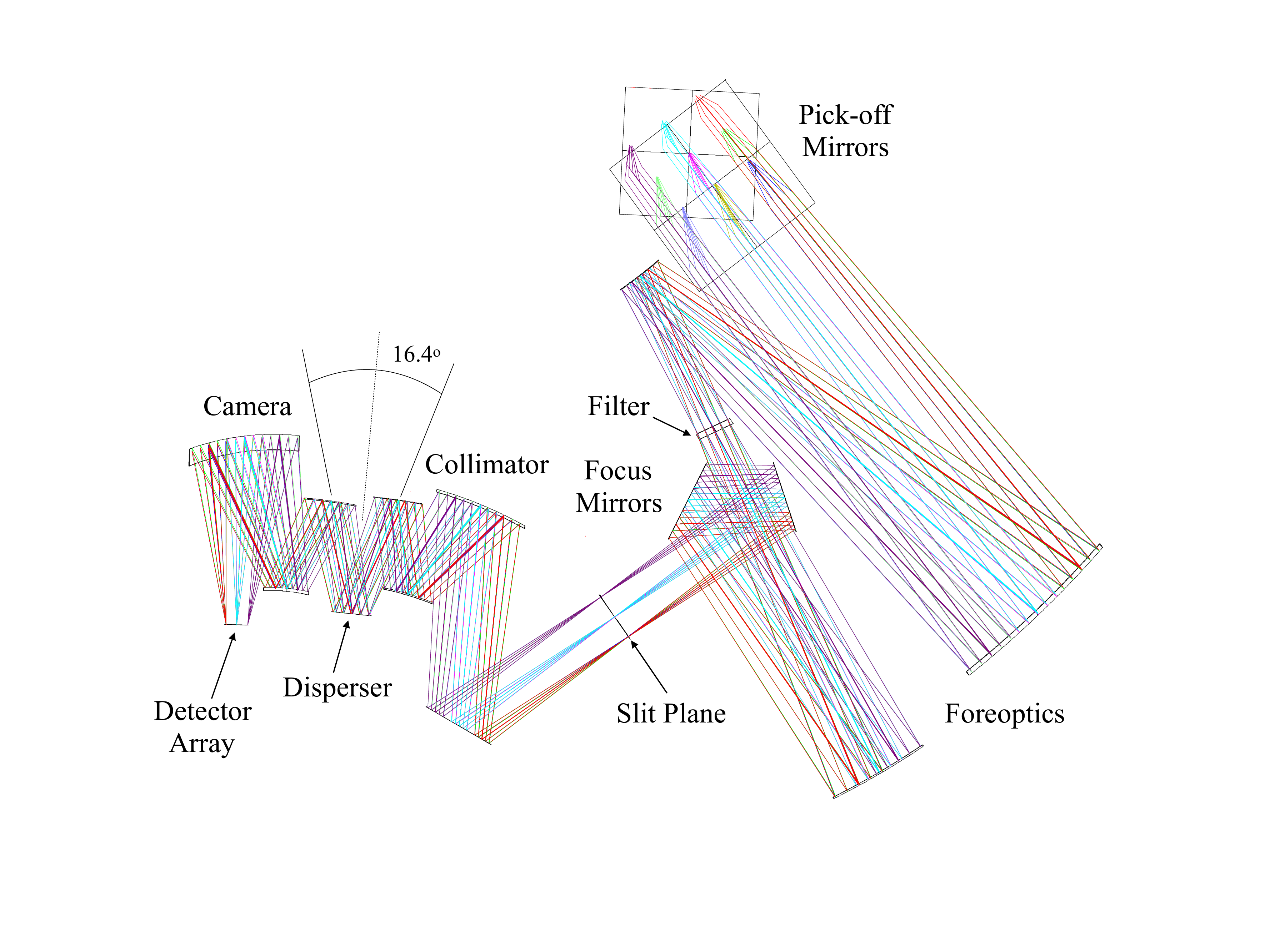}}
  \caption{Optical path through the NIRSpec instrument. The dispersion direction is out of the image.}
  \label{fig:path}
\end{figure}

The optical train of NIRSpec is shown in Fig.~\ref{fig:path}. The spectrograph design  is fairly conventional in concept, and consists of seven main components: The Foreoptics, the Filter Wheel, the Slit Plane, the Collimator Optics, the Grating Wheel, the Camera Optics and the Detector Array. An internal calibration light source completes the suite. These elements are all mounted on a common optical bench and preceded by a two-mirror periscope that steers the portion of the JWST focal surface sampled by its first pick-off mirror into the instrument. 

\begin{figure}
  \resizebox{\hsize}{!}{\includegraphics{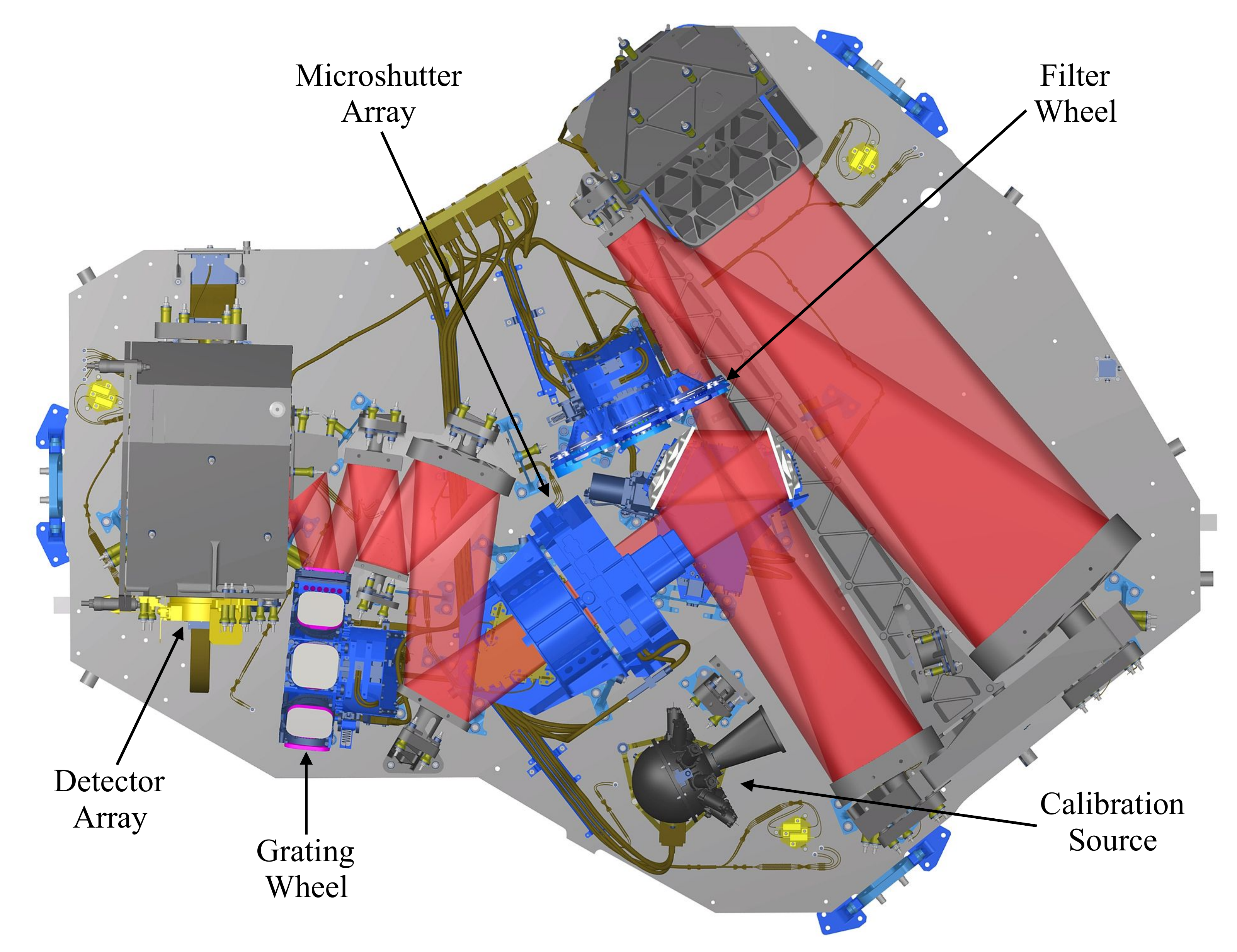}}
  \caption{CAD rendering of NIRSpec with its three major mechanisms and the calibration source identified.}
  \label{fig:cad}
\end{figure}

The light path is laid out along the plane of the optical bench (Fig.~\ref{fig:cad}), and the dispersion takes place  orthogonal to it (i.e., out of Figs.~\ref{fig:path} and \ref{fig:cad}). However, for reasons of convention, the Slit Plane (Sect.~\ref{sec:slitplane}) and the Detector Array (Sect.~\ref{sec:detector}) are throughout this paper depicted with the dispersion direction horizontal and the spatial direction vertical.


A cardinal feature of NIRSpec is the all-ceramic design of its structure and primary optics \citep{poss04,tepl05,bagn07}. The optical bench, all powered mirrors and their support structures are manufactured in Silicon-Carbide \citep[SiC,][]{brey12}. This assures that the entire instrument contracts in a nearly homologous manner during cool-down, thereby facilitating assembly and optical alignment at ambient. This choice of material is also highly polishable, and contributes to the overall instrument stability on orbit due to its high thermal conductivity and very low thermal expansion at cryogenic temperatures.

NIRSpec is passively cooled, as are the telescope and the other JWST instruments, and is held at its nominal operational temperature of {$\sim\!37$}\!~K by means of trim heaters fixed to the thermal strap connecting the optical bench with its dedicated instrument radiator. In order to reach this low a temperature, the instrument's three electronic boxes are all situated on a separate warmer bench well away from the instrument proper. The four moving mechanisms of NIRSpec are for the same reason extremely low power and only dissipate a few tens of mW when operated. The NIRSpec detector system (Sect.~\ref{sec:detector}) has its own separate radiator and thermal control system. A multilayer insulator enclosure fits over the entire optical bench to protect the optics from contamination and render the instrument light-tight. At a mass of 196\!~kg and a volume of 1.9\!~m $\times$ 1.4\!~m $\times$ 0.7\!~m NIRSpec is the largest of the four instruments on JWST.

The NIRSpec optical train is reflective throughout, except for the order-separation filters  and the low resolution dispersive prism (Sect.~\ref{sec:fwgw}). The three primary optical modules of NIRSpec are each implemented in the form of three-mirror anastigmats employing high-order aspherical surfaces \citep{geyl11}. Counting the five plane fold mirrors and the disperser, the light entering NIRSpec undergoes a total of 15 reflections before reaching the detector array. With the aim of achieving high optical throughput at the blue end, all mirrors are coated with protected silver. The IFU mode (Sect.~\ref{sec:ifu}) adds a further eight gold-coated reflections to the light path.

To complement the hardware, the NIRSpec team has developed and actively maintains a high fidelity instrument software model that allows the light path to be traced from the sky through the telescope and NIRSpec to the detector array in any observing mode to an accuracy of a fraction of a pixel \citep{dorn16}. The model has been tuned to closely match the as-built flight instrument during the NIRSpec ground calibration campaign \citep{birk16,giav16}, and forms the backbone of the NIRSpec team's observation planning and data reduction software \citep{boek12, alve18, pbs+2020}. However, the presently included mathematical description of the JWST telescope proper is of necessity the theoretical as-designed one. 

Throughout this paper, where appropriate, quoted sizes of key NIRSpec parameters of interest to the user will refer to angular sizes projected to the sky based on the latest instrument model, with the understanding that some of the quoted nominal values may need to be modified slightly once they are measured and verified on orbit with the properly phased and focused as-built telescope.

\subsection{Pick-off Mirrors and Foreoptics}\label{sec:fore}

\begin{figure}
  \resizebox{\hsize}{!}{\includegraphics[width=5cm]{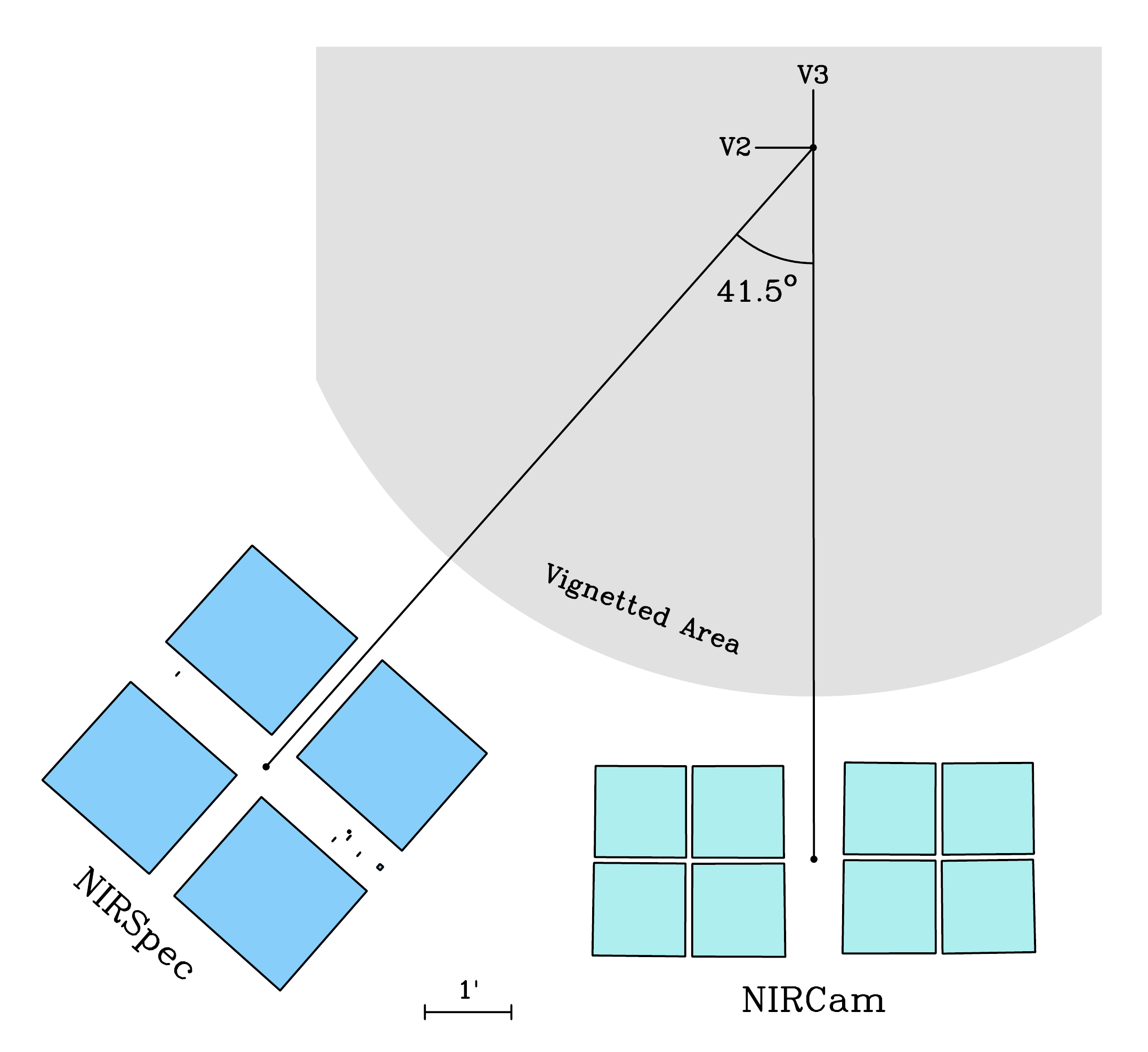}}
  \caption{Location and orientation of the NIRSpec field of view on the shared JWST focal sphere with respect to NIRCam and the telescope optical axis. The entrance apertures of the other two JWST instruments, MIRI and NIRISS, are situated to the right of and below NIRCam.}
  \label{fig:fov}
\end{figure}

The task of the NIRSpec Pick-off Mirrors and Foreoptics Module is to re-image a suitably demagnified section of the shared telescope focal sphere onto the Slit Plane containing three different choices for slit device (Sect.~\ref{sec:slitplane}). To ensure the best possible image quality, the NIRSpec field of view is rotated by $41.5^\circ$ with respect to the neighboring NIRCam instrument so that its midline along the dispersion direction intersects the telescope optical axis and origin of the spacecraft V2V3 coordinate system (Fig.\ref{fig:fov}). This makes the telescope focal sphere curvature symmetrical over the NIRSpec field of view. The Scheimpflug principle is then employed to image the best-fit tilted plane to the curved telescope focal surface onto the flat Slit Plane. The Foreoptics re-imaging takes place in a telecentric manner, such that the chief rays of the refocused f/12.5 beams from all points in the NIRSpec field of view are perpendicular to the  Slit Plane, thereby minimizing any field-dependent illumination variations in the slit transmission across the NIRSpec field of view, which is set by a rectangular Field Stop measuring 3\farcm6$\times$3\farcm4 located at the telescope focus inside the pick-off mirror periscope. The telecentric nature of the Foreoptics system also renders the image scale in the Slit Plane insensitive to changes in the telescope focus.

The Foreoptics module contains an eight-position filter wheel carrying five long-pass order-separation filters, and two finite-band target acquisition filters (Sect.~\ref{sec:fwgw}). An eighth opaque position serves as the instrument shutter, whose mirrored backside reflects light from the internal Calibration Assembly into the instrument (Sect.~\ref{sec:cal}). The 70~mm diameter filters are inserted in the $\simeq$56~mm beam at the Foreoptics pupil location, but are tilted by $5\fdg3$ in the spatial direction in order to avoid ghosting. Their use is further discussed in Sect.~\ref{sec:fwgw} below.
Following the Filter Wheel is a flat two-mirror focusing mechanism \citep{tacc08} to assure confocality with the other JWST instruments.

The wavefront error of the NIRSpec image in the focus of the Foreoptics module is specified to be below 185~nm rms, of which 150~nm stems from the telescope  \citep{light14}. It follows that the PSF in the NIRSpec Slit Plane will be diffraction-limited (Strehl ratio >0.8) at wavelengths above 2.46\!~$\mu$m (cf. Fig.~\ref{fig:psf_chain} below). The FWHM of the PSF in the Slit Plane at a wavelength of 2.5\!~$\mu$m is therefore 80~mas, which is comparable to the spatial resolution of the HST cameras at visible wavelengths. The focal length of the Foreoptics module was therefore chosen to provide a baseline NIRSpec slit width of $\simeq200$~mas in order to match the typical $\simeq$100-300~mas half-light radii of the highest redshift galaxies detected with HST \citep[e.g.,][]{bouw04}.  This choice of slit width was early on judged to be a suitable compromise between degrading the achievable faint-end sensitivity by making the slit too narrow, and at the other extreme degrading the signal-to-noise ratio by letting in too much sky background -- while keeping in mind that no single choice of slit width can be optimal over the more than three octaves in wavelength covered by NIRSpec. The requirement that the baseline slit width project to two pixels on the detector (Sect.~\ref{sec:detector}) in turn determines the focal lengths of the Collimator and Camera modules (Sects. \ref{sec:col} \& \ref{sec:cam}). This minimal sampling of the slit width is driven by the fact that NIRSpec is detector noise-limited in nearly all its modes (Sect.~\ref{sec:perf}, Appendix~A), which necessitates that the light in a spectral resolution element fall on as few detector pixels as possible. This choice of slit sampling also sets  the highest spectral resolution $R\simeq 2700$ for which one octave of spectrum will fit across two adjacent 2048$\times$2048 pixel detector arrays (Sect.~\ref{sec:detector}). The price to pay for a large pixel size of $\simeq$100~mas, however, is that the PSF is under-sampled in the spatial direction along the slit at all wavelengths within NIRSpec's operational range (Sect.~\ref{sec:detector}). 

The telescope and NIRSpec Foreoptics combination displays significant keystone-type optical field distortion, with deviations from the best-fit linear paraxial model predicted to reach $\simeq$1\farcs2 at the corners of the NIRSpec field of view. Since this level of target displacement significantly exceeds the 200~mas width of the baseline slit width, its accurate mathematical description, together with knowlege of the as-built metrology of the components of the NIRSpec slit plane, lie at the heart of the NIRSpec instrument model and observation preparation software (Paper~II). In fact, multiobject observations with NIRSpec will only be possible once the field distortion in the NIRSpec slit plane has been carefully measured and characterized in orbit during commissioning of the JWST observatory \citep{boek16}.

\subsection{Slit Plane}
\label{sec:slitplane}

The layout of the NIRSpec Slit Plane is shown in Fig.~\ref{fig:msa}. It supports three distinct observing modes: Multiobject Spectroscopy employing the Micro-Shutter Array (MSA), single object long slit spectroscopy employing one of five Fixed Slits, and spatially resolved two-dimensional spectroscopy over a 3\farcs1$\times$3\farcs2 field of view employing the Integral Field Unit (IFU).

\subsubsection{Micro-Shutter Array}\label{sec:msa}

The MSA is the most novel and complex of NIRSpec's subsystems. It was designed and procured by Goddard Space Flight Center as one of NASA's two hardware contributions to the instrument \citep{mose04,kuty08}. The MSA consists of four quadrants of  $365\times171$ individually addressable shutters, separated by 23\arcsec \ in the horizontal (dispersion) direction and 37\arcsec \ in the vertical (spatial) direction. The arrays were manufactured employing advanced MEMS technology. Briefly described, they consist of a 100\!~$\mu$m thick single crystal silicon honeycomb structure with a silicon nitride shutter door hinged to the top of each cell by  a flexible torsion bar. The shutter doors are lithographically coated with a mildly magnetic layer of molybdenum nitrite such that they can be pulled into their cells by means of a movable permanent magnet that is swept across the array along the dispersion direction. Shutters that are commanded to be held open are latched electrostatically to the inside wall of their cells by applying a voltage to the appropriate cell row and column electrodes deposited on the tops of the cell walls. 
The MoN coating also serves to make the doors opaque, and is carefully designed to assure that they are kept mechanically flat at the MSA cryogenic operating temperature.

\begin{figure*}
  \centering{\includegraphics[width=18cm]{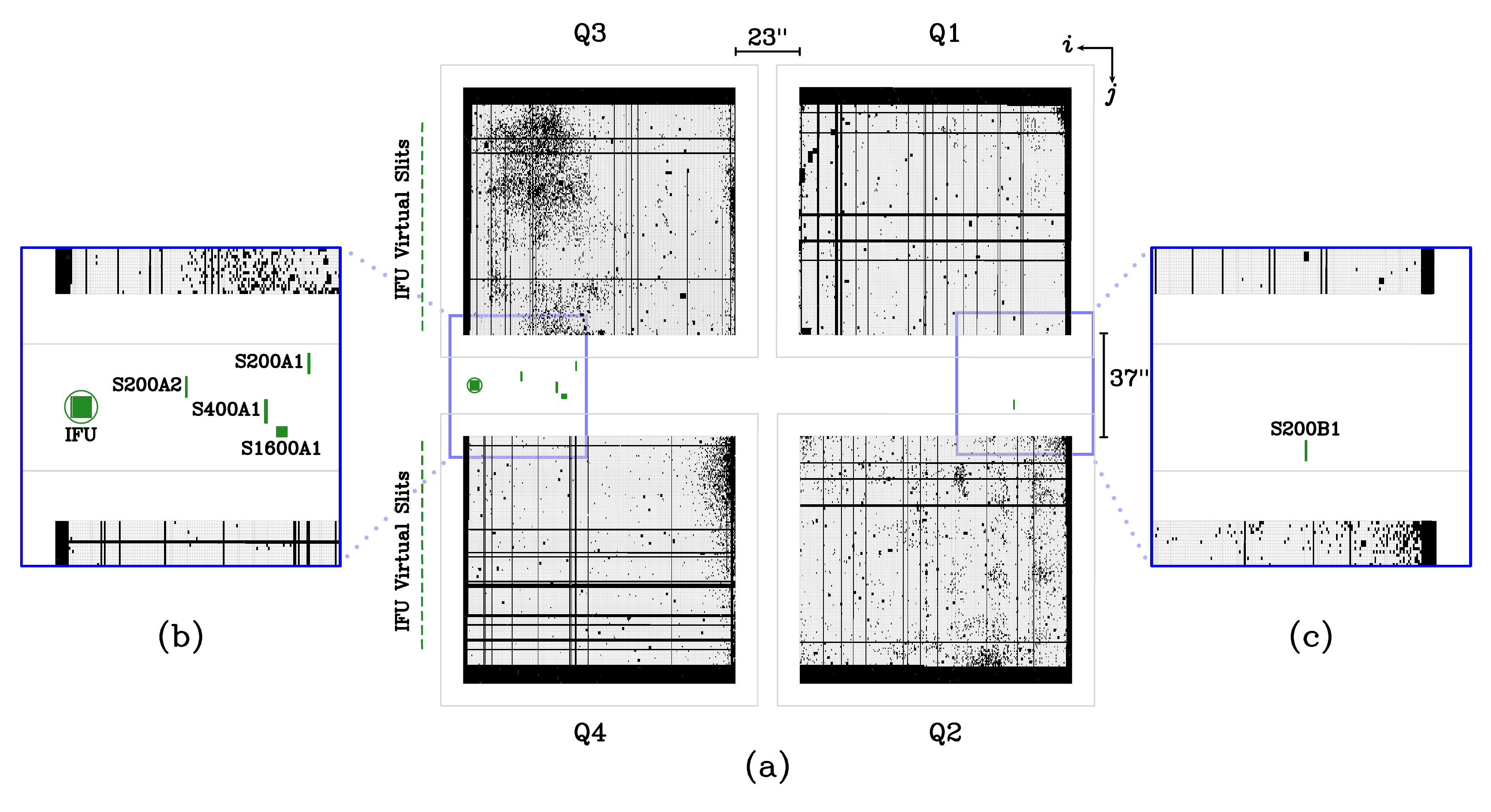}}
  \caption{Layout of the NIRSpec Slit Plane. (a) The four quadrants of the Microshutter Array, the five Fixed Slits, the entrance aperture of the Integral Field Unit, and the locations of its 30 virtual slit images. The dispersion direction is horizontal. Black areas in the MSA depict inoperable vignetted or permanently closed shutters. (b) Close-up of the left-most group of four Fixed Slits and the IFU entrance aperture. (c) Close-up of the right-most redundant Fixed Slit.}
  \label{fig:msa}
\end{figure*}

The arrays have a shutter pitch of 105\!~$\mu$m in the dispersion direction and 204\!~$\mu$m in the perpendicular spatial direction. Aluminum light shields deposited on the top of each cell cover the gap surrounding the shutter doors, yielding an open area measuring 78\!~$\mu$m $\times$ 178\!~$\mu$m, corresponding to a geometrical efficiency of 65\%. At the nominal platescale of the telescope and NIRSpec Foreoptics, the mean angular extent of the shutter open area on the sky is 199~mas $\times$ 461~mas. However, due to optical distortion and small metrology differences between the four MSA quadrants, these projected open sizes vary across the MSA field of view by $2.9$\% in width and $3.9$\% in height, such that the solid angle subtended by the open area of an MSA shutter varies by $6.8$\%. 

Individual shutters are designated $(k,i,j)$ where $k=1,4$ is the quadrant number, $i=1,365$ the shutter column and $j=1,171$ the shutter row. In the view depicted in Fig.~\ref{fig:msa}, $i$ and $j$ are measured from origins in the upper right corners of each quadrant, increasing to the left and downward, respectively. 
The shutter rows are mutually well-aligned in the critical spatial $j$ direction to within 5\% of a shutter height, and the four quadrant rows are clocked to the horizontal pixel rows of the NIRSpec detector array (Sect.~\ref{sec:detector}) to an accuracy of 39\arcsec or better. These residual misalignments nonetheless need to be taken into account when planning MSA observations, and are -- along with the relevant optical field distortions (Sects.~\ref{sec:fore}, \ref{sec:col} \& \ref{sec:cam}) -- captured in the NIRSpec instrument model \citep{dorn16}.

The MSA is not infinitely opaque -- typically one part in 5\,000 or less of the light incident upon a closed shutter can make its way through the shutter walls or around the light shield. The finite contrast of the MSA gives rise to spurious spectra from bright sources in the MSA field of view, as well as the build-up of a pervasive low level background signal due to sky background light leaking through the closed MSA shutters. Both these effects need to be taken into consideration in many applications \citep[Papers~II \& III]{desh18, alve18}.

As is evident from Fig.~\ref{fig:msa}, not all of the MSA's {$4 \times 365 \times 171 = 249\,660$} shutters are operational and available for use. In order to ease the optical alignment, the MSA is deliberately slightly oversized. Consequently, {$\simeq\!12$} shutter rows at the top and bottom and {$\simeq\!7$} columns along the outer sides of the four arrays are vignetted by the 3\farcm6$\times$3\farcm4 Field Stop at the instrument entrance in the telescope focal surface. Furthermore, a significant number of shutters are nonfunctional for other reasons \citep{rawl18}. Unintended electrical shorts occurring between the latching electrodes render entire rows and columns of shutters inoperable. These are therefore left in their default closed states. Individual shutters can also fail in both permanently closed and  open positions. A failed closed shutter is a nuisance in that an object of interest that happens to fall at its location in the NIRSpec Slit Plane cannot be observed at that particular pointing of the telescope. A failed open shutter on the other hand is a more serious matter since it will give rise to an always-present dispersed spectrum of the sky background and any sources that happen to be present within it, which takes up precious space on the NIRSpec detector array. For this reason during the screening of the MSA quadrants, identified failed open shutters were intentionally converted into failed closed ones by gluing a $3\times3$ shutter sized opaque patch over them. At the time of writing there are a total of nineteen failed open shutters, and a total of 32\,076 (14\%) inoperable or failed closed shutters present in the four unvignetted quadrants of the MSA. Although the MSA quadrant arrays were successfully subjected to accelerated life-testing during their manufacture, it is presently uncertain how the above shutter operability statistics will change after launch, let alone evolve during long-term scientific use under operational conditions in orbit. 

This state of affairs very much reflects the sheer novelty of the MSA device and the challenges encountered during its development. The manufacture of MSA quadrants was terminated for budgetary reasons in 2007, at a point where just over a dozen flight candidate quadrant arrays had been produced. Unfortunately the attrition rate of the arrays during their subsequent testing and assembly turned out to be considerably higher than had been anticipated when the decision to end the production run was made. As a result, candidate arrays initially considered to be not flight-worthy eventually had to be reconsidered for use. The situation was also not helped by the original integrated MSA mechanism needing to be exchanged with a replacement unit due the original flight device having degraded significantly during acoustic testing of the assembled instrument \citep{tepl16b,birk16}. The MSA quadrant manufacturing process was restarted in late 2012, but only one quadrant array (Q4) was produced in time to be incorporated in the replacement flight unit.

While the MSA quadrants depicted in Fig.~\ref{fig:msa} cannot be said to be impressive in cosmetic terms, it is worth stressing that the  195\,593 operational shutters in Fig.~\ref{fig:msa} span a total solid angle on the sky of 7.7~arcmin$^2$, which falls only 14\%  short of that available with a hypothetical perfect device. This allows for a considerable multiobject spectroscopic capability. The dominant bottle-neck determining the ultimate multiplexing capabilities of NIRSpec is not so much the MSA, but rather the finite area of the NIRSpec detector array and the number of spectra that it can accommodate without overlap. The multiobject capabilities of NIRSpec are discussed in further detail in Paper II.

\subsubsection{Fixed Slits}\label{sec:slits}

The NIRSpec slit plane also carries a total of five fixed slits. These fixed slits are cut into the 16\arcsec \ tall titanium plate separating the MSA quadrants Q1/Q3 and Q2/Q4 in the slit plane (Fig.~\ref{fig:msa}). The fixed slits serve a number of applications, and collectively provide a failsafe single-object spectroscopic observing mode should the MSA become inoperable in orbit.
  
An important aspect of the NIRSpec slit plane layout is that the spectra from shutters in the two topmost MSA quadrants Q1 and Q3 in Fig.~\ref{fig:msa} project solely to the topmost 43\%  of the pixel rows of the detector array, while the spectra from quadrants Q2 and Q4 occupy the lower 43\% rows of the array, thereby reserving the central rows of the detector for the exclusive use of the fixed slits. The five fixed slits are spaced in the spatial direction so that their dispersed spectra do not overlap, and remain open and functional in all operating modes of the instrument. The fixed slits also allow brighter targets to be observed without saturating the detector by only reading out the relevant rows of the arrays containing the dispersed image of the Fixed Slit employed in a rapid window (subarray) mode (Sect.~\ref{sec:detector}; Paper~IV).

The anticipated dimensions of the fixed slits projected to the sky are listed in Table~\ref{tab:slits}. The two workhorse slits are S200A1 and S200A2, which are offset in the dispersion direction by 19\farcs6  in order to bridge the 15\farcs9  gap between the two arrays of the NIRSpec detector in the highest resolution $R\!=\!2700$ modes (Sect.~\ref{sec:detector}). The matching S200B1 slit is included for redundancy to assure that full spectra can still be obtained in all modes in the event of a failure of the bluemost detector array.

The three S200 slits are primarily intended for kinematic studies of single extended objects in $R\!=\!2700$ mode. Compared to the MSA, they provide better contrast and a slit that is not interrupted by the MSA cell bars along the spatial direction. However, during the re-testing of NIRSpec that took place after the late changeout of the MSA and detector units, it was discovered that the fixed slits in the replacement MSA unit had not been cut sufficiently accurately and that the transmissions of the S200 slits vary by up to $\simeq$20\% along their lengths. This unwanted complication will need to be taken into account in the reduction of NIRSpec fixed slit spectra.

The lone wider S400A1 slit allows higher slit transmission at the expense of spectral resolution, and is intended to facilitate spectrophotometric calibrations. In the same vein, the large 1\farcs6$\times$1\farcs6 square slit was introduced in late 2008 to enable observations of transiting exoplanets, and has a size intended to keep variations in the slit transmission caused by pointing jitter and drift from masking the minute signal from the exoplanet atmosphere in the spectrum of its host star (Paper IV).

   \begin{table}
      \caption{NIRSpec fixed slits}
       \centering
         \label{tab:slits}
         \begin{tabular}{l c c}
            \hline\hline
            \noalign{\smallskip}
            \hfil Slit&  {Width}&  {Height}\\
              &  [mas]&  {[mas]}\\
            \noalign{\smallskip}
            \hline
            \noalign{\smallskip}
           S200A1& 193& 3274 \\
           S200A2& 194& 3303 \\
           S400A1& 395& 3759 \\
           S1600A1& 1603& 1599 \\
           S200B1& 201& 3301 \\
            \noalign{\smallskip}
            \hline
         \end{tabular}
   \end{table}

\subsubsection{Integral Field Unit}\label{sec:ifu}

The third slit device in NIRSpec is the Integral Field Unit (IFU), which provides the spectrum of a small contiguous 3\farcs1$\times$3\farcs2 area of the sky in the form of 30 separate long slit spectra of 103~mas wide and 3195~mas tall slices through the scene. The elaborate optics of the IFU are all-reflective and are manufactured entirely out of aluminum using advanced diamond turning techniques \citep{clos08,purl10}. 

The  IFU entrance aperture in Fig.~\ref{fig:msa} is normally blocked by a baffle attached to the MSA movable magnet when in its primary park position. 
When opened by moving the magnet slightly, a pickoff mirror sends the IFU field of view to two relay mirrors that magnify and re-image the field onto a reflective image slicer having 30 facets. Each reflected slice is then individually re-imaged and re-arranged into two spatial columns of 15 vertical slits in the slit plane by means of two sets of 30 individually shaped pupil and slit mirrors. The IFU re-imaging is anamorphic with the virtual slits magnified by a factor of two in the dispersion direction such that their 103~mas widths are sampled by two pixels on the NIRSpec detector. With the help of two additional flat folding mirrors, the entire IFU is packaged into a single compact unit measuring 140~mm $\times$ 71~mm $\times$ 204~mm that is mounted to the base plate of the MSA assembly. The eight reflective surfaces of the IFU are all gold coated, resulting in a net transmissive throughput of 60-90\% for the unit, depending on the wavelength.

An important feature of the IFU design is that the dispersed images of the virtual slits share the NIRSpec detector array with the MSA. In a manner mirroring how the two groups of two MSA quadrants are arranged, the two columns of 15 virtual slits are separated in the spatial direction so as to avoid the central pixel rows exclusively reserved for the Fixed Slits (Sect.~\ref{sec:slits}). The MSA in its all-closed configuration therefore acts as a light shield for the IFU. The dispersed image of any light making its way through the MSA due to its finite contrast or permanently failed-open shutters is therefore superimposed on the IFU spectra \citep{desh18}. The  capabilities of the NIRSpec IFU are further discussed in Paper~III.

\subsection{Collimator}
\label{sec:col}

The task of the Collimator optics is to convert the light exiting from any point in the Slit Plane into a $\simeq$56~mm diameter parallel beam aimed at the flat disperser situated in the second pupil plane of the instrument between the Collimator and the Camera.  The Collimator, the Dispersers and Camera Optics that follow the Slit Plane, are all oversized with the intent of limiting the light loss caused by the unavoidable angular broadening of the nominally f/12.5 exit beams emerging from the slits due to diffraction.
The Collimator focal length differs by 4\% between the spatial and dispersion directions, such that the 3\farcm6 $\times$ 3\farcm4 angular extent of the NIRSpec field of view on the sky is mapped to an angular beam spread of $\pm$3\fdg8 in the dispersion direction and $\pm$3\fdg4 in the spatial direction. 

In addition to its anamorphic magnification, the Collimator also displays significant keystone-type field distortion. This distortion complicates the mapping between the location of a given slit in the Slit Plane and the location of its dispersed spectrum in the detector focal plane.  The collimated beams from shutters situated along a given MSA row exit the Collimator at an angle that varies  by 0\farcm8 - 1\farcm3 in the spatial direction along the row. As a consequence, the spectra from MSA shutters located toward the middle of the MSA are imaged $\simeq4-6$ pixel rows higher on the detector relative to spectra from shutters near the two edges. Residual clocking errors in the mountings of the NIRSpec dispersers and the small misalignments among the MSA quadrants further complicate the nonlinear mapping between shutter and spectra locations, and make it disperser dependent. These mappings are taken into account by the MSA observation planning software in deciding whether any two candidate targets can be observed simultaneously without their spectra colliding on the detector \citep{kara14,pbs+2020}. Also here is the high fidelity NIRSpec instrument modul central to the task \citep{dorn16}.

\subsection{Filter and Grating Wheels}
\label{sec:fwgw}

The NIRSpec Grating Wheel carries a total of seven dispersive elements, any of which can be used in combination with any of the Fixed Slits, the MSA or the IFU described in Sect.~\ref{sec:slitplane}. The eighth position carries a flat nondispersive mirror that is used for Target Acquisition (Sect.~\ref{sec:ta}). Although the NIRSpec Filter Wheel for reasons of straylight control is physically located in the Foreoptics module (Sect.~\ref{sec:fore}), the seven filters it carries (Table~\ref{tab:filters}) are all designed for use with specific elements on the Grating Wheel. 

 \begin{table}
    \caption{NIRSpec filters}
     \centering
       \label{tab:filters}
       \begin{tabular}{c c c c}
          \hline\hline
          \noalign{\smallskip}
          Filter&  Bandpass&  {Substrate}& {Primary Use}\\
                &  [$\mu$m]&  & \\
          \noalign{\smallskip}
          \hline
          \noalign{\smallskip}
         CLEAR& >0.6&   CaF$_2$& PRISM \& TA\\
         F070LP& >0.7&   CaF$_2$& G140M \& G140H\\
         F100LP& >1.0&   CaF$_2$& G140M \& G140H\\
         F170LP& >1.7&   CaF$_2$& G235M \& G235H\\
         F290LP& >2.9&   CaF$_2$& G395M \& G395H\\
         F140X& 0.8 - 2.0& BK7& Target Acquisition\\
         F110W& 1.0 - 1.3& BK7& Target Acquisition\\
          \noalign{\smallskip}
          \hline
       \end{tabular}
 \end{table}

   \begin{table}
      \caption{NIRSpec gratings}
       \centering
         \label{tab:gratings}
         \begin{tabular}{c c c c}
            \hline\hline
            \noalign{\smallskip}
           Disperser&  Groove Density&  Blaze Angle& Blaze $\lambda$\\
           &  [mm$^{-1}$]&  [deg]& [$\mu$m]\\
            \noalign{\smallskip}
            \hline
            \noalign{\smallskip}
 G140M& 95.40& 3.5& 1.3\\
 G235M& 56.88& 3.7& 2.2\\
 G395M& 33.82& 3.7& 3.7\\
 G140H& 252.29& 9.5& 1.3\\
 G235H& 150.29& 9.7& 2.2\\
 G395H& 89.41& 9.9& 3.7\\
            \noalign{\smallskip}
            \hline
         \end{tabular}
   \end{table}

NIRSpec employs two sets of three diffraction gratings providing spectral resolutions of  $R{\simeq}1000$ and $R{\simeq}2700$, respectively. The two sets of gratings are formally specified to cover the 1.0-5.0\!~$\mu$m wavelength region in each of three overlapping bands: 1.0-1.8\!~$\mu$m (Band~I); 1.7-3.0\!~$\mu$m (Band~II) and 2.9-5.0\!~$\mu$m (Band~III), with the blaze wavelength of each grating set to the harmonic mean of the above band endpoints to balance the blaze function across the band. These three nominal bands are common to the $R\!=\!1000$ and $R\!=\!2700$ gratings, whose properties are listed in Table~\ref{tab:gratings}. The gratings measure 70~mm $\times$ 70~mm in active area and were manufactured by Carl Zeiss Optronics by mechanically ruling triangular grooves having the desired blaze angle directly onto a gold layer deposited on a Zerodur substrate. The resulting gratings all have very high peak efficiencies (>80\%) and low straylight. Their dispersion directions are clocked to coincide with the pixel rows of the detector array (Sect.~\ref{sec:detector}) to an accuracy of 5\farcm9 or better.

The six gratings are all used in the inside (negative) first order. As is the case for most wide field multiobject spectrographs, the NIRSpec detector array is sufficiently wide in the dispersion direction that both second-order spectra and  zero-order images will occasionally also be registered in exposures employing the $R\!=\!1000$ gratings (Sect.~\ref{sec:wcoverage}). In particular, since the presence of in-band second-order spectra appearing as extensions at the red ends of all dispersed first order grating spectra is unavoidable, the decision was made early on to employ long-pass filters rather than finite-band filters for order separation in NIRSpec, on the grounds that the former tend to have both higher transmission and fewer and shallower transmission gradients compared to corresponding finite-band filters. 

Table~\ref{tab:filters} lists the four long-pass (LP) filters carried on the NIRSpec filter wheel. They are all interference filters with multilayered TiO$_2$ and SiO$_2$ coatings deposited on CaF$_2$ substrates. The F100LP, F170LP and F290LP filters are common to the two grating sets, and are respectively designed for use with the G140M/G140H (Band~I), G235M/G235H (Band~II) and G395M/G395H (Band~III) gratings.
Spectra taken in these combinations will be free of second-order contamination over exactly one octave of wavelength, starting at the cut-on wavelength of the filter being employed and up to twice that wavelength, beyond which the red extensions of the primary first order spectra will be superposed by in-band second-order spectra at twice the first order dispersion. Depending on the slit option being used, the registered uncontaminated portions of the Band I, II and III $R\!=\!1000$ grating spectra will often extend a good deal further to the red than indicated by their official bandpasses (Sect.~\ref{sec:wcoverage}), albeit at reduced blaze efficiency. 

The price to be paid for employing long-pass order separation filters is that the zero order images that fall on the detector with the $R\!=\!1000$ gratings will be quite strong due to their containing the poorly-attenuated integrated red light from the source out to the $\simeq$5.3\!~$\mu$m red cut-off wavelength of the detector (Sect.~\ref{sec:detector}). The intensity of the zero order images therefore diminishes going from Band~I to Band~III.

Because of the presence of these zero- and second-order spectra, all NIRSpec  
grating spectra effectively occupy the entire width of the NIRSpec detector (Fig.~\ref{fig:spectrace}). This limits how many spectra can be accommodated on the detector without overlap, and thereby places a strong constraint on the level of multiplexing achievable when using the MSA in any grating mode (Paper~II). 

The shorter wavelength F070LP long-pass filter listed in Table~\ref{tab:filters} is included to enable observations below 1\!~$\mu$m with the G140M and G140H gratings, which will then be free of contaminating second-order light only up to a wavelength of $\simeq$1.3\!~$\mu$m. This observing mode was included to enable $R\!=\!1000$ and $R\!=\!2700$ observations of Ly$\alpha$ absorption and emission at redshifts down to {$z\simeq 4.8$}. Without this mode NIRSpec would only be capable of studying the Ly$\alpha$ line at these spectral resolutions at redshifts {$z>7.3$}, and thereby run the risk of unintentionally leapfrogging an important portion of the epoch of reionization \citep[e.g.,][]{daya18}.  

\begin{figure}
  \resizebox{\hsize}{!}{\includegraphics{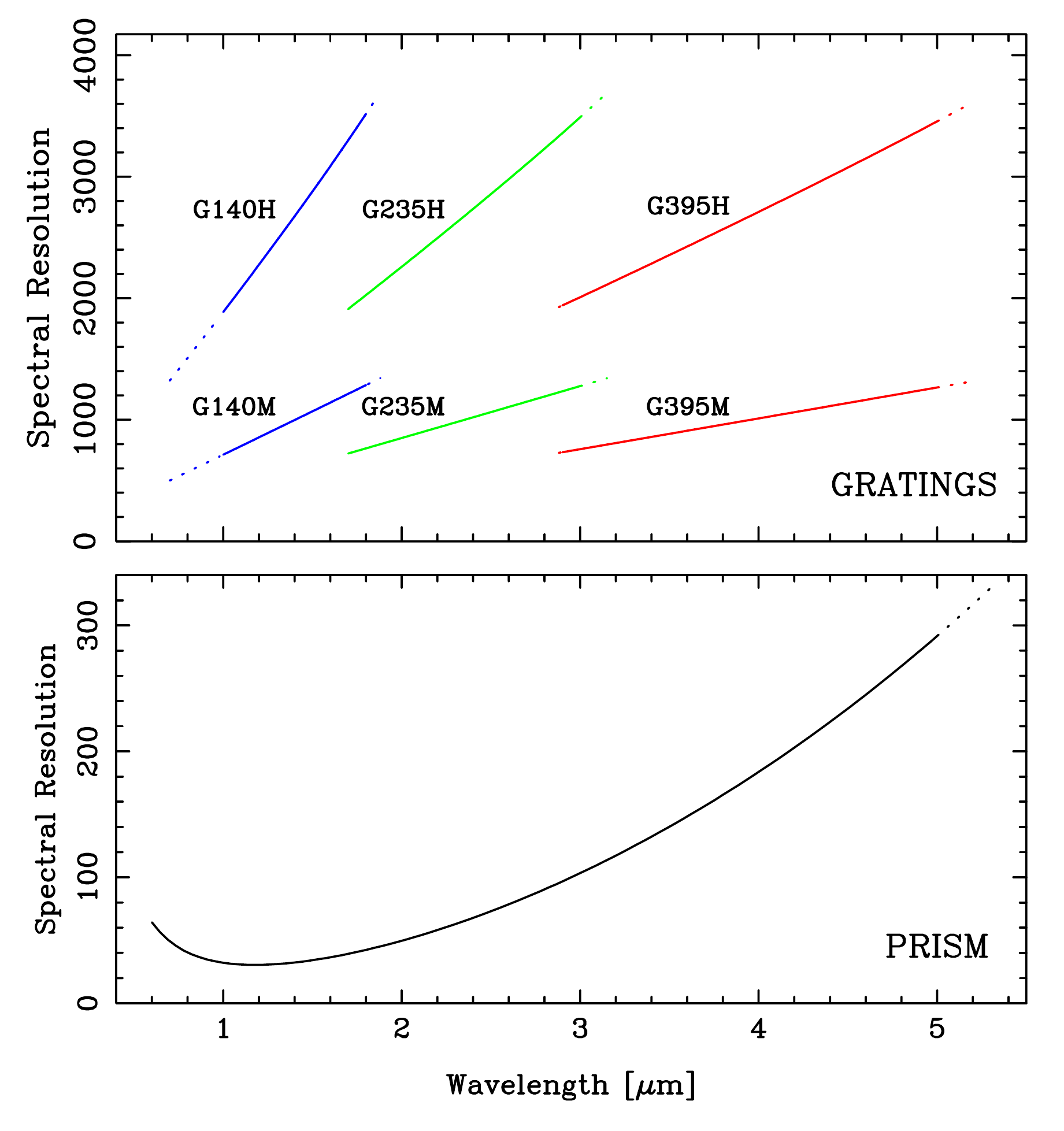}}
  \caption{Wavelength ranges and spectral resolutions achievable with the seven dispersers of NIRSpec. The nominal wavelength range covered by each disperser is shown as a full line, and the anticipated actual usable range as a dotted lines, including the blue extensions of the G140M and G140H gratings when used with the F070LP filter.
}
\label{fig:res}
\end{figure}

The entry labeled CLEAR in Table~\ref{tab:filters} refers to an uncoated CaF$_2$ filter with an effective cut-on wavelength of $\simeq$0.6\!~$\mu$m set by the gold coatings of the telescope mirrors. This filter is primarily intended for use with the lowest resolution $R\!=\!100$ disperser carried by the Grating Wheel. The double-pass prism consists of an 80~mm $\times$ 80~mm, 16\fdg5 wedge of CaF$_2$ with a protected silver reflective coating deposited on its backside. It produces low resolution {$R\!\simeq\!100$} spectra over the complete 0.6-5.3\!~$\mu$m range of the instrument. The index of refraction of CaF$_2$ varies strongly with wavelength in the near-IR \citep{trop95}. As a consequence, the rather uneven dispersion and spectral resolution of the prism spectra ranges from a low of {$R\!\simeq\!31$} near 1.1\!~$\mu$m and steadily rises to a high of {$R\!\simeq\!330$} at 5.3\!~$\mu$m (Fig.~\ref{fig:res}). In contrast to the grating spectra, the prism spectra are free of multiple orders and are only between 390 and 420~pixels long. The width of the MSA allows up to four nonoverlapping prism spectra to be fit side-by-side on the NIRSpec detector, thereby enabling a correspondingly higher level of multiplexing to be achieved (Fig.~\ref{fig:spectrace}; Paper~II).

   \begin{table}
            \caption{Baseline disperser/filter combinations}
      \centering 
         \label{tab:modes}
         \begin{tabular}{c c c c}
            \hline\hline
            \noalign{\smallskip}
           Disperser/Filter&  $\lambda$ range&  $R$& Length\\
           &  [$\mu$m]&  & [pixels]\\
            \noalign{\smallskip}
            \hline
            \noalign{\smallskip}
 PRISM/CLEAR&  0.6 - 5.3& 30 - 330& 411\\ 
 G140M/F070LP& 0.70 - 1.26& 500 - 898& 878\\
 G140M/F100LP& 0.98 - 1.88& 699 - 1343& 1411\\
 G235M/F170LP& 1.70 - 3.15& 722 - 1342& 1356\\
 G395M/F290LP& 2.88 - 5.20& 728 - 1317& 1290\\
 G140H/F070LP& 0.70 - 1.26& 1321 - 2395& 2327\\
 G140H/F100LP& 0.98 - 1.87& 1849 - 3675& 3753\\
 G235H/F170LP& 1.70 - 3.15& 1911 - 3690& 3646\\
 G395H/F290LP& 2.88 - 5.20 & 1927 - 3613& 3467\\
            \noalign{\smallskip}
            \hline
         \end{tabular}
   \end{table}

The nine disperser and filter combinations that make up the baseline spectroscopic modes of NIRSpec are summarized in Table~\ref{tab:modes}, which also lists their anticipated actual extended uncontaminated wavelength range -- which, however, is not always captured by the detector in all slit options (Sect.~\ref{sec:wcoverage}).  Since the longpass filters are not perfect step functions in wavelength, the blue ends of the spectra are not perfectly sharp, nor does the second order contamination at the red end set in abruptly. 
The usable wavelength ranges listed in Table~\ref{tab:modes} are at their blue ends defined as the wavelength at which the total instrument throughput of the filter and disperser combination (Sect.~\ref{sec:throughput}) reaches 50\% of its peak value, and at their red ends as the wavelength at which the throughput of the contaminating second order light reaches 0.1\% of the throughput of the underlying first order spectrum \citep{pf2016}. Table~\ref{tab:modes} also lists the spectral resolution achievable over the given wavelength ranges and the lengths of the spectra calculated for the S200A2 fixed slit nearest the middle of the NIRSpec field, and an assumed spectral resolution element spanning 2.2 detector pixels in the dispersion direction. These values are also displayed graphically in Fig.~\ref{fig:res}. Since the gratings project a nearly constant wavelength interval per pixel, the plotted $R(\lambda)$ curves for the gratings are nearly linear in wavelength and vary by a factor of two over an octave of spectrum. The spectral resolution at a given wavelength varies with slit location by $\pm$0.9\% in the $R\!=\!1000$ grating modes, $\pm$2.3\% in the $R\!=\!2700$ grating modes and $\pm$3.4\% in prism mode.

The last element of the Grating Wheel is a plane mirror, which provides an undispersed image of the NIRSpec field of view on the sky, albeit modulated by the opaque grid of the all-open MSA quadrants. This mode is intended for taking images of the field to be observed in one of the two finite band F140X or F110W filters or the CLEAR filter carried by the filter wheel for automatic processing by the onboard target acquisition software (Sect.~\ref{sec:ta}).

As depicted in Fig.~\ref{fig:cad}, the NIRSpec Grating Wheel takes the form of a carousel with the optical elements mounted around its periphery. The rotation occurs around an axis  parallel to the Optical Bench, which shifts the outgoing beams along the dispersion direction. A spring-loaded ratchet mechanism holds the wheel at each of its eight positions \citep{weid08}. The worst case angular repeatability of the mechanism is $\simeq$25\arcsec in both tip and tilt, corresponding to image displacements between Grating Wheel activations of up to $\simeq$4~pixels as measured on the detector. In order to achieve a higher level of accuracy in the wavelength calibration of NIRSpec spectra and during target acquisition, the Grating Wheel mechanism also includes two magneto-resistive position sensors that allow the precise orientation of the elements of the device in any of its positions to be determined to an accuracy of $\simeq$300~mas. Extensive testing carried out during the NIRSpec ground calibration campaign \citep{dema12} has verified that the readings of these devices allow NIRSpec image locations to be predicted to an accuracy of $\simeq$0.04 pixel or better, corresponding to $\simeq{1}/{50}$th of a spectral resolution element or $\simeq$4~mas on the sky. The readings from these sensors are therefore included in both the NIRSpec data reduction pipeline and the on-board target acquisition software \citep{alve18, keye18}.

As in any wide-field imaging spectrograph in which the light from different objects in the field of view illuminates the disperser at slightly different angles of incidence, the length and spectral resolution of NIRSpec spectra taken with any of its dispersers varies depending on which MSA shutter or slit option is used. Moreover, as indicated in Fig.~\ref{fig:path}, in order to achieve clearance between the incoming beam from the Collimator and the outgoing light coming off the dispersers, the Collimator output is tilted in bulk along the spatial direction parallel to the optical bench by an angle of $\gamma_0=$16\fdg4, such that the beams from different field points illuminate the dispersers at spatial incidence angles $\gamma=$16\fdg4$\pm$3\fdg6 away from the dispersion plane perpendicular to the grating grooves. Illuminating the gratings in this oblique $\gamma \ne 0$ manner gives rise to slit curvature, which manifests itself locally by the monochromatic image of the slit not being perpendicular to the dispersion direction \citep[cf.][]{schroed00}.  The amount of slit-tilt is significant in NIRSpec spectra and varies both with field position and along any given spectrum
(Fig.~\ref{fig:spectrace}). Tilt values range from 1\fdg4-3\fdg3 in $R\!=\!1000$ mode to 3\fdg6-7\fdg1 in $R\!=\!2700$ mode. An equivalent effect holds for the double-pass $R\!=\!100$ prism, but its dispersed spectra display a  slit-tilt in the range  5\fdg8-9\fdg6 that is in the opposite sense of that of the gratings. A worst-case slit-tilt of 9\fdg6 implies that the wavelength difference between the top and bottom detector rows of the $\simeq$32 pixel tall S200A slits in Table~\ref{tab:slits} are offset horizontally in wavelength by $\simeq$5.4~pixels or $\simeq$2.4 resolution elements. This systematic mismatch in wavelength between pixel rows needs to be taken into account in the data processing if the spectral resolution of the extracted spectra is not to be compromised \citep{boek12}.

\subsection{Camera}\label{sec:cam}

The task of the Camera Optics Module is to focus the dispersed parallel beams coming off the Grating Wheel onto a sequence of points on the detector array. The focal length of the Camera is matched to the preceding optics such that the 200~mas nominal slit width  projects to two pixels on the detector array.  The wide Camera field of view captures the large $\pm$7\fdg7 angular extent of the beams output by the gratings in the dispersion direction and $\pm$3\fdg6 in the spatial direction without vignetting. The imaging on the detector array is not telecentric; the converging f/5.7 beams deviate from the detector normal by up to $\pm$9\degr along the dispersion direction.

The Camera optics display a considerable amount of optical field distortion, again in the form of  anamorphic magnification overlaid by keystone distortion. The variation of the spatial distortion along the dispersion direction introduces a noticeable (14-24~pixel) upward curvature to the ends of all NIRspec spectra (Fig.~\ref{fig:spectrace}), whose dispersed images come off the plane dispersers perfectly straight in angular space. This optical distortion-induced spectral curvature is distinct from, and comes on top of, the slit-tilt discussed in Sect.~\ref{sec:fwgw}. Needing to take into account both these effects significantly complicates the NIRSpec data reduction process, which also here will rely on the NIRSpec instrument model  to rectify and wavelength-calibrate the extracted raw spectra \citep[Paper~II]{dorn16}.

The anamorphic magnification and keystone field distortion pattern of the Camera Optics is intentionally clocked in the opposite sense of that of the Collimator Optics. As a result, the net distortion in imaging mode with the Grating Wheel flat mirror in place is  substantially reduced, thereby easing the onboard target acquisition processing (Sect.~\ref{sec:ta}).

\subsection{Detector Array}\label{sec:detector}

The detector subsystem is NASA's second hardware contribution to NIRSpec.
It consists of two Teledyne 2048$\times$2048 H2RG arrays controlled and read by SIDECAR ASICs \citep{loos07}. The HgCdTe devices have a long wavelength cut-off at 5.3\!~$\mu$m, and are the same as those employed in the long wavelength channel of NIRCam and in NIRISS  on JWST \citep{rau14}. The detector subsystem is isolated thermally from the rest of NIRSpec and is held at its operating temperature of $\simeq$43\!~K to an accuracy of $\simeq$0.1\!~K by means of its own radiator and thermal control system. 

The original NIRSpec flight detectors suffered from degradation after instrument delivery and were exchanged with a superior improved set at the same time that the original flight MSA was replaced \citep{rau14,tepl16b}. The replacement arrays have excellent quantum efficiency, lower dark current, and much improved cosmetics compared to the original pair. 

The 18\!~$\mu$m $\times$ 18\!~$\mu$m pixels of the detector arrays project to an average extent on the sky of 103~mas in the dispersion direction and 105~mas in the spatial direction. The wavefront error of the NIRSpec PSF in the Camera focus is specified to be less than 238~nm rms. The image on the detector array is therefore   diffraction-limited (Strehl ratio > 0.8) only at wavelengths above 3.17\!~$\mu$m. The inferior image quality at shorter wavelengths is, however, masked by the large angular pixel size, which undersamples even the 170~mas FWHM of the PSF at the 5.3\!~$\mu$m long wavelength cut-off in response of the detectors (Fig.~\ref{fig:psf_chain}).

The four outer pixel rows and columns of each array are so-called reference pixels that do not respond to light. The two arrays, imaginatively named NRS1 and NRS2,  are mounted adjacent to each other in the dispersion direction leaving a 2.8~mm or 15\farcs9 horizontal gap between the last and first responsive columns of the two arrays. The presence of this gap results in portions of NIRSpec spectra being missed if they happen to project onto it (Sect.~\ref{sec:wcoverage}).

The pixels of the two NIRSpec arrays are read out sequentially using four simultaneous outputs at a fixed clocking rate of 10\!~$\mu$s per pixel, corresponding to full frame read time of 10.73\!~s. A user-selectable sequence of nondestructive capacitative reads of the accumulated electron contents of each pixel are carried out during each integration, and the registered photon rate of interest is determined in the up-the-ramp fashion with or without frame-averaging groups of subsequent reads \citep{raus07, raus10}. Aside from the averaging of grouped frames, NIRSpec ramps are not processed onboard, but are telemetered to the ground in their entirety. 

The bright-end photometric limit of NIRSpec is set by the detector full well capacity of ${\simeq}10^5$\!~e per pixel, with saturation effects setting in at about half this accumulated signal. In order to increase the dynamic range, rectangular subsets of the NIRSpec arrays can be read out in a much faster windowed mode allowing a much higher frame rate and shorter time between pixel resets. Windowed readout modes are the defaults for Fixed Slit and Target Acquisition images, and are crucial for NIRSpec spectroscopy of Exoplanets (Paper~IV).

A key performance parameter of vital importance  to the NIRSpec faint-end sensitivity, is the detector readout noise. As discussed in detail in Appendix~A, NIRSpec is at its faint end detector noise-limited in all grating modes and partially so in prism mode. The 1/f noise in the detector electronics leads to significant artifacts in the detector output and places a floor on how far the read noise can be averaged down through repeated nondestructive reads. In order to address these issues, a unique readout scheme named IRS$^2$ has been developed for NIRSpec that makes more frequent excursions to the outer reference pixels during each full frame read, thereby increasing the frame-time to 14.59\!~s \citep{mose10, rau17}. Use of this readout scheme significantly reduces the correlated noise in the images \citep{birk18}.

A further uncertainty impacting the faint-end sensitivity is the rate of particle hits due to galactic cosmic rays, and the scarcer but much more intense Coronal Mass Ejections that arise in Solar flares, that the HgCdTe devices will experience in orbit at L2. Such shielded particle event rates are notoriously difficult to predict beforehand accurately, but determine the longest integration time that can be undertaken before too many pixels in an exposure are adversely affected by particle hits. This exposure limit has long been assumed to be of order $\simeq$\!1000\!~s, which necessitates that longer NIRSpec exposures be broken into multiple subexposures of approximately this duration and combined after the fact. However, given that the effective read noise does not increase significantly with longer integration time \citep{birk18}, there is potentially a  sensitivity gain to be had if on-orbit experience shows that subexposures longer than 1000\!~s are in fact feasible. The primary motivation for employing up-the-ramp sampling -- as opposed to more traditional Fowler sampling -- is that it offers partial recovery from particle hits since the slope of the accumulating signal in an affected pixel can in principle be measured up to the last read before the particle hit occurred and, depending on the severity of the event, often afterwards as well \citep{giar19}. These topics are discussed in more detail in Appendix~A.

Some further quirks of the NIRSpec detectors are also worth mentioning. Both arrays possess $\simeq$\!1.5\% randomly distributed pixels that are inoperable for science observations for various reasons \citep{birk18}. Such bad pixels drive the motivation for \emph{dithering} or \emph{nodding} NIRSpec exposures, that is changing the telescope pointing between subexposures so as to shift the spectra locations on the detector \citep[][Appendix~A]{boek12}. For wavelengths below 1.4\!~$\mu$m there is also a finite probability that a detected photon will release two electrons rather than  a single one \citep{rau14}. The probability of this occurring rises with increasing photon energy, and makes the detection statistics a compounded Poisson process that needs to be accounted for in both the radiometric calibration of affected NIRSpec spectra and the associated error propagation calculation. The same holds for the signal-dependent inter-pixel cross-talk that occurs between adjacent pixels at the $\simeq$\!1\% level, and which introduces correlations between neighboring pixels that affect the variance when summing the signal over adjacent pixels \citep{giar13}. Lastly, the NIRSpec arrays are prone to persistence, meaning that they hold a residual signal for some time after having been exposed to a bright source \citep{rau14}. Fortunately, the persistence signal decays inversely with the lapsed recovery time, and can if necessary be dealt with through prudent scheduling of NIRSpec observations of bright targets.

\subsection{Wavelength coverages}\label{sec:wcoverage}

Due to the varied lengths of the dispersed spectra, and the finite size of the NIRSpec detector array, the wavelengths covered in a NIRSpec spectrum taken with a given disperser depend on the choice of slit or MSA shutter employed. The two NIRSpec detector arrays are by design positioned in the Camera focal plane such that 0.6 - 5.3\!~$\mu$m $R\!=\!100$ prism spectra and one full octave of Band~I \& II $R\!=\!1000$ spectra taken with any of the A side Fixed Slits in Table~\ref{tab:slits} or the IFU fall entirely on NRS1. Complete Band~III $R\!=\!1000$ spectra up to the detector cut-off at 5.3\!~$\mu$m taken with the IFU or the S200A2 Fixed Slits are also fully captured by the NRS1 array, whereas the red extensions of such spectra taken in the S200A1, S400A1 and S1600A1 slits fall into the detector gap at wavelengths beyond 5.10\!~$\mu$m, 5.22\!~$\mu$m, and 5.18\!~$\mu$m, respectively. 

\begin{figure*}
  \centering{\includegraphics[width=17cm]{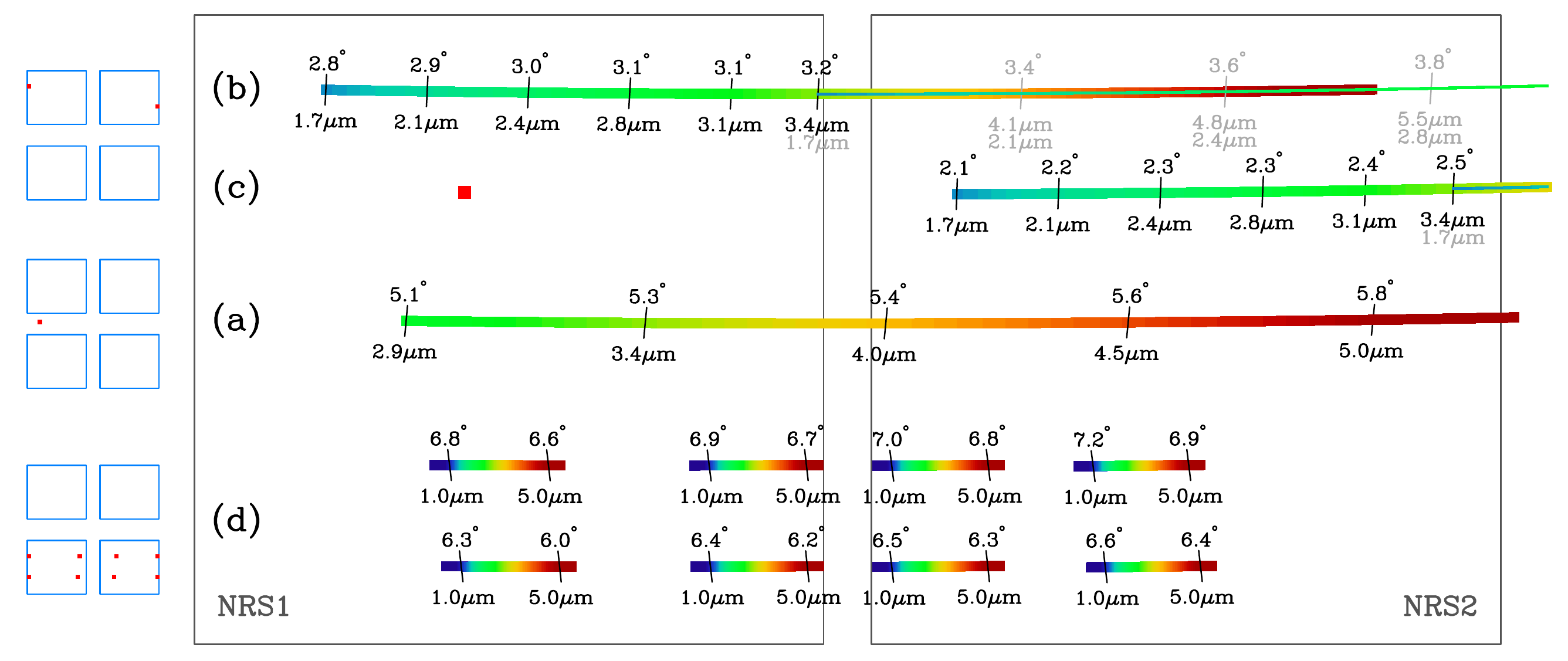}}
  \caption{Representative traces of spectra on the NIRSpec detector. The two large squares mark the outer boundaries of the active areas of the two detector arrays. The orientation of this figure matches that of the Slit Plane shown in Fig.~\ref{fig:msa}, such that spectra shift left and right and up and down in the same sense as the Fixed Slit or MSA shutter employed. The traces span the full wavelength band covered by each disperser and are widened for clarity. Fiducial wavelengths and the magnitude of the slit tilt are indicated along each trace. For grating spectra their red extensions are also shown  with the portions contaminated by second-order light overdrawn with a thinner line. (a) G395H/F290LP, $R\!=\!2700$ Band~III spectrum from the S200A2 Fixed Slit. (b) G235M/F170LP $R\!=\!1000$ Band~II spectrum from shutter (3,355,50) on the extreme left side of MSA quadrant Q3. (c) G235M/F170LP, $R\!=\!1000$ Band~II spectrum from shutter (1,10,115) at the opposite side of the opposing MSA quadrant Q1. The filled red square shows the location of the zero order image. (d) Eight $R\!=\!100$ prism spectra from shutters (2,10,50), (2,265,50), (2,10,115) and (2,280,115) of quadrant Q2 and shutters (4,40,50), (4,355,50), (4,53,115) and (4,355,115) of the opposing quadrant Q4.}
  \label{fig:spectrace}
\end{figure*}

$R\!=\!2700$ grating spectra are more than 3500~pixels long, and will therefore invariably miss a $\simeq156$~pixel or $\Delta\lambda\simeq 0.037$, 0.062 or $0.104~\mu$m wide wavelength segment due to the gap between the active areas of NRS1 and NRS2. This is illustrated in Fig.~\ref{fig:spectrace} (a), which shows the detailed trace of a G235H/F290LP spectrum taken with the S200A2 Fixed Slit. By design, the nominal 2.9-5.0\!~$\mu$m spectral region fits comfortably across the two arrays, and continues out to 5.27\!~$\mu$m before it drops off NRS2. However, wavelengths between 3.81\!~$\mu$m and 3.92\!~$\mu$m are lost to the detector gap. If desired, this gap can be filled by taking a second G235H/F290LP spectrum in the offset S200A1 slit, in which case the wavelength interval 3.69 - 3.79\!~$\mu$m will be absent.  The equivalent $R\!=\!2700$ wavelength gaps in the other three Fixed Slits and the IFU mode are fixed, and cannot be recovered. However, the 30 spectra from the IFU virtual slits are horizontally offset on the detector due to the slit tilt (Sect.~\ref{sec:fwgw}), hence the missing wavelengths also shift between the slices (Paper~III). The situation for the two other $R\!=\!2700$ gratings, G140H and G395H, is entirely analogous.

In the case of a failure of NRS1, complete single-source PRISM and $R\!=\!1000$ spectra can be still obtained with the S200B1 slit projecting entirely onto NRS2. Complete $R\!=\!2700$ spectra can also still be obtained in two exposures by capturing their blue portions on NRS2 with S200B1 and their red portions on NRS2 using S200A1 or S200A2. The IFU mode, however will be lost except for the red portions of the $R\!=\!2700$ spectra. Conversely, should NRS2 fail, then complete single source PRISM and $R\!=\!100$ spectra can still be obtained with NRS1 using any of the A side slits and the IFU, but the red portions of the $R\!=\!2700$ spectra will then be lost.

The wavelength coverage is more complicated in MSA mode, which is first and foremost intended for use with the $R\!=\!100$ prism and the $R\!=\!1000$ gratings. By design, no PRISM or $R\!=\!1000$ spectrum taken through any MSA shutter will encounter the extreme outer edges of the two detector arrays  over its full octave extended band. This is demonstrated in Fig.~\ref{fig:spectrace} (b) which traces the G235M/F170LP, Band~II spectrum from one of the first un-vignetted shutters $(3,355,50)$ located at the extreme left edge of row 50 of MSA quadrant Q3 in Fig.~\ref{fig:msa}. A full 1.7-3.4\!~$\mu$m octave of useful spectrum is captured on NRS1, with the in-band second-order extension extending across NRS2. Trace (c) shows the matching spectrum taken in an  un-vignetted shutter $(1,10,115)$ in row 115 at the opposite extreme right edge of the opposing quadrant Q1. The full spectral octave is now projected onto NRS2, with most of the second-order contaminated spectrum beyond 3.4\!~$\mu$m falling off the detector edge. However, in this case the matching strong zero-order image appears on NRS1. 

It is evident that if the employed shutters producing traces (b) and (c) are gradually shifted toward the opposing sides of the MSA, the two spectra will move inward in opposite directions and soon encounter and move across the detector gap, which will leave an up to $\Delta\lambda=0.28~\mu$m wide gap in the spectra. This occurs for 61\% of the functional shutters located in a broad vertical swath running  down the middle of  the MSA (skewed by 
$\simeq5\deg$ due to the slit tilt). The situation for the other two $R\!=\!1000$ gratings 
G140M and G395M is closely similar, except that the detector gap takes out, respectively,  $\Delta\lambda=0.10~\mu$m and $0.17~\mu$m of the spectra.

Zero-order images from $R\!=\!1000$ gratings such as that in spectral trace (c) of Fig.~\ref{fig:spectrace} appear $\simeq$1600 pixels to the blue of the starting cut-on wavelength of the first-order spectra, and are therefore registered on NRS1 in $R\!=\!1000$ spectra taken with nearly all MSA shutters located in quadrants Q1 and Q2. Permanent zero-order images are also seen in $R\!=\!1000$ spectra taken with the S200B1 Fixed slit and the at present 5 failed-open shutters in Q1 and the 3 failed-open shutters in Q2. The zero-order images in higher resolution  $R\!=\!2700$ spectra appear more than $\simeq$4100 pixels to the blue of their cut-on wavelengths, and therefore miss the NIRSpec detectors in all slit options. The second order $R\!=\!2700$ spectra also fall off the detector for all slit and grating options, except the G140H/F070LP combination.

${R\!=\!100}$ PRISM spectra display neither zero- nor second-order spectra. Fig.~\ref{fig:spectrace} (d) traces eight PRISM spectra belonging to four shutters located at the outer edges of the MSA and four shutters toward the center in the same rows, but in columns chosen to lie just outside the slanted central vertical region containing shutters that result in  PRISM spectra affected by the detector gap. This gap has a much more serious impact on the shorter PRISM spectra since it can take out more than half of their wavelength range. In the interest of avoiding that such truncated PRISM spectra take up precious detector space that could be occupied by complete PRISM spectra, the central 18\% of functional shutters that lead to incomplete PRISM spectra are a priori excluded from use by the NIRSpec team's MSA preparation software \citep{pbs+2020}. As demonstrated by traces (d), even with this restriction it is still possible to accommodate up to four complete PRISM spectra side by side along the same shutter row without overlap.

The three $R\!=\!2700$ gratings may also be used in MSA mode, but for these the wavelength coverage is clearly more complicated. Only between 23\% and 30\% of the extreme leftmost shutter columns in Fig.~\ref{fig:msa} yield $R\!=\!2700$ spectra whose nominal wavelength ranges are captured in their entirety by NRS1 and NRS2 (save for the unavoidable gap). As the shutter employed is moved to the right, more and more of the red portions of the $R\!=\!2700$ spectra drop off the outer right edge of NRS2. By the time the right edge of the MSA is reached, only the bluemost $\simeq$50\% of the nominal bandpass remains. While less straightforward to use, multiobject spectroscopy employing the MSA in combination with the $R\!=\!2700$ gratings is still scientifically useful, especially when dealing with sources at known redshifts suspected of having strong emission lines in their spectra, in which case the NIRSpec instrument model can be employed to verify beforehand that the wavelengths of interest are captured for each target.

\subsection{Calibration Assembly}\label{sec:cal}

The last subsystem of NIRSpec is the internal calibration source, referred to as the Calibration Assembly (CAA), which consists of a 13~cm diameter gold-plated integrating sphere fed by 11 filtered Tungsten filament lamps which provides uniform illumination of the slit plane at the correct f-ratio. This is achieved by means of a folding flat and two powered mirrors, one of which is located on the back side of the Filter Wheel closed position (Sect.~\ref{sec:fwgw}), that insert  the light from the 26.5~mm diameter exit pupil of the integrating sphere into the light path of the Foreoptics just after the Filter Wheel \citep{tepl05}.

The available calibration lamps include five continuum sources intended for spectral flat-fielding that employ filters matching the five long-pass order-separation filters listed in Table~\ref{tab:filters}. Conventional atomic line sources dissipate too much power for inclusion in NIRSpec. Another four sources therefore employ Fabry-Perot type interference filters to produce 5-6 Lorentzian-profiled wavelength calibration lines in each of NIRSpec bands and over the full 0.6-5.3\!~$\mu$m spectral range of the low resolution prism. The temperature dependencies of these sources were accurately calibrated during ground testing \citep{birk11}. To supplement the Fabry-Perot-type line sources, another source employing an Erbium-doped filter provides an absolute wavelength reference near 1.4\!~$\mu$m. The final source provides a low-flux broadband continuum intended for use in imaging mode employing the plane target acquisition mirror of the Grating Wheel. 
Taken together, the suite of calibration light sources allows spectral flat-fielding, verification of the stability of the photometric and wavelength calibrations, as well as monitoring of the state of the MSA shutters, to be performed based exclusively on internal exposures.

\subsection{Target acquisition}\label{sec:ta}

The absolute pointing accuracy of the JWST observatory following a slew to a target field is not expected to be sufficient to accurately position NIRSpec targets within its $\simeq$200~mas wide slits. A robust and accurate  autonomous procedure for fine-tuning the NIRSpec pointing is therefore crucial for the operation of the instrument. Two such Target Acquisition procedures are foreseen for NIRSpec \citep{keye18}. 

The first procedure is intended for observations employing the Fixed Slits, and involves offsetting the initial pointing such that either the target itself or a nearby reference star falls within the large 1\farcs6$\times$1\farcs6 Fixed Slit (see Table~\ref{tab:slits}). An undispersed image of the square aperture is then taken with the choice of F140X, F110W or CLEAR filter and the exposure time matched beforehand to the brightness of the target or reference star. The onboard target acquisition software then locates and centroids the target or reference star in this image, and calculates the appropriate small slew in pitch and yaw required to place the target in the center of the desired aperture, taking into account the relevant optical distortions and the readings of the Grating Wheel fine position sensors (Sect.~\ref{sec:fwgw}).

The second procedure is a more elaborate variant of the same overall approach, and is intended for multiobject observations using the MSA. When preparing an MSA observation, once the user has determined how the MSA needs to be configured and oriented on the sky so that the targets of interest can be observed, he or she will be asked to identify up to eight suitably bright reference stars present in the MSA field of view at the given pointing. When the program is executed, images will then be taken of the specified reference stars, and used by the onboard software to calculate the pitch and yaw (and roll if needed) corrections needed to place the MSA at the desired pointing and orientation. An extra complication is that the reference stars will of necessity be imaged through the opaque grid of the all-open MSA, which will perturb their centroid locations on the detector array. With the aim of  minimizing this bias, the locations in two reference star images taken with the telescope pointing offset by half the shutter pitch in both directions will be used, and outlier reference stars affected by their proximity to failed closed shutters or detector blemishes will be iteratively eliminated in the least squares fit. 

The choice of target acquisition filter and exposure duration can be matched to the available reference stars, which ideally should lie within $\simeq$2.6~magnitudes of each other within the overall brightness range of $19.5 \la AB \la 25.7$ in order to achieve the best possible signal-to-noise ratio without saturation. Sufficiently sharp profiled galaxies may be substituted for stars in fields where there is a scarcity of suitable reference stars. It is anticipated that a final positional accuracy of ${\simeq}$25~mas -- or better, corresponding to $<$13\% of the nominal slit width -- will be achievable through this approach, provided that the relative positions of the reference stars and the targets themselves are known sufficiently accurately and referenced to the same (GAIA) astrometric coordinate system. In practice, this invariably requires prior HST or JWST imaging of the field to be available.

\section{Anticipated performance}\label{sec:perf}

The assembled NIRSpec instrument was subjected to an extensive ground-testing and calibration campaign prior to delivery to NASA \citep{birk12, birk16, giav16, tepl18}. These campaigns have validated that the as-built instrument meets all its formal requirements and is expected to function as intended in the areas that it is possible to verify beforehand. 

\subsection{Photometric throughput}
\label{sec:throughput}

\begin{figure}
  \resizebox{\hsize}{!}{\includegraphics{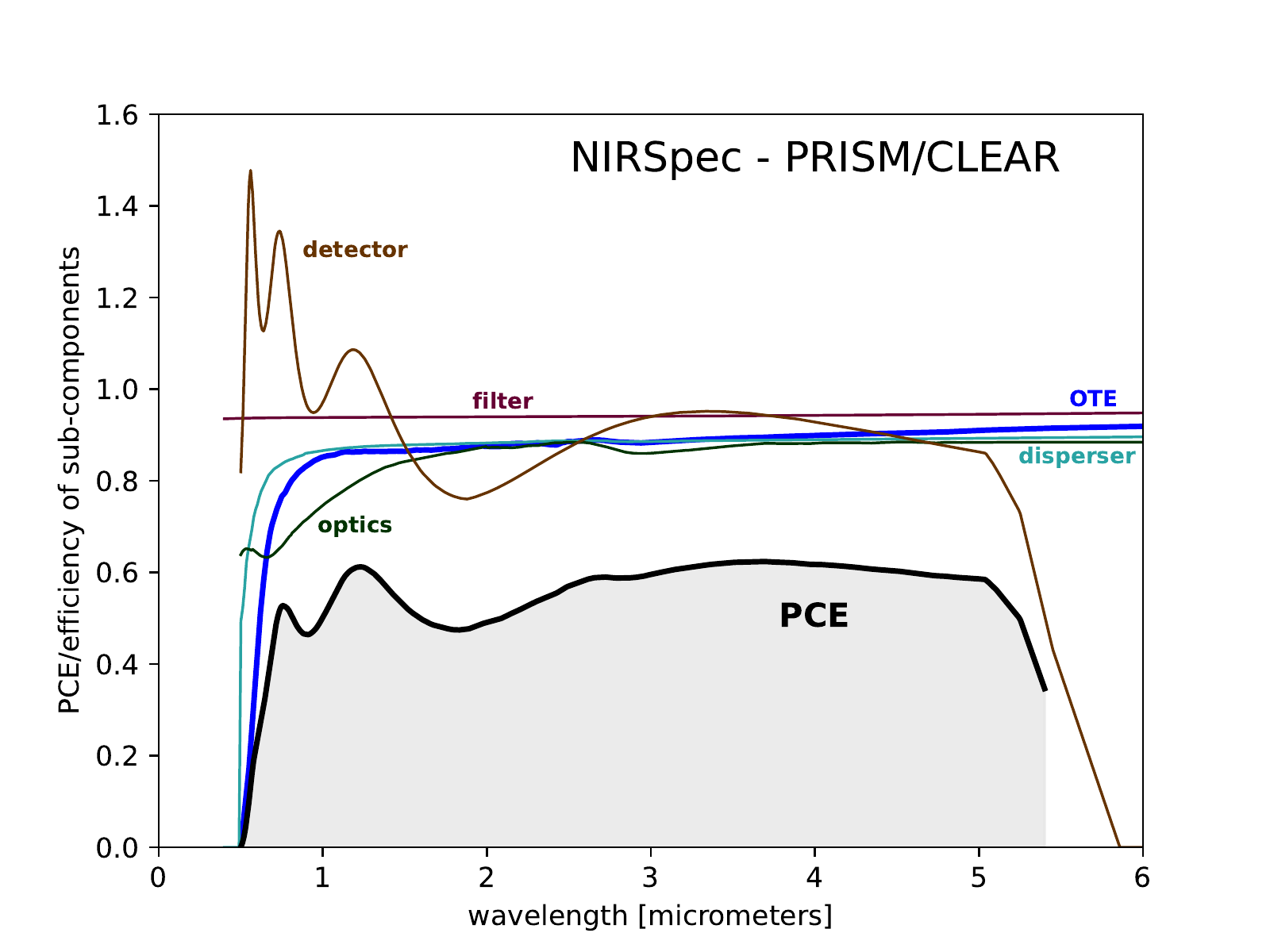}}
  \caption{Photon conversion efficiency (PCE) as a function of wavelength for NIRSpec in PRISM mode. The shown PCE curve includes the reflectivity of the telescope optics (marked OTE) and the responsive quantum efficiency of the NIRSpec detector array, but excludes  the light losses occurring in the slit device employed. The individual contributions making up the total PCE are also shown. The PCE contribution from the detector array exceeds unity at wavelengths below 1.4~$\mu$m due to multiple electrons being released for sufficiently energetic photons.}
\label{fig:pce_prism}
\end{figure}

\begin{figure}
  \resizebox{\hsize}{!}{\includegraphics{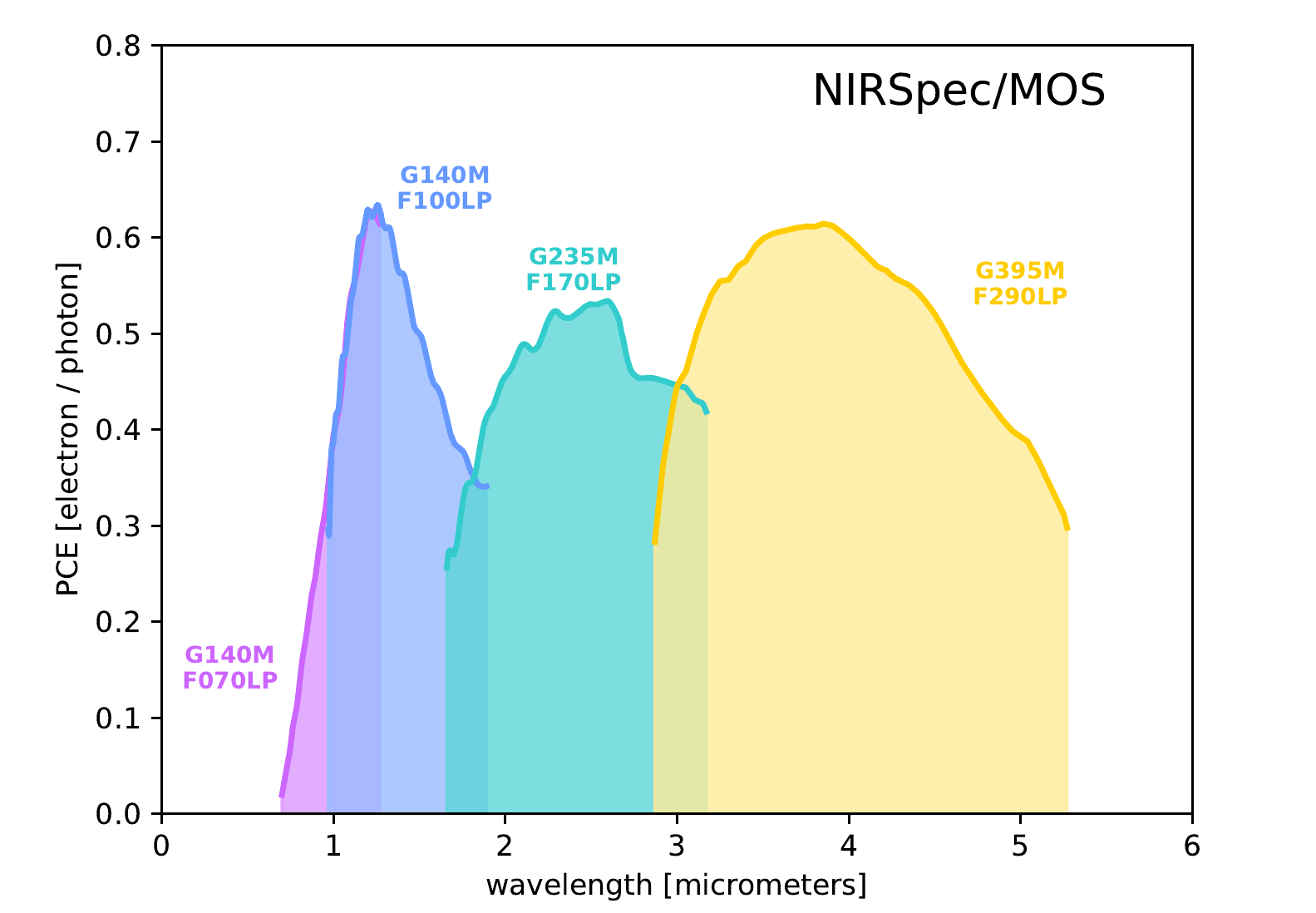}}
  \caption{Photon conversion efficiency of NIRSpec as a function of wavelength when employing the three $R\!=\!1000$ gratings and their order-separation filters.}
\label{fig:pce_medium}
\end{figure}

\begin{figure}
  \resizebox{\hsize}{!}{\includegraphics{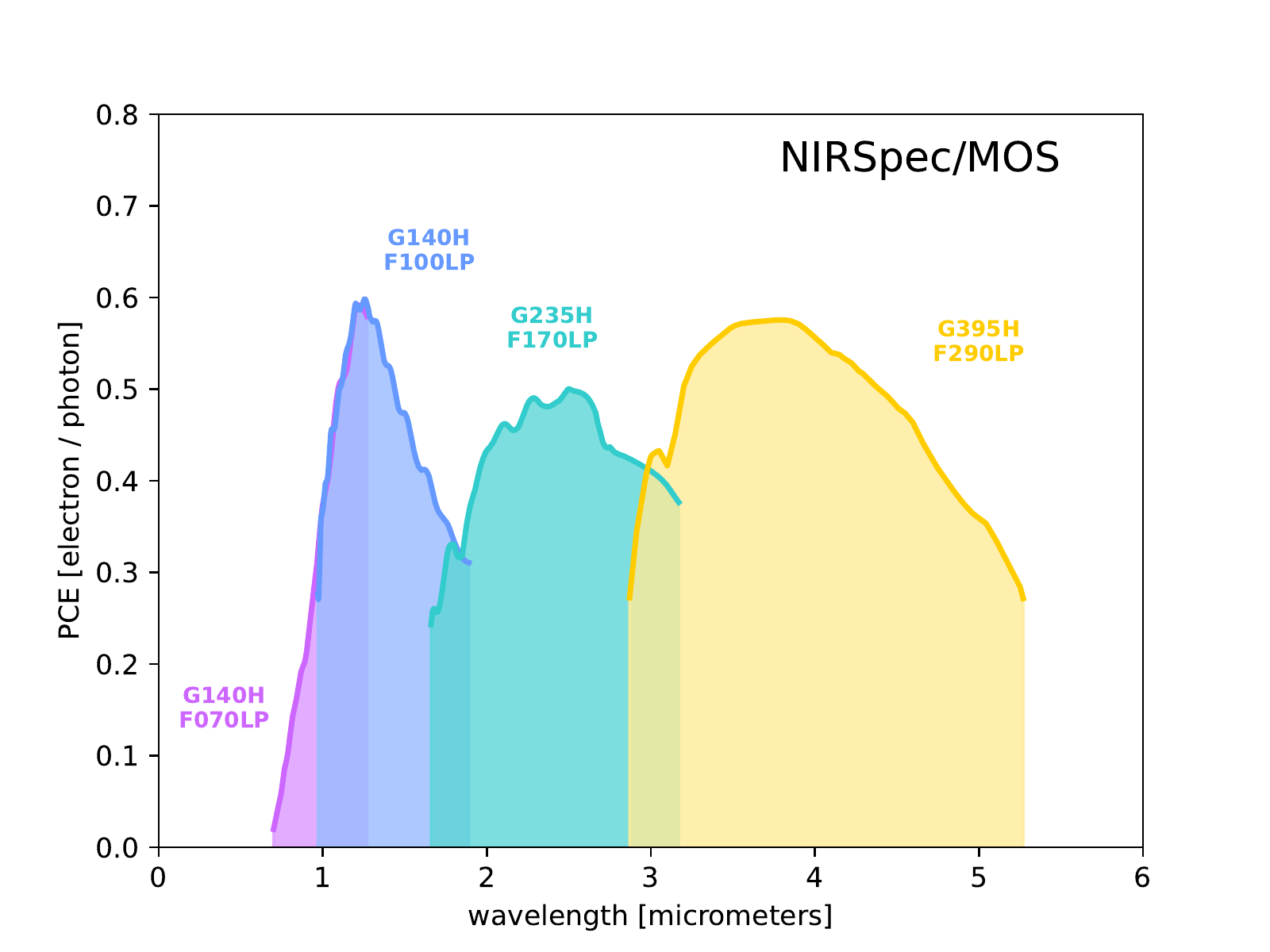}}
  \caption{Photon conversion efficiency of NIRSpec as a function of wavelength when employing  the three $R\!=\!2700$ gratings and their order-separation filters.}
\label{fig:pce_high}
\end{figure}

A key factor in determining the photometric sensitivity of NIRSpec is the total optical throughput of the instrument. Given the large number of reflections in the NIRSpec optical train, particular attention was devoted to assuring that all optical surfaces received the highest quality optical coatings and are kept clean during instrument assembly \citep{geyl11}. This diligence has seemingly paid off. Figure~\ref{fig:pce_prism} shows the net Photon conversion efficiency (PCE) of NIRSpec as function of wavelength in $R\!=\!100$ CLEAR/PRISM mode inferred from the individual measurements made on each optical component or appropriate witness samples. The shown total PCE curve includes the nominal reflectivity of the telescope optics \citep{light14} and the measured absolute quantum efficiency of the NIRSpec detector arrays, but excludes light losses caused by the selected slit device. The equivalent PCE curves for the three $R=\!\!1000$ and $R=\!\!2700$ gratings are shown in Figs.~\ref{fig:pce_medium} and \ref{fig:pce_high}.

The PCE curves predicted from the component measurements were verified independently during the NIRSpec ground campaign by means of an absolute calibrated external source \citep{birk12}. However, the results presented here are of necessity updated component level predictions that are anchored to the measured absolute quantum efficiency of the replacement flight detector array that was installed after the ground calibration \citep{rau14}. Nonetheless, it is clear that the anticipated optical throughput of NIRSpec excluding slit losses is very high, reaching in excess of $50$\% in PRISM mode, and at peak blaze in all six gratings. These levels of throughput are to our knowledge unprecedented for such a complex instrument employing such a high number of reflective surfaces in its light path.

\subsection{Slit losses and image quality}
\label{sec:slit_image}

\begin{figure*}
  \centering{\includegraphics[width=16cm]{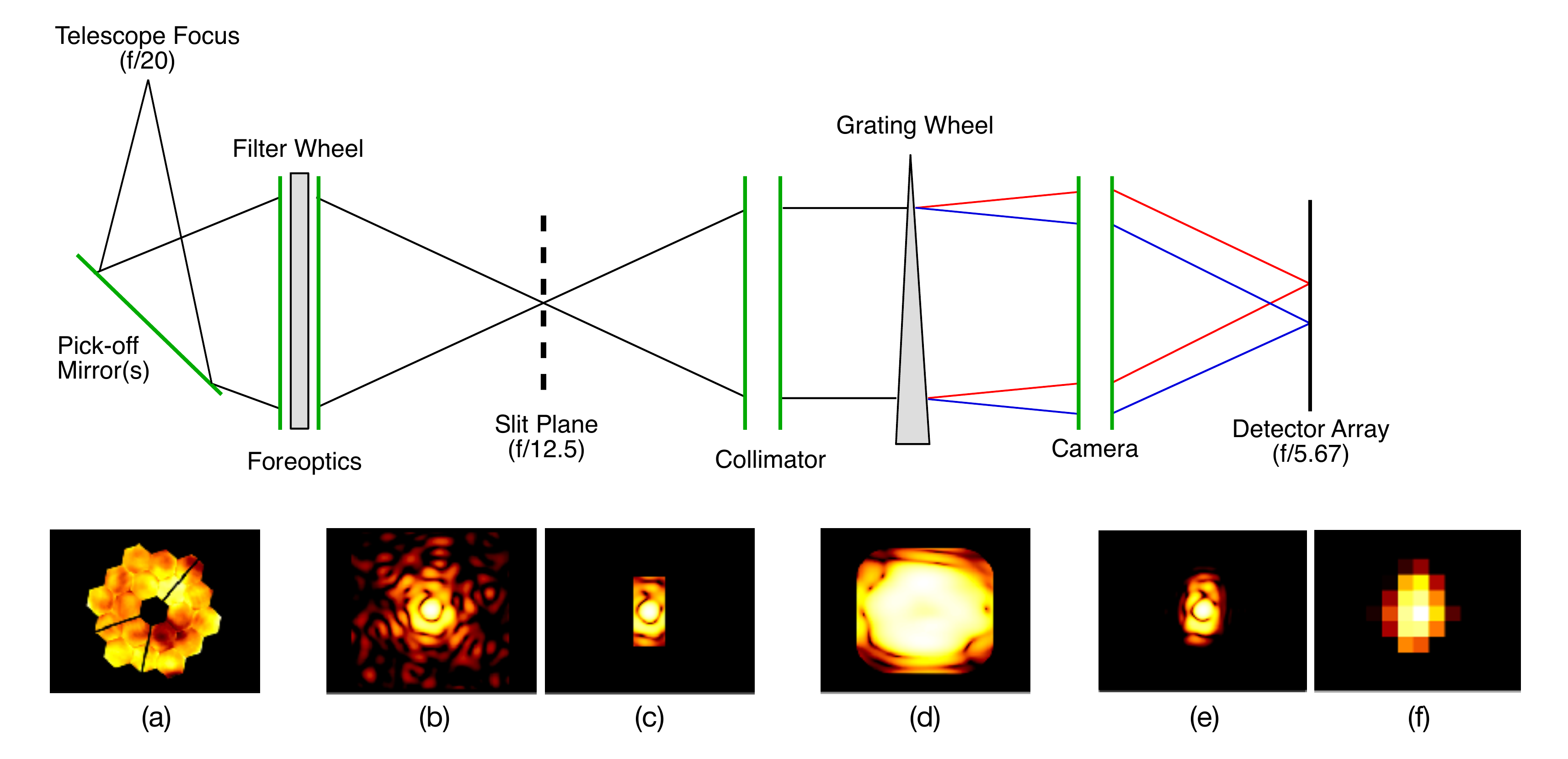}}
  \caption{Schematic of the optical chain of NIRSpec showing Fourier optical simulations of the expected Point Spread Function and beam footprint in the two pupil and focal planes of the system (see text). The simulations refer to a monochromatic point source  at a wavelength of $\lambda\!=\!2$\!~$\mu$m. A logarithmic look-up table is employed in order to show the full range of the light distributions.}
  \label{fig:psf_chain}
\end{figure*}

The total amount of light registered by NIRSpec from a object being observed is also affected by light losses occurring in the slit device employed. These losses depend on a variety of factors: the nature and shape of the object in question, the size and shape of the slit employed, the positioning of the object within the slit, and the optical quality of the image formed in the NIRSpec slit plane and elsewhere within the instrument. It is therefore natural to separate the slit losses from the PCE curve when discussing the total optical throughput of NIRSpec in a given mode. 

The task of radiometrically correcting NIRSpec spectra for slit losses is a challenging problem, even for point sources. The area spanned by the nearly diffraction-limited PSF in the slit plane changes by almost an order of magnitude over the 0.6-5.3\!~$\mu$m wavelength range of the instrument. This makes the slit transmission strongly wavelength dependent. Since the magnitude and wavelength dependency of the slit losses vary with the precise source location within the slit, the problem is especially complicated in the case of  multiobject observations carried with the MSA, where most targets cannot be expected to be located at the centers of their respective shutters (Paper~II). 

In addition to the high fidelity geometrical instrument model referred to above \citep{dorn16}, the NIRSpec Team has also developed the so-called Instrument Performance Simulator (IPS) to provide guidance in these complex issues \citep{Piqueras2008, Piqueras2010}. The IPS is capable of performing detailed Fourier optical simulations of the JWST and NIRSpec combination based on the available measured wavefront error maps of all surfaces in the path. Figure~\ref{fig:psf_chain} shows an illustrative series of simulations produced by the IPS in the case of a $\lambda\!=2\!~\mu$m monochromatic point source centered within a single shutter of the MSA. From left to right: (a): The wavefront error map of the telescope and Foreoptics at the Filter Wheel pupil position. (b): The intensity distribution of the resulting image incident on the Slit Plane. (c): The same image truncated by the open area of a single open shutter of the MSA. The light lost in this truncation gives rise to the dominant ``geometrical'' part of the slit loss.  (d): The intensity distribution of the beam projected by the Collimator over the surface of the 70~mm~$\times$~70~mm grating. Diffraction occurring at the MSA shutter causes the beam exiting the MSA to broaden beyond its incident f/12.5 f-ratio. The light missing the over-sized grating gives rise to the second  ``diffraction'' contribution to the total slit loss. (e): The image of the now twice truncated and dispersed beam focussed by the NIRSpec Camera onto the detector array where it is sheared slightly by the slit tilt. (f): The final image coarsely sampled by the 103\!~mas$\times$105\!~mas pixels of the detector array.

\begin{figure}
  \resizebox{\hsize}{!}{\includegraphics{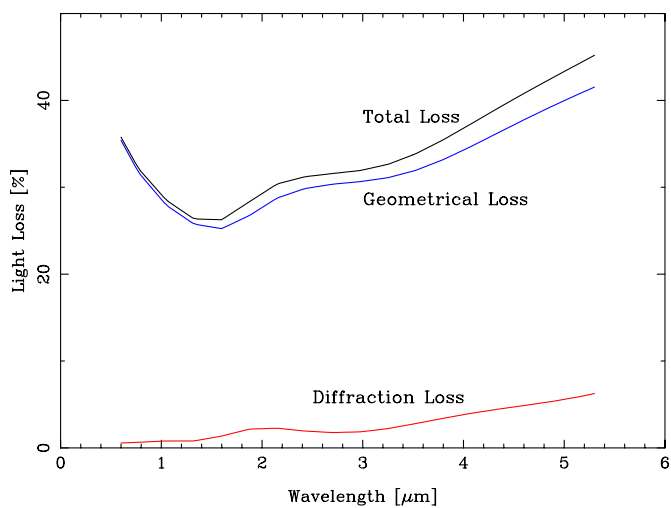}}
  \caption{Fourier optical prediction for the total slit loss experienced by a point source perfectly centered within a single shutter of the MSA as a function of wavelength. The geometrical loss caused by the truncation of the point spread function by the open area of the shutter and the subsequent diffraction loss caused by light in the spread-out beam missing the disperser are shown individually.}
\label{fig:slit_loss_point}
\end{figure}

\begin{figure}
\centering
\resizebox{0.55\hsize}{!}{\includegraphics{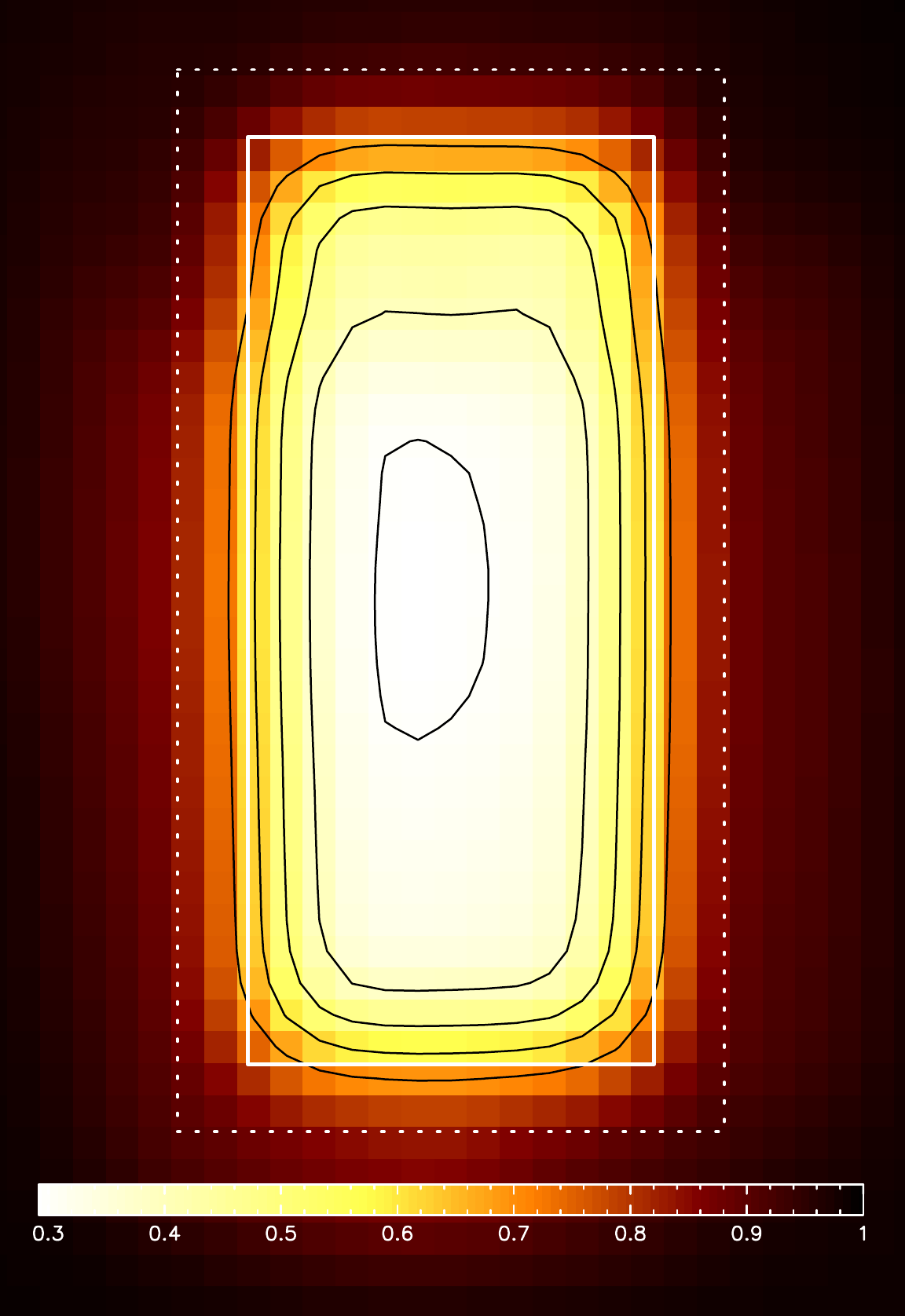}}
\caption{Fourier optical prediction for how the total slit loss from a point source varies with its placement within a single shutter at a wavelength of $\lambda\!=\!2.0$\!~$\mu$m. The shutter open area and pitch are indicated by the solid and dotted white lines.}
\label{fig:slit_loss_map}
\end{figure}

\begin{figure}
  \resizebox{\hsize}{!}{\includegraphics{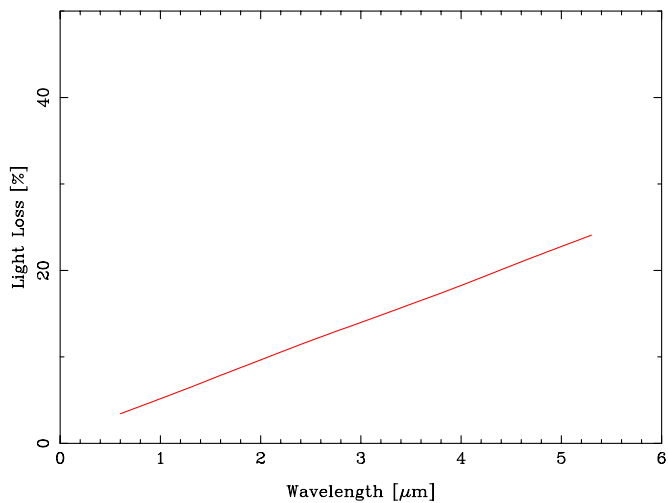}}
  \caption{Fourier optical prediction for the diffraction loss occurring in the amount of light captured by a single shutter from a uniform extended source as a function of wavelength.}
\label{fig:slit_loss_extend}
\end{figure}

Figure~\ref{fig:slit_loss_point} plots the corresponding slit losses for a perfectly centered point source as a function of wavelength predicted by the IPS. The increase in the light loss toward shorter wavelengths below 1.5~$\mu$m reflects the decreasing Strehl ratio and the diminishing portion of the light contained in the central image core of the 185~nm rms wfe PSF. Above the minimum near 1.5~$\mu$m the losses grow steadily larger with increasing wavelength as the increasingly diffraction-limited PSF grows in size w.r.t. the shutter open area, and the diffraction occurring at the shutter exit strengthens. The total single-shutter slit transmission for a centered point source is predicted to vary between 55\% and 74\%.  However, an inherent feature of the NIRSpec MSA and its fixed regular grid of slits is that most targets in any given MSA pointing cannot be expected to be well-centered in their respective shutters. Figure~\ref{fig:slit_loss_map} plots the Fourier optical prediction for how the total slit loss at a wavelength of $\lambda\!=\!2.0\!~\mu$m changes with intra-shutter position. As expected, the point source slit loss increases notably as the shutter edges are approached.

This can be contrasted to the slit losses experienced by an extended source.  
In the limit of a large target that displays a uniform surface brightness over the entire open area of a shutter, there is an obvious geometrical slit loss occurring w.r.t. the integrated total magnitude of the object (which can be quite substantial), but none w.r.t. the light entering NIRSpec since the effective light-collecting solid angle extended by the PSF-convolved shutter open area remains the same irrespective of wavelength. In this case only diffraction losses occur. Figure~\ref{fig:slit_loss_extend} plots the diffraction losses predicted by the IPS for a uniform extended source illuminating a single shutter as a function of wavelength. In this case the net light loss is notably lower, and increases with wavelength in a nearly linear fashion since the interplay between the fine structure in the PSF and the shutter aperture is effectively smoothed out. 

Considering that most remote galaxies fall between these two extremes, it is evident that accurate spectrophotometric calibration of galaxy spectra obtained with the NIRSpec MSA will be challenging. This topic is discussed further in Paper~II.

\subsection{Limiting sensitivity and multiplexing}\label{sec:sens_and_multi}

In programmatic terms, the faint-end sensitivity of NIRSpec is captured  by two so-called Level 2 formal requirements that were placed on the instrument since its inception. The first requirement is that NIRSpec in $R\!=\!100$ prism mode be able to detect the $\lambda=3~\mu$m continuum flux from a point source of brightness 132~nJy ($AB=26.1$) at a signal-to-noise ratio of $S/N=10$ in an integration time of 10\,000~s. The second requirement is that NIRSpec in $R\!=\!1000$ grating mode be able to detect a line flux of $5.72\times10^{-19}$~erg~s$^{-1}$~cm$^{-2}$ at 2~{$\mu$}m from a point source at a signal-to-noise ratio of $S/N=10$ in an integration time of 100\,000~s. 

In addition to its light gathering power, the ultimate faint-end photometric sensitivity of NIRSpec depends on several external factors that will only be known accurately once the instrument is exposed to the environment at L2, and views the sky through the deployed JWST. In Appendix~A we revisit the detailed signal-to-noise calculation required to demonstrate that the as-built NIRSpec is very likely to exceed these two benchmark requirements given the throughput estimates presented above, the measured characteristics of the replacement detector arrays, and the nominal energetic particle environment that its detector arrays are expected to be subjected to at L2. There it is shown that in spite of NIRSpec being partially detector noise-limited in the $R\!=\!100$ case, and heavily so in the $R\!=\!1000$ case, the two Level~2 requirements are expected to be met with margins in the predicted signal-to-noise ratio in excess of $\simeq{60}$\% and $\simeq{100}$\%, respectively. It is also shown that these sensitivity margins can be further increased by 15-20\% if master sky background subtraction is developed and employed for MSA observations. Although these signal-to-noise calculations do not include more systematic contributions due to bad pixels, inaccuracies in the detector dark current subtraction, flat-fielding \citep[which is quite complex on NIRSpec;][]{rawl16}, these secondary sources of error are not expected to degrade the situation substantially. 

This comfortable margin in faint-end sensitivity with respect to the formal requirements owes much to the two new state-of-the-art detector arrays that were installed in place of the of the original pair, especially now that the IRS$^2$ readout scheme has been developed and implemented \citep{rau17}. Although this approach has not quite succeeded in averaging down the read noise as efficiently as was originally hoped, the two replacement devices nonetheless produce much cleaner data and display higher quantum detection efficiencies and lower dark currents compared to the original devices \citep{rau14,birk18}.

A remaining uncertainty, however, is the cosmic ray flux that the NIRSpec detector arrays  will be subjected to in orbit during periods of quiescent Solar activity. In Appendix~A we therefore also present a statistical model for how the limiting sensitivity of NIRSpec is impacted by the cosmic ray event rate, and use the model to quantify how the achievable signal-to-noise is expected to degrade with increasing event rate, and explore the potential gain that could be won by breaking long duration exposures into fewer but longer subexposures compared to the baseline $\simeq1000$\!~s maximum duration presently assumed.

In addition to its photometric sensitivity (further details of which are given in Papers II-IV), the other key performance metric anticipated of NIRSpec is its capability to obtain the spectra of many objects simultaneously with the MSA. This topic is discussed in detail in Paper~II, where it is shown that provided there are enough sources available on the sky, the MSA in its present state is capable of accommodating up to $\sim$230 nonoverlapping spectra in $R\!=\!100$ PRISM mode and $\sim$57 spectra in $R\!=\!1000$ grating mode when all targets are observed in three shutter tall slitlets. These multiplexing limits are also in excess of the formal requirements. 

\section{Summary and conclusions}

NIRSpec, the first slit-based multiobject spectrograph developed for a space-based astronomical observatory, is of necessity a rather complex instrument that will require time and  effort to learn to operate in an optimal manner once in orbit. The as-built flight instrument has been subjected to extensive ground testing, and is expected to perform as intended in all its modes at the sensitivity needed to meet its long-standing primary scientific objective of obtaining diagnostic spectra of the most remote galaxies that can only be imaged with present day facilities. What is more, present indications are that the highest redshift galaxies possess extremely rich and strong emission line spectra in the near infrared \citep[e.g.,][]{smit16}, to the point where randomly placed `empty sky' MSA shutters are expected to pick up serendipitous [OIII] and H$\alpha$ lines from low luminosity high-$z$ objects in deep NIRSpec exposures \citep{mase19}. Cautiously high expectations for the scientific return from NIRSpec would appear to be in order. 

\begin{acknowledgements}
NIRSpec owes its existence to the dedication and years of hard work of a great number of engineers, scientists and managers scattered across European and US industry, academia, and the ESA and NASA science programs. The contributions of these colleagues and institutions are too numerous to list here, but are greatly appreciated all the same.
Alan Dressler and Marcia Rieke are thanked for providing helpful comments on the  condensed historical narrative of Sect.~\ref{sec:intro}. We also thank the anonymous referee for insightful suggestions that have improved the presentation of this paper.
\end{acknowledgements}

\bibliographystyle{aa}
\bibliography{nirspec_refs} 

\begin{thebibliography}{73}
\expandafter\ifx\csname natexlab\endcsname\relax\def\natexlab#1{#1}\fi

\bibitem[{{Alves de Oliveira} {et~al.}(2018){Alves de Oliveira}, Birkmann,
  Böker, Ferruit, Giardino, Lützgendorf, Puga, Rawle, Sirianni, \&
  te~Plate}]{alve18}
{Alves de Oliveira}, C., Birkmann, S.~M., Böker, T., {et~al.} 2018, Proc.
  SPIE, 10704, 282

\bibitem[{Bagnasco {et~al.}(2007)Bagnasco, Kolm, Ferruit, Honnen, Koehler,
  Lemke, Maschmann, Melf, Noyer, Rumler, Salvignol, Strada, \& Plate}]{bagn07}
Bagnasco, G., Kolm, M., Ferruit, P., {et~al.} 2007, Proc. SPIE, 6692, 174

\bibitem[{{Bahcall} {et~al.}(1990){Bahcall}, {Guhathakurta}, \&
  {Schneider}}]{bahcall90}
{Bahcall}, J.~N., {Guhathakurta}, P., \& {Schneider}, D.~P. 1990, Science, 248,
  178

\bibitem[{Barth {et~al.}(2000)Barth, Isaacs, \& Poivey}]{barth2000}
Barth, J., Isaacs, J., \& Poivey, C. 2000, The Radiation Environment for the
  Next Generation Space Telescope, Tech. rep., NASA GSFC

\bibitem[{Beckwith {et~al.}(2006)Beckwith, Stiavelli, Koekemoer, Caldwell,
  Ferguson, Hook, Lucas, Bergeron, Corbin, Jogee, Panagia, Robberto, Royle,
  Somerville, \& Sosey}]{beck06}
Beckwith, S. V.~W., Stiavelli, M., Koekemoer, A.~M., {et~al.} 2006, \aj, 132,
  1729

\bibitem[{{Birkmann} {et~al.}(2021){Birkmann}, Ferruit, Giardino, Nielsen,
  Garc\'ia Mu\~noz, Kendrew, Rauscher, \& at~al.}]{birk20}
{Birkmann}, S., Ferruit, P., Giardino, G., {et~al.} 2021, \aap, in press (Paper
  IV)

\bibitem[{Birkmann {et~al.}(2011)Birkmann, Böker, Ferruit, Giardino, Jakobsen,
  de~Marchi, Sirianni, te~Plate, Savignol, Gnata, Wettemann, Dorner, Cresci,
  Rosales-Ortega, Stuhlinger, Cole, Tandy, \& Brockley-Blatt}]{birk11}
Birkmann, S.~M., Böker, T., Ferruit, P., {et~al.} 2011, Proc. SPIE, 8150, 102

\bibitem[{Birkmann {et~al.}(2012)Birkmann, Ferruit, Böker, Marchi, Giardino,
  Sirianni, Stuhlinger, Jensen, te~Plate, Rumler, Dorner, Gnata, \&
  Wettemann}]{birk12}
Birkmann, S.~M., Ferruit, P., Böker, T., {et~al.} 2012, Proc. SPIE, 8442, 1204

\bibitem[{Birkmann {et~al.}(2016)Birkmann, Ferruit, Rawle, Sirianni,
  de~Oliveira, Böker, Giardino, Lützgendorf, Marston, Stuhlinger, te~Plate,
  Jensen, Rumler, Dorner, Karl, Mosner, Wright, \& Rapp}]{birk16}
Birkmann, S.~M., Ferruit, P., Rawle, T., {et~al.} 2016, Proc. SPIE, 9904, 92

\bibitem[{Birkmann {et~al.}(2018)Birkmann, Sirianni, Ferruit, Willott,
  Maiolino, Rauscher, de~Oliveira, Böker, Giardino, Lützgendorf, Marston,
  Puga, Rawle, te~Plate, Jensen, \& Rumler}]{birk18}
Birkmann, S.~M., Sirianni, M., Ferruit, P., {et~al.} 2018, Proc. SPIE, 10709,
  708

\bibitem[{{Bonaventura} {et~al.}(2022){Bonaventura}, {Jakobsen}, {Ferruit},
  {et~al.}}]{pbs+2020}
{Bonaventura}, N., {Jakobsen}, P., {Ferruit}, P., {et~al.} 2022, in preparation

\bibitem[{Bonneville {et~al.}(2003)Bonneville, Prieto, Ferruit, Henault,
  Lemonnier, Prost, Bacon, \& Fevre}]{bonn03}
Bonneville, C., Prieto, E., Ferruit, P., {et~al.} 2003, Proc. SPIE, 4842, 162

\bibitem[{{Bouwens} {et~al.}(2004){Bouwens}, {Illingworth}, {Blakeslee},
  {Broadhurst}, \& {Franx}}]{bouw04}
{Bouwens}, R.~J., {Illingworth}, G.~D., {Blakeslee}, J.~P., {Broadhurst},
  T.~J., \& {Franx}, M. 2004, \apjl, 611, L1

\bibitem[{Breysse {et~al.}(2012)Breysse, Castel, \& Bougoin}]{brey12}
Breysse, J., Castel, D., \& Bougoin, M. 2012, Proc. SPIE, 10564, 982

\bibitem[{{Böker} {et~al.}(2021){Böker}, {Arribas}, {L\"utzgendorf}, {Alves
  de Oliveira}, {Beck}, {Birkmann}, {Bunker}, \& et~al.}]{boek20}
{Böker}, T., {Arribas}, S., {L\"utzgendorf}, N., {et~al.} 2021, \aap, in press
  (Paper III)

\bibitem[{Böker {et~al.}(2012)Böker, Birkmann, de~Marchi, Ferruit, Giardino,
  Sirianni, \& Beck}]{boek12}
Böker, T., Birkmann, S., de~Marchi, G., {et~al.} 2012, Proc. SPIE, 8442, 1213

\bibitem[{Böker {et~al.}(2016)Böker, Muzerolle, Bacinski, de~Oliveira,
  Birkmann, Ferruit, Karl, Lemke, Lützgendorf, Marston, Mosner, Rawle, \&
  Sirianni}]{boek16}
Böker, T., Muzerolle, J., Bacinski, J., {et~al.} 2016, Proc. SPIE, 9904, 1338

\bibitem[{{Chevallard} {et~al.}(2019){Chevallard}, {Curtis-Lake}, {Charlot},
  {Ferruit}, {Giardino}, {Franx}, {Maseda}, {Amorin}, {Arribas}, {Bunker},
  {Carniani}, {Husemann}, {Jakobsen}, {Maiolino}, {Pforr}, {Rawle}, {Rix},
  {Smit}, \& {Willott}}]{chev19}
{Chevallard}, J., {Curtis-Lake}, E., {Charlot}, S., {et~al.} 2019, \mnras, 483,
  2621

\bibitem[{Closs {et~al.}(2008)Closs, Ferruit, Lobb, Preuss, Rolt, \&
  Talbot}]{clos08}
Closs, M.~F., Ferruit, P., Lobb, D.~R., {et~al.} 2008, Proc. SPIE, 7010, 331

\bibitem[{{Dahlen} {et~al.}(2013){Dahlen}, {Mobasher}, {Faber}, {Ferguson},
  {Barro}, {Finkelstein}, {Finlator}, {Fontana}, {Gruetzbauch}, {Johnson},
  {Pforr}, {Salvato}, {Wiklind}, {Wuyts}, {Acquaviva}, {Dickinson}, {Guo},
  {Huang}, {Huang}, {Newman}, {Bell}, {Conselice}, {Galametz}, {Gawiser},
  {Giavalisco}, {Grogin}, {Hathi}, {Kocevski}, {Koekemoer}, {Koo}, {Lee},
  {McGrath}, {Papovich}, {Peth}, {Ryan}, {Somerville}, {Weiner}, \&
  {Wilson}}]{dahl13}
{Dahlen}, T., {Mobasher}, B., {Faber}, S.~M., {et~al.} 2013, \apj, 775, 93

\bibitem[{{Dayal} \& {Ferrara}(2018)}]{daya18}
{Dayal}, P. \& {Ferrara}, A. 2018, \physrep, 780, 1

\bibitem[{{de Marchi} {et~al.}(2012){de Marchi}, Birkmann, Böker, Ferruit,
  Giardino, Sirianni, Stuhlinger, te~Plate, Salvignol, Barho, Gnata, Lemke,
  Kosse, \& Mosner}]{dema12}
{de Marchi}, G., Birkmann, S.~M., Böker, T., {et~al.} 2012, Proc. SPIE, 8442,
  1222

\bibitem[{Deshpande {et~al.}(2018)Deshpande, Lützgendorf, Ferruit, Giardino,
  de~Oliveira, Birkmann, Böker, Puga, Rawle, Sirianni, \& te~Plate}]{desh18}
Deshpande, A., Lützgendorf, N., Ferruit, P., {et~al.} 2018, Proc. SPIE, 10698,
  1612

\bibitem[{{Dorner} {et~al.}(2016){Dorner}, {Giardino}, {Ferruit}, {Alves de
  Oliveira}, {Birkmann}, {B{\"o}ker}, {De Marchi}, {Gnata}, {K{\"o}hler},
  {Sirianni}, \& {Jakobsen}}]{dorn16}
{Dorner}, B., {Giardino}, G., {Ferruit}, P., {et~al.} 2016, \aap, 592, A113

\bibitem[{Ferruit(2016)}]{pf2016}
Ferruit, P. 2016, ESA-JWST–SCI-NRS-TN-2016-015, Tech. rep., ESA ESTEC

\bibitem[{{Ferruit} {et~al.}(2021){Ferruit}, {Jakobsen}, {Giardino}, {Rawle},
  {Alves de Oliveira}, {Arribas}, {Beck}, {Birkmann}, {B\"{o}ker}, \&
  et~al.}]{ferr20}
{Ferruit}, P., {Jakobsen}, P., {Giardino}, G., {et~al.} 2021, \aap, in press
  (Paper II)

\bibitem[{Fixsen {et~al.}(2000)Fixsen, Offenberg, Hanisch, Mather,
  Nieto-Santisteban, Sengupta, \& Stockman}]{fixs00}
Fixsen, D.~J., Offenberg, J.~D., Hanisch, R.~J., {et~al.} 2000, \pasp, 112,
  1350

\bibitem[{{Gardner} {et~al.}(2006){Gardner}, {Mather}, {Clampin}, {Doyon},
  {Greenhouse}, {Hammel}, {Hutchings}, {Jakobsen}, {Lilly}, {Long}, {Lunine},
  {McCaughrean}, {Mountain}, {Nella}, {Rieke}, {Rieke}, {Rix}, {Smith},
  {Sonneborn}, {Stiavelli}, {Stockman}, {Windhorst}, \& {Wright}}]{gard06}
{Gardner}, J.~P., {Mather}, J.~C., {Clampin}, M., {et~al.} 2006, \ssr, 123, 485

\bibitem[{Geyl {et~al.}(2011)Geyl, Ruch, Vayssade, Leplan, \& Rodolfo}]{geyl11}
Geyl, R., Ruch, E., Vayssade, H., Leplan, H., \& Rodolfo, J. 2011, Proc. SPIE,
  8146, 130

\bibitem[{{Giardino} {et~al.}(2019){Giardino}, {Birkmann}, {Robberto},
  {Ferruit}, {Rauscher}, {Sirianni}, {Alves de Oliveira}, {Boeker},
  {Luetzgendorf}, {te Plate}, {Puga}, \& {Rawle}}]{giar19}
{Giardino}, G., {Birkmann}, S., {Robberto}, M., {et~al.} 2019, \pasp, 131,
  094503

\bibitem[{Giardino {et~al.}(2016)Giardino, Luetzgendorf, Ferruit, Dorner,
  de~Oliveira, Birkmann, Boeker, Rawle, \& Sirianni}]{giav16}
Giardino, G., Luetzgendorf, N., Ferruit, P., {et~al.} 2016, Proc. SPIE, 9904,
  1345

\bibitem[{Giardino {et~al.}(2013)Giardino, Sirianni, Birkmann, Rauscher,
  Lindler, Böker, Ferruit, Marchi, Stuhlinger, Jensen, \& Strada}]{giar13}
Giardino, G., Sirianni, M., Birkmann, S.~M., {et~al.} 2013, Optical
  Engineering, 52, 1

\bibitem[{{Giavalisco} {et~al.}(2004){Giavalisco}, {Ferguson}, {Koekemoer},
  {Dickinson}, {Alexander}, {Bauer}, {Bergeron}, {Biagetti}, {Brandt},
  {Casertano}, {Cesarsky}, {Chatzichristou}, {Conselice}, {Cristiani}, {Da
  Costa}, {Dahlen}, {de Mello}, {Eisenhardt}, {Erben}, {Fall}, {Fassnacht},
  {Fosbury}, {Fruchter}, {Gardner}, {Grogin}, {Hook}, {Hornschemeier}, {Idzi},
  {Jogee}, {Kretchmer}, {Laidler}, {Lee}, {Livio}, {Lucas}, {Madau},
  {Mobasher}, {Moustakas}, {Nonino}, {Padovani}, {Papovich}, {Park},
  {Ravindranath}, {Renzini}, {Richardson}, {Riess}, {Rosati}, {Schirmer},
  {Schreier}, {Somerville}, {Spinrad}, {Stern}, {Stiavelli}, {Strolger},
  {Urry}, {Vandame}, {Williams}, \& {Wolf}}]{giav04}
{Giavalisco}, M., {Ferguson}, H.~C., {Koekemoer}, A.~M., {et~al.} 2004, \apjl,
  600, L93

\bibitem[{{Graham}(2000)}]{grah00}
{Graham}, J.~R. 2000, ASP Conference Series, 207, 240

\bibitem[{{Grogin} {et~al.}(2011){Grogin}, {Kocevski}, {Faber}, {Ferguson},
  {Koekemoer}, {Riess}, {Acquaviva}, {Alexander}, {Almaini}, {Ashby}, {Barden},
  {Bell}, {Bournaud}, {Brown}, {Caputi}, {Casertano}, {Cassata}, {Castellano},
  {Challis}, {Chary}, {Cheung}, {Cirasuolo}, {Conselice}, {Roshan Cooray},
  {Croton}, {Daddi}, {Dahlen}, {Dav{\'e}}, {de Mello}, {Dekel}, {Dickinson},
  {Dolch}, {Donley}, {Dunlop}, {Dutton}, {Elbaz}, {Fazio}, {Filippenko},
  {Finkelstein}, {Fontana}, {Gardner}, {Garnavich}, {Gawiser}, {Giavalisco},
  {Grazian}, {Guo}, {Hathi}, {H{\"a}ussler}, {Hopkins}, {Huang}, {Huang},
  {Jha}, {Kartaltepe}, {Kirshner}, {Koo}, {Lai}, {Lee}, {Li}, {Lotz}, {Lucas},
  {Madau}, {McCarthy}, {McGrath}, {McIntosh}, {McLure}, {Mobasher},
  {Moustakas}, {Mozena}, {Nandra}, {Newman}, {Niemi}, {Noeske}, {Papovich},
  {Pentericci}, {Pope}, {Primack}, {Rajan}, {Ravindranath}, {Reddy}, {Renzini},
  {Rix}, {Robaina}, {Rodney}, {Rosario}, {Rosati}, {Salimbeni}, {Scarlata},
  {Siana}, {Simard}, {Smidt}, {Somerville}, {Spinrad}, {Straughn}, {Strolger},
  {Telford}, {Teplitz}, {Trump}, {van der Wel}, {Villforth}, {Wechsler},
  {Weiner}, {Wiklind}, {Wild}, {Wilson}, {Wuyts}, {Yan}, \& {Yun}}]{grog11}
{Grogin}, N.~A., {Kocevski}, D.~D., {Faber}, S.~M., {et~al.} 2011, \apjs, 197,
  35

\bibitem[{Henein {et~al.}(2004)Henein, Spanoudakis, Schwab, Giriens, Lisowski,
  Onillon, \& Myklebust}]{hene04}
Henein, S., Spanoudakis, P., Schwab, P., {et~al.} 2004, Proc. SPIE, 5487, 765

\bibitem[{{Illingworth} {et~al.}(2013){Illingworth}, {Magee}, {Oesch},
  {Bouwens}, {Labb{\'e}}, {Stiavelli}, {van Dokkum}, {Franx}, {Trenti},
  {Carollo}, \& {Gonzalez}}]{illi13}
{Illingworth}, G.~D., {Magee}, D., {Oesch}, P.~A., {et~al.} 2013, \apjs, 209, 6

\bibitem[{Karakla {et~al.}(2014)Karakla, Shyrokov, Pontoppidan, Beck, Gilbert,
  Valenti, Kassin, \& Soderblom}]{kara14}
Karakla, D., Shyrokov, A., Pontoppidan, K., {et~al.} 2014, Proc. SPIE, 9149,
  735

\bibitem[{Keyes {et~al.}(2018)Keyes, Beck, Peña-Guerrero, de~Oliveira,
  Ferruit, Jakobsen, Giardino, Sirianni, Boeker, Birkmann, \&
  Proffitt}]{keye18}
Keyes, C.~D., Beck, T.~L., Peña-Guerrero, M., {et~al.} 2018, Proc. SPIE,
  10704, 587

\bibitem[{{Koekemoer} {et~al.}(2007){Koekemoer}, {Aussel}, {Calzetti}, {Capak},
  {Giavalisco}, {Kneib}, {Leauthaud}, {Le F{\`e}vre}, {McCracken}, {Massey},
  {Mobasher}, {Rhodes}, {Scoville}, \& {Shopbell}}]{koek07}
{Koekemoer}, A.~M., {Aussel}, H., {Calzetti}, D., {et~al.} 2007, \apjs, 172,
  196

\bibitem[{Kutyrev {et~al.}(2008)Kutyrev, Collins, Chambers, Moseley, \&
  Rapchun}]{kuty08}
Kutyrev, A.~S., Collins, N., Chambers, J., Moseley, S.~H., \& Rapchun, D. 2008,
  Proc. SPIE, 7010, 1025

\bibitem[{Lightsey(2016)}]{light16}
Lightsey, P.~A. 2016, Proc. SPIE, 9904, 81

\bibitem[{Lightsey {et~al.}(2012)Lightsey, Atkinson, Clampin, \&
  Feinberg}]{light12}
Lightsey, P.~A., Atkinson, C.~B., Clampin, M.~C., \& Feinberg, L.~D. 2012,
  Optical Engineering, 51, 1

\bibitem[{Lightsey {et~al.}(2014)Lightsey, Knight, \& Golnik}]{light14}
Lightsey, P.~A., Knight, J.~S., \& Golnik, G. 2014, Proc. SPIE, 9143, 14

\bibitem[{Loose {et~al.}(2007)Loose, Beletic, Garnett, \& Xu}]{loos07}
Loose, M., Beletic, J., Garnett, J., \& Xu, M. 2007, Proc. SPIE, 6690, 124

\bibitem[{{MacKenty} {et~al.}(2004){MacKenty}, {Green}, {Greenhouse}, \&
  {Ohl}}]{mack04}
{MacKenty}, J.~W., {Green}, R.~F., {Greenhouse}, M.~A., \& {Ohl}, R.~G. 2004,
  Proc. SPIE, 5492, 1105

\bibitem[{{Maseda} {et~al.}(2019){Maseda}, {Franx}, {Chevallard}, \&
  {Curtis-Lake}}]{mase19}
{Maseda}, M.~V., {Franx}, M., {Chevallard}, J., \& {Curtis-Lake}, E. 2019,
  \mnras, 486, 3290

\bibitem[{Moseley {et~al.}(2004)Moseley, Arendt, Boucarut, Jhabvala, King,
  Kletetschka, Kutyrev, Li, Meyer, Rapchun, \& Silverberg}]{mose04}
Moseley, S.~H., Arendt, R.~G., Boucarut, R.~A., {et~al.} 2004, Proc. SPIE,
  5487, 645

\bibitem[{Moseley {et~al.}(2010)Moseley, Arendt, Fixsen, Lindler, Loose, \&
  Rauscher}]{mose10}
Moseley, S.~H., Arendt, R.~G., Fixsen, D.~J., {et~al.} 2010, Proc. SPIE, 7742,
  404

\bibitem[{Nielsen {et~al.}(2016)Nielsen, Ferruit, Giardino, Birkmann, Muñoz,
  Valenti, Isaak, de~Oliveira, Böker, Lützgendorf, Rawle, \&
  Sirianni}]{niel16}
Nielsen, L.~D., Ferruit, P., Giardino, G., {et~al.} 2016, Proc. SPIE, 9904,
  1218

\bibitem[{{Oesch} {et~al.}(2016){Oesch}, {Brammer}, {van Dokkum},
  {Illingworth}, {Bouwens}, {Labb{\'e}}, {Franx}, {Momcheva}, {Ashby}, {Fazio},
  {Gonzalez}, {Holden}, {Magee}, {Skelton}, {Smit}, {Spitler}, {Trenti}, \&
  {Willner}}]{oesc16}
{Oesch}, P.~A., {Brammer}, G., {van Dokkum}, P.~G., {et~al.} 2016, \apj, 819,
  129

\bibitem[{{Piqu{\'e}ras} {et~al.}(2008){Piqu{\'e}ras}, {Legay}, {Legros},
  {Ferruit}, {et~al.}}]{Piqueras2008}
{Piqu{\'e}ras}, L., {Legay}, P., {Legros}, E., {Ferruit}, P., {et~al.} 2008,
  Proc. SPIE, 7017, 317

\bibitem[{{Piqu{\'e}ras} {et~al.}(2010){Piqu{\'e}ras}, {Legros}, {Pons},
  {Legay}, {et~al.}}]{Piqueras2010}
{Piqu{\'e}ras}, L., {Legros}, E., {Pons}, A., {Legay}, P., {et~al.} 2010, Proc.
  SPIE, 7738, 407

\bibitem[{Pontoppidan {et~al.}(2016)Pontoppidan, Pickering, Laidler, Gilbert,
  Sontag, Slocum, Jr., Hanley, Earl, Pueyo, Ravindranath, Karakla, Robberto,
  Noriega-Crespo, \& Barker}]{pont16}
Pontoppidan, K.~M., Pickering, T.~E., Laidler, V.~G., {et~al.} 2016, Proc.
  SPIE, 9910, 381

\bibitem[{Posselt {et~al.}(2004)Posselt, Holota, Kulinyak, Kling, Kutscheid,
  Fevre, Prieto, \& Ferruit}]{poss04}
Posselt, W., Holota, W., Kulinyak, E., {et~al.} 2004, Proc. SPIE, 5487, 688

\bibitem[{Purll {et~al.}(2010)Purll, Lobb, Barnes, Talbot, Rolt, Robertson,
  Closs, \& te~Plate}]{purl10}
Purll, D.~J., Lobb, D.~R., Barnes, A.~R., {et~al.} 2010, Proc. SPIE, 7739, 410

\bibitem[{{Rauscher} {et~al.}(2017){Rauscher}, {Arendt}, {Fixsen},
  {Greenhouse}, {Lander}, {Lindler}, {Loose}, {Moseley}, {Mott}, {Wen},
  {Wilson}, \& {Xenophontos}}]{rau17}
{Rauscher}, B.~J., {Arendt}, R.~G., {Fixsen}, D.~J., {et~al.} 2017, \pasp, 129,
  105003

\bibitem[{{Rauscher} {et~al.}(2014){Rauscher}, {Boehm}, {Cagiano}, {Delo},
  {Foltz}, {Greenhouse}, {Hickey}, {Hill}, {Kan}, {Lindler}, {Mott},
  {Waczynski}, \& {Wen}}]{rau14}
{Rauscher}, B.~J., {Boehm}, N., {Cagiano}, S., {et~al.} 2014, \pasp, 126, 739

\bibitem[{{Rauscher} {et~al.}(2007){Rauscher}, {Fox}, {Ferruit}, {Hill},
  {Waczynski}, {Wen}, {Xia-Serafino}, {Mott}, {Alexander}, {Brambora}, {Derro},
  {Engler}, {Garrison}, {Johnson}, {Manthripragada}, {Marsh}, {Marshall},
  {Martineau}, {Shakoorzadeh}, {Wilson}, {Roher}, {Smith}, {Cabelli},
  {Garnett}, {Loose}, {Wong-Anglin}, {Zandian}, {Cheng}, {Ellis}, {Howe},
  {Jurado}, {Lee}, {Nieznanski}, {Wallis}, {York}, {Regan}, {Hall}, {Hodapp},
  {B{\"o}ker}, {De Marchi}, {Jakobsen}, \& {Strada}}]{raus07}
{Rauscher}, B.~J., {Fox}, O., {Ferruit}, P., {et~al.} 2007, \pasp, 119, 768

\bibitem[{{Rauscher} {et~al.}(2010){Rauscher}, {Fox}, {Ferruit}, {Hill},
  {Waczynski}, {Wen}, {Xia-Serafino}, {Mott}, {Alexander}, {Brambora}, {Derro},
  {Engler}, {Garrison}, {Johnson}, {Manthripragada}, {Marsh}, {Marshall},
  {Martineau}, {Shakoorzadeh}, {Wilson}, {Roher}, {Smith}, {Cabelli},
  {Garnett}, {Loose}, {Wong-Anglin}, {Zandian}, {Cheng}, {Ellis}, {Howe},
  {Jurado}, {Lee}, {Nieznanski}, {Wallis}, {York}, {Regan}, {Hall}, {Hodapp},
  {B{\"o}ker}, {De Marchi}, {Jakobsen}, \& {Strada}}]{raus10}
{Rauscher}, B.~J., {Fox}, O., {Ferruit}, P., {et~al.} 2010, \pasp, 122, 1254

\bibitem[{Rawle {et~al.}(2016)Rawle, de~Oliveira, Birkmann, Boeker, de~Marchi,
  Ferruit, Giardino, Luetzgendorf, \& Sirianni}]{rawl16}
Rawle, T.~D., de~Oliveira, C.~A., Birkmann, S.~M., {et~al.} 2016, Proc. SPIE,
  9904, 1353

\bibitem[{Rawle {et~al.}(2018)Rawle, Giardino, de~Oliveira, Birkmann, Böker,
  Ferruit, Lützgendorf, Ogle, Puga, Sirianni, \& Plate}]{rawl18}
Rawle, T.~D., Giardino, G., de~Oliveira, C.~A., {et~al.} 2018, Proc. SPIE,
  10698, 1101

\bibitem[{{Schroeder}(2000)}]{schroed00}
{Schroeder}, D.~J. 2000, {Astronomical Optics} (Academic Press), p356

\bibitem[{Smit {et~al.}(2016)Smit, Bouwens, Labb{\'{e}}, Franx, Wilkins, \&
  Oesch}]{smit16}
Smit, R., Bouwens, R.~J., Labb{\'{e}}, I., {et~al.} 2016, \apj, 833, 254

\bibitem[{{Sosey} {et~al.}(2012){Sosey}, {Hanley}, {Laidler}, {Lindsay},
  {Sienkiewicz}, {Greenfield}, \& {Barker}}]{sose12}
{Sosey}, M., {Hanley}, C., {Laidler}, V., {et~al.} 2012, ASP Conf. Series, 461,
  221

\bibitem[{Taccola {et~al.}(2008)Taccola, Bagnasco, Barho, Caprini, Giampietro,
  Gaillard, Mondello, Salvignol, Plate, \& Tonetti}]{tacc08}
Taccola, M., Bagnasco, G., Barho, R., {et~al.} 2008, Proc. SPIE, 7018, 699

\bibitem[{te~Plate {et~al.}(2016)te~Plate, Birkmann, Rumler, Jensen, Eder,
  Ehrenwinkler, Merkle, Mosner, Roedel, Speckmaier, Johnson, Mott, \&
  Snodgrass}]{tepl16b}
te~Plate, M., Birkmann, S., Rumler, P., {et~al.} 2016, Proc. SPIE, 9904, 115

\bibitem[{te~Plate {et~al.}(2018)te~Plate, Birkmann, Sirianni, Rawle,
  de~Oliveira, Böker, Puga, Lützgendorf, Marston, Rumler, Jensen, Giardino,
  Ferruit, Ehrenwinkler, Mosner, Karl, Altenburg, Maschmann, Rapp, Smith, Ogle,
  Guerrero, Proffitt, Wu, Kanarek, \& Muzerolle}]{tepl18}
te~Plate, M., Birkmann, S., Sirianni, M., {et~al.} 2018, Proc. SPIE, 10698, 69

\bibitem[{te~Plate {et~al.}(2005)te~Plate, Holota, Posselt, Koehler, Melf,
  Bagnasco, \& Marenaci}]{tepl05}
te~Plate, M. B.~J., Holota, W., Posselt, W., {et~al.} 2005, Proc. SPIE, 5904,
  185

\bibitem[{{Tropf}(1995)}]{trop95}
{Tropf}, W.~J. 1995, Optical Engineering, 34, 1369

\bibitem[{Weidlich {et~al.}(2008)Weidlich, Fischer, Ellenrieder, Gross,
  Salvignol, Barho, Neugebauer, Königsreiter, Trunz, Müller, \&
  Krause}]{weid08}
Weidlich, K., Fischer, M., Ellenrieder, M.~M., {et~al.} 2008, Proc. SPIE, 7018,
  720

\bibitem[{{Williams} {et~al.}(2000){Williams}, {Baum}, {Bergeron}, {Bernstein},
  {Blacker}, {Boyle}, {Brown}, {Carollo}, {Casertano}, {Covarrubias}, {de
  Mello}, {Dickinson}, {Espey}, {Ferguson}, {Fruchter}, {Gardner}, {Gonnella},
  {Hayes}, {Hewett}, {Heyer}, {Hook}, {Irwin}, {Jones}, {Kaiser}, {Levay},
  {Lubenow}, {Lucas}, {Mack}, {MacKenty}, {Madau}, {Makidon}, {Martin},
  {Mazzuca}, {Mutchler}, {Norris}, {Perriello}, {Phillips}, {Postman}, {Royle},
  {Sahu}, {Savaglio}, {Sherwin}, {Smith}, {Stiavelli}, {Suntzeff}, {Teplitz},
  {van der Marel}, {Walker}, {Weymann}, {Wiggs}, {Williger}, {Wilson},
  {Zacharias}, \& {Zurek}}]{will00}
{Williams}, R.~E., {Baum}, S., {Bergeron}, L.~E., {et~al.} 2000, \aj, 120, 2735

\bibitem[{{Williams} {et~al.}(1996){Williams}, {Blacker}, {Dickinson}, {Dixon},
  {Ferguson}, {Fruchter}, {Giavalisco}, {Gilliland }, {Heyer}, {Katsanis},
  {Levay}, {Lucas}, {McElroy}, {Petro}, {Postman}, {Adorf}, \& {Hook}}]{will96}
{Williams}, R.~E., {Blacker}, B., {Dickinson}, M., {et~al.} 1996, \aj, 112,
  1335

\end{thebibliography}

\begin{appendix}
\section{Benchmark NIRSpec signal-to-noise calculations and the impact of cosmic ray events}

This appendix is concerned with the detailed statistical calculations needed to verify that the as-built NIRSpec is likely to meet the two top-level formal sensitivity requirements described in Sec.~\ref{sec:perf}. Our approach is to first present the statistical recipe used to evaluate the signal-to-noise ratios expected to be reached in the two benchmark observations based on our present knowledge of the relevant as-built instrument and observatory performance parameters, and thereafter augment the calculations with a statistical model for how the NIRSpec sensitivity will be adversely affected by cosmic ray particle impacts on its detector arrays. Our approach is similar to those adopted in the more general JWST exposure time calculators that are available \citep{sose12, pont16, niel16}. It is nevertheless highly instructive to revisit the assumptions that enter the calculations for the two cases at hand and examine some of the intermediate results.

\subsection{Benchmark signal-to-noise calculation}
\label{sec:sn}

\begin{figure}
  \resizebox{\hsize}{!}{\includegraphics{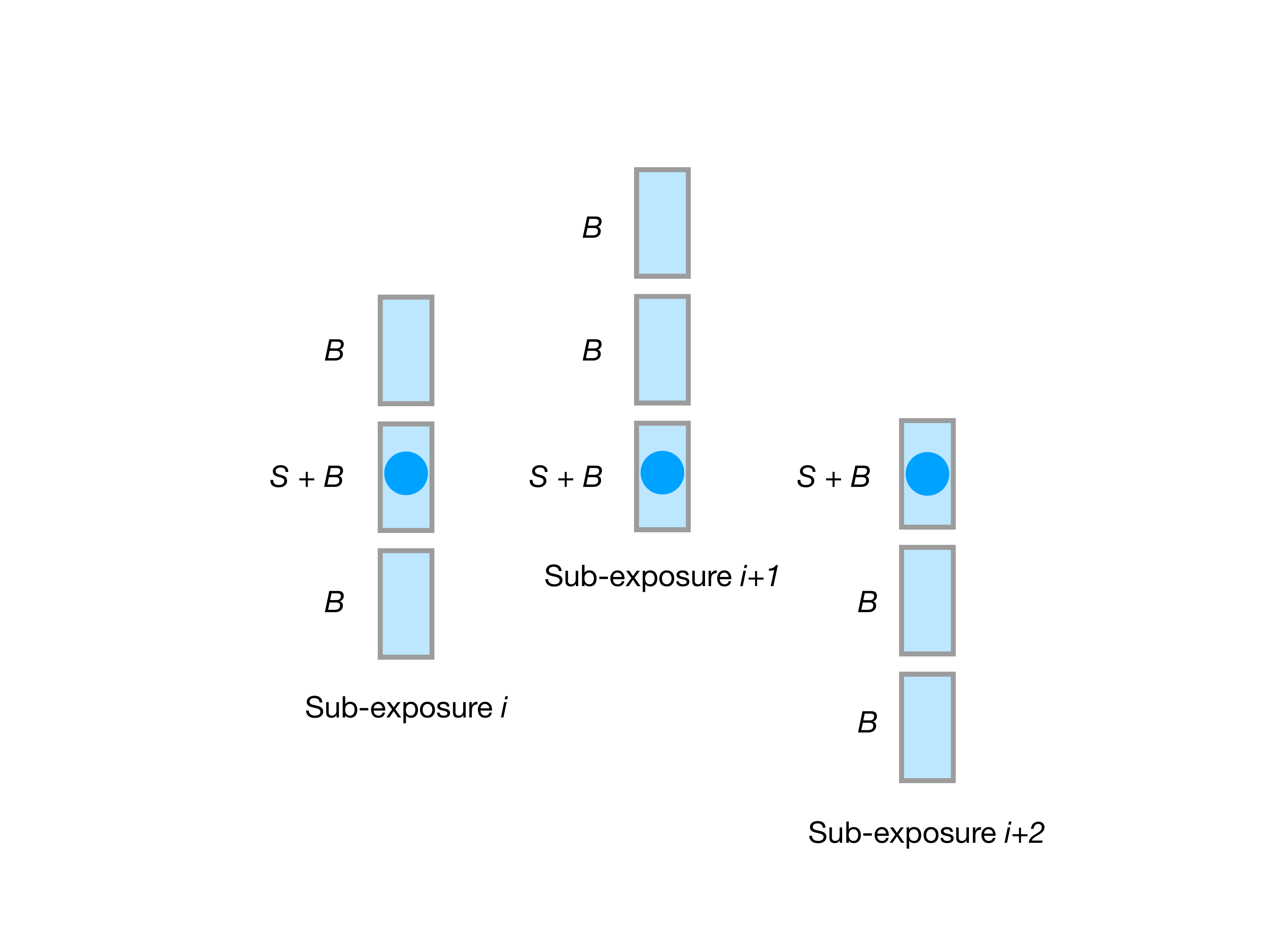}}
  \caption{Schematic illustration of the default MSA observing scheme where each target is assigned a three shutter tall slitlet, and the telescope is nodded between subexposures such that the target is cycled between the three shutters of the slitlet.}
\label{fig:nodding}
\end{figure}

As the starting point, both benchmark observations are assumed to be carried out employing the MSA in the presently default manner with a three shutter tall slitlet assigned to the target, and the telescope pointing nodded between each subexposure such that the target is cycled between the three shutters making up the slitlet (Fig.\ref{fig:nodding}). In the case of perfectly centered point sources, each subexposure yields in this scheme one spectrum of the combined target signal $S$ plus the sky background $B$, and $n_\mathrm{B}=2$ matching spectra of the sky background $B$ only. 
It is also assumed that the signal-to-noise requirements above refer to a single resolution element in the one-dimensional spectrum produced by the single shutter containing the target, which we take to extend $n_\mathrm{pix}=8$ pixels on the detector (i.e., two pixels along the dispersion direction and four pixels in the spatial direction). It is also assumed that the resolution element captures the uniform sky background contained within the full shutter open area.

In the $R\!=\!100$ continuum source observation the average number of photons from the source detected in the resolution element centered on $3\!~\mu$m in a subexposure of duration $t_\mathrm{sub}$ is:
\begin{equation}
S = f_\lambda \ \Delta\lambda \ A \ s_{\scriptscriptstyle P} (\lambda) \ \epsilon_{\scriptscriptstyle PCE}(\lambda)   \ t_\mathrm{sub}; 
\label{eq:signalS}
\end{equation}
where $f_\lambda=6.64\times10^{-5}$~ph~s$^{-1}$\!~cm$^{-2}$\!~$\mu$m$^{-1}$ is the benchmark 132~nJy target continum flux expressed in suitable photon units; $\Delta\lambda$ is the wavelength band spanned by the two pixel wide resolution element; $A=25$\!~m$^2$ the required minimum geometrical collecting area of the JWST primary; $s_{\scriptscriptstyle P}(\lambda)$ the centered point source slit transmission; $\epsilon_{\scriptscriptstyle PCE}(\lambda)$ the total photon conversion efficiency of the telescope and NIRSpec combination, all evaluated at a wavelength of $\lambda=3\!~\mu$m. 

In the $R\!=\!1000$ instance, we assume that the target emission line at 2\!~$\mu$m is unresolved in the sense that  the line flux falls entirely within one resolution element in the NIRSpec spectrum. Assuming that the underlying continuum flux from the source at 2\!~$\mu$m can be ignored, the signal in the resolution element then becomes:
\begin{equation}
S = f_{l} \ A \ s_{\scriptscriptstyle P}(\lambda) \ \epsilon_{\scriptscriptstyle PCE}(\lambda) \  t_\mathrm{sub}; 
\label{eq:signalE}
\end{equation}
where $f_\mathrm{l}=5.67\times10^{-7}$\!~ph~s$^{-1}$\!~cm$^{-2}$ is the target line flux expressed in photon units, and the other parameters are as before, but now evaluated at $\lambda=2\!~\mu$m.

Table~\ref{tab:snparameters} lists the instrument parameters adopted for the calculation. The listed values of $\Delta\lambda$ sampled by the two pixel wide resolution element were calculated using the instrument model. The listed values for the source slit transmissions $s_{\scriptscriptstyle P}(\lambda)$ stem from Fig.~\ref{fig:slit_loss_point}. The photon conversion efficiencies of $\epsilon_{\scriptscriptstyle PCE}(\lambda)=0.59$ in PRISM/CLEAR mode at $3\!~\mu$m and $\epsilon_{\scriptscriptstyle PCE}(\lambda)=0.46$ in  G235M/F170LP grating mode at $2\!~\mu$m are from Figs.~\ref{fig:pce_prism} and \ref{fig:pce_medium}\footnote{We note that the photon signals at the benchmark wavelengths of 2~$\mu$m and 3~$\mu$m are not affected by the stochastic release of multiple electrons per detected photon that occurs at lower wavelengths (Sect.~\ref{sec:detector})}. The JWST  collecting area is $A=25$\!~m$^2$ \citep{light12}.

The signal $S$ calculated from (\ref{eq:signalS}) or (\ref{eq:signalE}) will be detected on top of the signal from the underlying sky background spectrum captured by the shutter containing the target:
\begin{equation}
B = I_\lambda \ \Delta\lambda \ \Omega_\mathrm{sh}  \ A\ s_{\scriptscriptstyle E}(\lambda) \ \epsilon_{\scriptscriptstyle PCE}(\lambda) \ t_\mathrm{sub} 
\label{eq:signalB};
\end{equation}
where $I_\lambda$ is the continuum surface brightness of the sky background at the relevant wavelength expressed in photon units, $\Omega_\mathrm{sh}=0.199\times0.461$~arcsec$^2$ is the mean solid angle extended by the shutter open area and $s_{\scriptscriptstyle E}(\lambda)$ is the slit transmission for a uniform extended source from Fig.~\ref{fig:slit_loss_extend}. The benchmark observations both refer to targets located in directions of low sky background. This occurs near the ecliptic poles, where the sky brightness experienced by NIRSpec is predicted to be $I_\lambda=0.149$\!~MJy\!~sr$^{-1} = 2.64\times10^{-3}$\!~ph~\!s$^{-1}$\!~cm$^{-2}$\!~$\mu$m$^{-1}$\!~arcsec$^{-2}$ at 2~$\mu$m and $I_\lambda=0.094$~MJy~sr$^{-1} = 1.11\times10^{-3}$\!~ph~\!s$^{-1}$\!~cm$^{-2}$\!~$\mu$m$^{-1}$\!~arcsec$^{-2}$ at 3~$\mu$m, in both cases made up of roughly two thirds Zodiacal Light and one third scattered starlight \citep{light16}.

   \begin{table}
      \caption{Parameters assumed in signal-to-noise calculation}
       \centering
         \label{tab:snparameters}
         \begin{tabular}{c c c}
            \hline\hline
            \noalign{\smallskip}
           Parameter&  $R\!=\!100$& $R\!=\!1000$\\
            \noalign{\smallskip}
            \hline
            \noalign{\smallskip}
$\lambda$& 3.0~$\mu$m& 2.0~$\mu$m\\
$\Delta\lambda$& 0.0265~$\mu$m& 0.00214~$\mu$m\\
$s_{\scriptscriptstyle P}(\lambda)$&  0.68& 0.71\\
$s_{\scriptscriptstyle E}(\lambda)$&  0.86& 0.90\\
$\epsilon_{\scriptscriptstyle PCE}(\lambda)$& 0.59& 0.46\\
$A$&  \multicolumn{2}{c}{$2.5\times 10^5$~cm$^2$}\\
$\Omega_\mathrm{sh}$& \multicolumn{2}{c}{$9.17\times 10^{-2}$~arcsec$^2$} \\
$n_\mathrm{pix}$& \multicolumn{2}{c}{8} \\
$n_\mathrm{g}$& \multicolumn{2}{c}{70} \\
$m$& \multicolumn{2}{c}{1}\\
$t_\mathrm{f}$& \multicolumn{2}{c}{14.59~s}\\
$t_\mathrm{sub}$& \multicolumn{2}{c}{1006.7~s}\\
%$t_\mathrm{c}$& \multicolumn{2}{c}{$7.72\times10^3$\!~s}\\
          \noalign{\smallskip}
            \hline
         \end{tabular}
   \end{table}

   \begin{table}
      \caption{Detector noise performance parameters}
       \centering
         \label{tab:scafits}
         \begin{tabular}{c c c }
            \hline\hline
            \noalign{\smallskip}
           Parameter& NRS1&  NRS2\\
            \noalign{\smallskip}
            \hline
            \noalign{\smallskip}
$r_o$ & 6.8112& 9.1075\\
$p_1$ & 0.9801& 1.5403\\
$p_2$ & -0.2151& 1.8992\\
$d_c$ & 0.007055~s$^{-1}$& 0.003989~s$^{-1}$\\
          \noalign{\smallskip}
            \hline
         \end{tabular}
   \end{table}

In addition to the photon noise inherent in the two photon signals $S$ and $B$ above, the NIRSpec detector will also contribute significantly to the total noise in the form of read noise and  Poisson fluctuations in the detector dark current. 
The lowest noise IRS$^2$ advanced read-out scheme is implemented in two ways on NIRSpec, both of which are non-windowed and sample the full area of the arrays. In so-called NRSIRS2RAPID mode all frames are read out, saved onboard and eventually telemetered to the ground.  In  NRSIRS2 mode, $m\!=\!5$ consecutive frames are read-out and averaged onboard into a group before saving.  
For simplicity, we assume that the detector arrays are operated in NRSIRS2RAPID mode at unity gain, and that both benchmark observations are broken up into a suitable number of identical subexposures each spanning $n_\mathrm{g}\!=\!70$ ungrouped $m\!=\!1$ reads. The IRS$^2$ frame time is $t_\mathrm{f}=14.59$\!~s and since the first group is treated as zero in the up-the-ramp processing approach, the net integration time of each subexposure is $t_\mathrm{sub}=(n_\mathrm{g}-1)\ m\ t_\mathrm{f}=1006.7$\!~s. The two benchmark observations  then result from combining $n_\mathrm{sub}\!=\!10$ separate subexposures in the $R\!=\!100$ case, and $n_\mathrm{sub}\!=\!100$ subexposures in the $R\!=\!1000$ case, such that the total integration times of all subexposures:
\begin{equation}
t_\mathrm{int} =  n_\mathrm{sub} \ (n_\mathrm{g}-1) \ m \ t_\mathrm{f},
\label{eq:exptime}
\end{equation}
are $t_\mathrm{int}\!=\!10\,067$\!~s and $t_\mathrm{int}\!=\!100\,671$\!~s.

\citet{raus07,raus10} have shown that in up-the-ramp processing, the  variance contribution due to the read noise on the source signal extracted by least-squares fitting a slope to a ramp consisting of $n_\mathrm{g}$ grouped reads each averaged over $m$ frames, equals the read noise in a single frame $r^2$ multiplied by the numerical factor: 
\begin{equation}
\psi(n_\mathrm{g},m)=\frac{12(n_\mathrm{g}-1)}{(n_\mathrm{g}+1)}\frac{1}{m\ n_\mathrm{g}}
\label{eq:readfac};
\end{equation}
where the last $(m\,n_\mathrm{g})^{-1}$ factor is the reduction in total read noise expected statistically if the same quantity were read $(m\ n_\mathrm{g})$ times at constant single-frame read noise, and the preceding factor reflects that the accumulated signals in the $n_\mathrm{g}$ individual up-the-ramp reads are mutually correlated. 
Similarly, the inherent Poisson variance of the total photon and dark current signal accumulated in the exposure time $t_\mathrm{sub}=(n_\mathrm{g}-1)\ m \ t_\mathrm{f}$ contributes to the variance on the extracted signal amplified by the numerical factor: 
\begin{equation}
\phi(n_\mathrm{g},m)={\frac{6(n_\mathrm{g}^2+1)}{5 n_\mathrm{g} (n_\mathrm{g}+1)}} -  {\frac{2 (m^2-1)}{ n_\mathrm{g} (n_\mathrm{g}+1) \ m^2}};
\label{eq:poifac}
\end{equation}
where $\phi(n_\mathrm{g},m)$ increases slowly for $n_\mathrm{g} > 10$ and takes on the value $\phi=1.183$ for $n_\mathrm{g} = 70$, $m=1$. This slight amplification of the photon noise is the statistical penalty to be paid for maintaining the temporal resolution required by the up-the-ramp approach to estimating the Poisson parameter of the registered photons.

\citet{birk18} have as part of the NIRSpec ground-testing campaign measured the detailed noise performance of the two flight detector arrays in IRS$^2$ mode, and obtained good fits to eqs. (\ref{eq:readfac}) and (\ref{eq:poifac}) augmented with an empirical model in which the effective pixel single-read noise increases with exposure duration $n_\mathrm{g}$ as:
\begin{equation}
r^2(n_\mathrm{g},m) = r_o^2 + m\ n_\mathrm{g} \ (p_1^2 - {\frac{1}{12}} {\frac{(n_\mathrm{g}+1)}{(n_\mathrm{g}-1)}}\ \frac{p_2}{\sqrt{m}});
\label{eq:readnoise}
\end{equation}
where $r_o$, $p_1$ and $p_2$ are fitted constants whose values in electron units, valid over the sampled range $2 \le n_\mathrm{f} \le 200$, are listed in Table~\ref{tab:scafits}. When multiplied by eq. (\ref{eq:readfac}), the two $p_1$ and $p_2$ terms are designed to capture the fact that $1/f$ noise in the electronics causes the total variance due to read noise to level out at moderate values of $n_\mathrm{g}$.

For $n_\mathrm{g}\!=\!70$, and $m\!=\!1$ the NRS1 and NRS2 arrays display effective single-read noise values of $r\!=\!10.7~e$ per pixel and $r\!=\!15.4~e$ per pixel together with a variance reduction factor of $\psi\!=\!0.167$. This difference in read noise is partly compensated by NRS1 having nearly twice the dark current of NRS2. 

Ignoring pixel cross-talk, the total single-read noise variance in a resolution element employing $n_\mathrm{pix}$ pixels is:
\begin{equation}
RN^2 = n_\mathrm{pix} \ r^2;
\label{eq:totreadnoise}
\end{equation}
where the per pixel read noise $r^2$ is given by (\ref{eq:readnoise}).
Similarly, the variance due to the Poisson noise in the detector dark current summed over the resolution element is:
\begin{equation}
DC =  n_\mathrm{pix} \ d_c  \ t_\mathrm{sub}; 
\label{eq:background}
\end{equation}
where $d_c$ is the measured detector dark current per pixel listed in Table~\ref{tab:scafits}.

The estimate of the target signal $S$ in each subexposure results from subtracting the background signal $B\!+\!DC$  from  the target plus background signal $S\!+\!B\!+\!DC$ of interest. Assuming that the background signal is obtained by averaging the signal in $n_\mathrm{B}$ suitable sky-only shutters, the signal-to-noise ratio of the resulting background-corrected estimate of the target signal $S$ is: 
\begin{equation}
{S\!N}_\mathrm{sub} = \frac{S}{\sqrt{ \ \phi \ S + (1 + \frac{1} {n_\mathrm{B}})(\phi \ (B+DC)+\psi\ RN^2)}};
\label{eq:snsub}
\end{equation}
where the denominator is the total noise on the measured value of $S$ obtained by summing the variances due to the amplified photon noise in the source, the sky backgrounds, the detector dark current, and the total detector read noise over the pixels carrying these signals. Note that when $n_\mathrm{B}=1$ the Poisson noise in the sky background and dark current and the read noise all appear \emph{twice}; once when extracting the signal shutter itself, and again as the noise on the subtracted adjacent background signal. Only if the background subtraction results from averaging  {$n_\mathrm{B}>1$} measurements of different sky-only shutters will the noise introduced by the background subtraction step decrease accordingly.

Finally, the net signal-to-noise ratio on $S$ that results from the coaddition of $n_\mathrm{sub}$ identical but statistically independent subexposures is obtained as the square root of $n_\mathrm{sub}$ times the subexposure signal-to-noise ratio:
\begin{equation}
{S\!N}_\mathrm{tot} = \sqrt{n_\mathrm{sub}} \ {S\!N}_\mathrm{sub}.
\label{eq:sntot0}
\end{equation}

A key point enters the discussion here. The official NIRSpec sensitivity projections have historically been calculated and reported using eq. (\ref{eq:snsub}) assuming a perfectly centered point source and the value $n_\mathrm{B}\!=\!2$, reflecting the assumption that the background is to be measured by  averaging the signal in the two nonoccupied shutters of the target's three shutter tall slitlet, either within the same subintegration, or in the same shutter in two dithered subintegrations. However, this `local' background subtraction approach is not viable in practice for many targets\footnote{We note that the longstanding choices $n_\mathrm{g}\!=\!10$ and  $n_\mathrm{g}\!=\!100$ employed in the two benchmark calculations are not multiples of three and are therefore strictly speaking not entirely consistent with literal three shutter nodding.}. Very few of the objects to be observed in an actual multiobject MSA observation will be perfectly centered within their shutters. Moreover, even point sources located near the top or bottom edges of their shutters will tend to spill over some of their flux into the adjacent open shutter at the longer wavelengths where the PSF is larger. This leads to highly detrimental self-subtraction issues if the local background subtraction is applied blindly. Clearly, these issues become even more pronounced when observing extended sources.

In order to address this matter, the NIRSpec team is actively pursuing an alternative ``global'' approach to determining the background spectrum by incorporating the sky signals in \emph{all} suitably empty shutters in each subexposure (including intentionally placed sky-only slitlets), and applying the resulting  ``master background'' spectrum to all source-containing shutters in the subexposure. While the validity of this approach rests on  the premise  that the intensity and spectrum of the background is spatially constant over the NIRSpec field of view, and faces a number of detailed technical challenges due to the variation in sampling, plate-scale and dispersion across the NIRSpec field of view, the expectation is that these can be dealt with employing the  NIRSpec instrument model. If the number of sky-only shutters contributing to determining the master background spectrum is sufficiently large, the background subtraction step becomes essentially noiseless, resulting in a sensitivity gain corresponding to setting $1/n_\mathrm{B} = 0$ in eq. (\ref{eq:snsub}). Note also that in the master background approach it is in principle immaterial whether one, two or three shutter tall slitlets are assigned to each target, as long as an adequate number of sky-only shutters are present throughout the subimages.

  \begin{table}
      \caption{S/N calculation for benchmark $R\!=\!100$ observation}
       \centering
         \label{tab:snr100}
         \begin{tabular}{c c c}
            \hline\hline
            \noalign{\smallskip}
           Parameter&  NRS1& NRS2\\
            \noalign{\smallskip}
            \hline
            \noalign{\smallskip}
$n_\mathrm{g}$& \multicolumn{2}{c}{70}\\
$m$& \multicolumn{2}{c}{1}\\
$n_\mathrm{sub}$& \multicolumn{2}{c}{10}\\
$r(n_\mathrm{g},m)$& 10.7& 15.4\\
$\psi(n_\mathrm{g},m)$& \multicolumn{2}{c}{0.167}\\
$\phi(n_\mathrm{g},m)$& \multicolumn{2}{c}{1.18}\\
         \noalign{\smallskip}
$S$& \multicolumn{2}{c}{178}\\
$B$& \multicolumn{2}{c}{345}\\
         \noalign{\smallskip}
$\psi RN^2$& 153& 317\\
$DC$& 56.8& 32.1\\
         \noalign{\smallskip}\hline\noalign{\smallskip}
$1/n_\mathrm{B}$& \multicolumn{2}{c}{0.5}\\
${S\!N}_\mathrm{sub}$& 5.23& 4.82\\
${S\!N}_\mathrm{tot}$& 16.5& 15.3\\
         \noalign{\smallskip}\hline\noalign{\smallskip}
$1/n_\mathrm{B}$& \multicolumn{2}{c}{0}\\
${S\!N}_\mathrm{sub}$& 6.13& 5.70\\
${S\!N}_\mathrm{tot}$& 19.4& 18.0\\
         \noalign{\smallskip}
            \hline
         \end{tabular}
   \end{table}

  \begin{table}
      \caption{S/N calculation for benchmark $R\!=\!1000$ observation}
       \centering
         \label{tab:snr1000}
         \begin{tabular}{c c c}
            \hline\hline
            \noalign{\smallskip}
           Parameter&  NRS1& NRS2\\
            \noalign{\smallskip}
            \hline
            \noalign{\smallskip}
$n_\mathrm{g}$& \multicolumn{2}{c}{70}\\
$m$& \multicolumn{2}{c}{1}\\
$n_\mathrm{sub}$& \multicolumn{2}{c}{100}\\
$r(n_\mathrm{g},m)$& 10.7& 15.4\\
$\psi(n_\mathrm{g},m)$& \multicolumn{2}{c}{0.167}\\
$\phi(n_\mathrm{g},m)$& \multicolumn{2}{c}{1.18}\\
         \noalign{\smallskip}
$S$& \multicolumn{2}{c}{47.3}\\
$B$& \multicolumn{2}{c}{54.1}\\
         \noalign{\smallskip}
$\psi RN^2$& 153& 318\\
$DC$& 56.8& 31.7\\
         \noalign{\smallskip}\hline\noalign{\smallskip}
$1/n_\mathrm{B}$& \multicolumn{2}{c}{0.5}\\
${S\!N}_\mathrm{sub}$& 2.15& 1.81\\
${S\!N}_\mathrm{tot}$& 21.5& 18.1\\
         \noalign{\smallskip}\hline\noalign{\smallskip}
$1/n_\mathrm{B}$& \multicolumn{2}{c}{0}\\
${S\!N}_\mathrm{sub}$& 2.57& 2.17\\
${S\!N}_\mathrm{tot}$& 25.7& 21.7\\
         \noalign{\smallskip}
           \hline
         \end{tabular}
   \end{table}

Tables~\ref{tab:snr100} and \ref{tab:snr1000} list the outcomes of the two benchmark signal-to-noise calculations for NRS1 and NRS2 for  $n_\mathrm{f}\!=\!70$ and $m\!=\!1$ and the two choices $1/n_\mathrm{B}=0.5$ and $1/n_\mathrm{B}=0$. Useful insight into the anticipated faint-end photometric performance of NIRSpec can be gleaned from these tables. First of all, the calculations indicate that both benchmark observations should be achievable with NIRSpec with comfortable margins in ${S\!N}_\mathrm{tot}$: 62\% and 115\% if NRS1 is used, and 50\% and 80\% if NRS2 is used in the $1/n_\mathrm{B}=0.5$ approach. The gain in ${S\!N}_\mathrm{tot}$ achieved by going to the noiseless master background approach is 14-21\%, which is clearly worth pursuing.  These healthy margins notwithstanding, it is nonetheless evident that the $R\!=\!1000$ benchmark, with its lengthy coadded sequence of $n_\mathrm{sub}=100$ noisy nodded subexposures each having {${S\!N}_\mathrm{sub}\!\simeq\!2$}, clearly falls in the category of ``heroic'' NIRSpec observations that will rarely, if ever, be attempted in practice.

\begin{figure}
  \resizebox{\hsize}{!}{\includegraphics{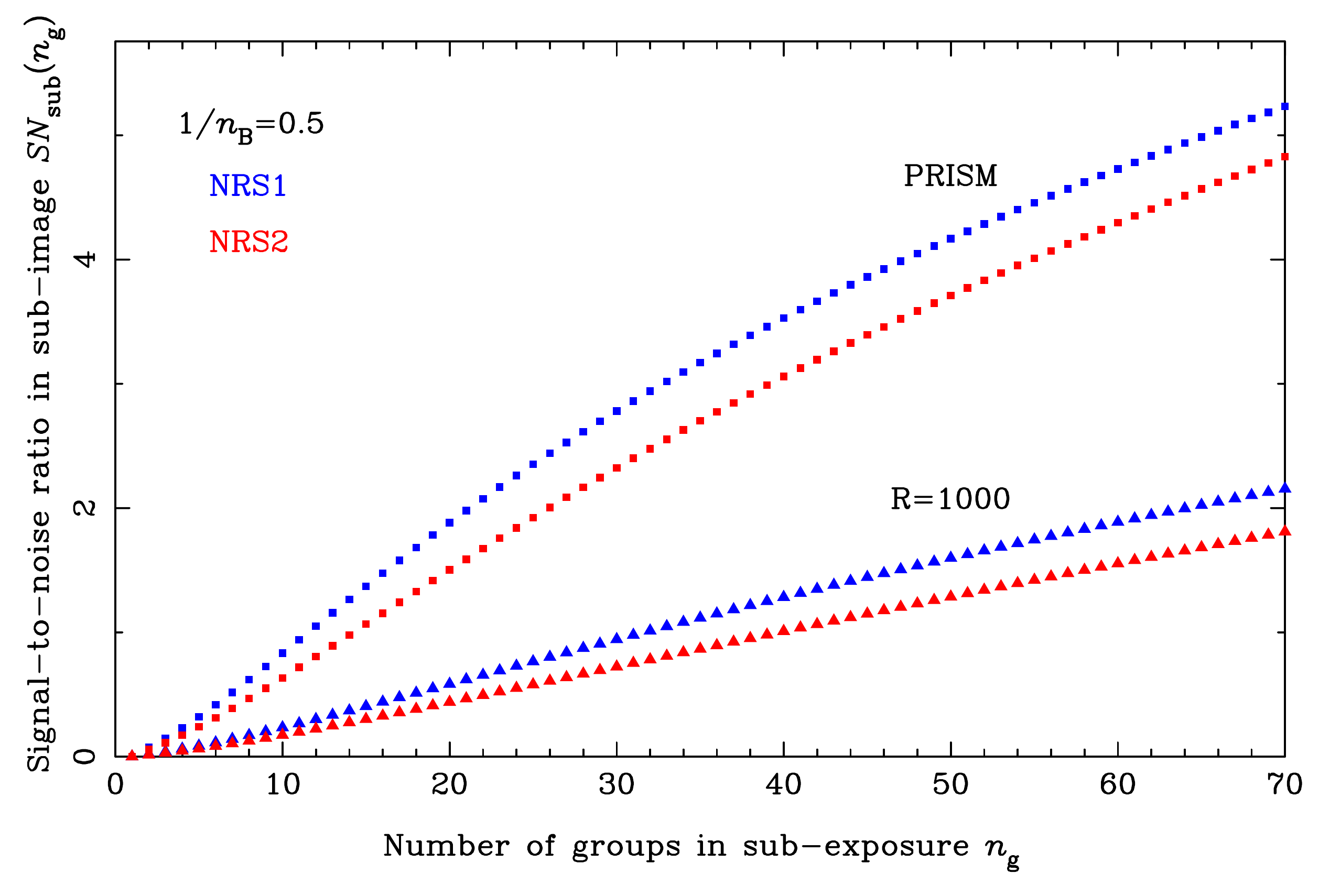}}
  \caption{Build-up of the signal-to-noise ratio in a subimage with increasing subexposure duration $n_\mathrm{g}$ in NRSIRS2RAPID mode for the two benchmark observations.}
\label{fig:snvsnf}
\end{figure}

A closer examination of the table entries also reveals that both benchmark observations are sky background dominated ($S\!<\!B$). It follows that accurate sky-subtraction is vital for both observations lest the extracted target flux measures be systematically biased. The $R\!=\!1000$ line detection is furthermore strongly detector noise-dominated ($\phi S\!+\!\phi B\!<\!\psi RN^2\!+\!\phi DC$), while the $R\!=\!100$ continuum source detection case is only partially so ($\phi S\!+\!\phi B\!\sim\!\psi RN^2\!+\!\phi DC$). This is also borne out by Fig.~\ref{fig:snvsnf} which plots the signal-to-noise ratios calculated from eqs.~\ref{eq:signalS}-\ref{eq:snsub} as a function of the subexposure duration $n_\mathrm{f}$. It is evident that the signal-to-noise ratio in the $R\!=\!1000$ case grows linearly with $n_\mathrm{f}$ for both arrays as expected when the total noise is wholly dominated by the nearly constant detector noise. The two $R\!=\!100$ curves on the other hand do display a slight curvature, thereby signaling that the 
$\sqrt{n_\mathrm{g}}$ dependency of the photon noise is not entirely overwhelmed by the detector noise in the PRISM case.

One way of quantifying the detector noise-limited nature of NIRSpec for the two benchmark observations is to compare to the signal-to-noise ratios that would be achievable at $n_\mathrm{B}=2$ with a hypothetical noiseless detector having the same detection efficiency, but zero dark current and read noise (and {$\phi\!=\!1$}):  ${S\!N}_\mathrm{tot}=21.3$ in the $R\!=\!100$ case and ${S\!N}_\mathrm{tot}=41.8$ in the $R\!=\!1000$ case. If measured by the ratio between the ideal photon Poisson variance to the total variance \citep{fixs00},  NIRSpec can in the two benchmark cases  be thought of as operating at its faint end at $\simeq$56\% of the ideal photon-limited efficiency in $R\!=\!100$ mode, and at $\simeq$22\% ideal efficiency in the higher resolution $R\!=\!1000$ mode.

It is also evident from Tables~\ref{tab:snr100} and \ref{tab:snr1000} that the dominant source of detector noise in both detector arrays for the choice $n_\mathrm{f}=70$ is the read noise ($\psi RN^2\gg \phi DC$). Furthermore, the NRS1 array is noticeably less noisy than NRS2, in spite of the former having the larger dark current. It will be recalled from Sect.~\ref{sec:wcoverage} that the $R\!=\!100$ prism and $R\!=\!1000$ grating spectra from the A-side fixed slits and the IFU all project intentionally onto the more sensitive NRS1 array.

\subsection{The impact of cosmic ray events}
\label{sec:cr}

The sensitivity calculations so far have ignored the impact of energetic particle radiation hits to the detectors, beyond assuming limited duration subexposures spanning only $n_\mathrm{g}\!=\!70$ $m=1$ grouped frames or $t_\mathrm{sub}=1006.7$\!~s. The sole reason for employing up-the-ramp sampling on all JWST detectors is to record the temporal build-up of the signal in each pixel, such that the partial signal accumulated by the pixel in the time interval up to a particle hit occurring can be extracted.
Thus particle events add a stochastic element to the exposures by effectively shortening the integration times of random pixels by random amounts, thereby leading to a net loss in the signal-to-noise ratio achieved in a given integration.

The baseline energetic particle flux that the JWST instruments are expected to be subjected to at L2 during quiescent Solar periods is $\simeq5.0$~cm$^{-2}$\!~s$^{-1}$ \citep{barth2000,giar19}. If we for simplicity assume that a spectral resolution element is impacted if any one of its $n_\mathrm{pix}=8$ pixels measuring $18\times 18~\mu$m$^2$ suffers a direct hit, the mean time between such particle events is $t_\mathrm{c} = (5.0 \times 0.0018^2 \times n_\mathrm{pix})^{-1} = 7.72 \times 10^3$~s. The number of particle hits occurring in a given time $t$ is a Poisson process with parameter $t/t_\mathrm{c}$, and the elapsed waiting time between pixel resets and the first particle hit follows an Exponential distribution with the rate parameter $t_\mathrm{c}^{-1}$. This last fact makes it straightforward to construct a statistical model for the impact of cosmic rays under the conservative assumptions that the impact of a particle event in a given ramp is readily recognizable by the processing software, and that the signal in any of the $n_\mathrm{pix}=8$ pixels making up a given resolution element under consideration can only be recovered up to and including the last grouped read completed prior to the first pixel being struck by a cosmic ray.

For an integration starting with a reset frame followed by $n_\mathrm{g}$ groups of $m$ averaged frames, let $\tilde{n}_\mathrm{g} \le n_\mathrm{g}$ denote the cosmic ray-shortened duration of the ramp, and let $p_\mathrm{c}(i,m)$ denote the probability that the $\tilde{n}_\mathrm{g}$ takes on the value $i$ where $0 \le i \le n_\mathrm{g}$.
Since two or more valid groups are needed to extract a slope from a ramp, the integration will need to be rejected (for the resolution element under consideration) if a cosmic ray hit takes place during the initial reset frame or during one of the first two groups. The probability of this occurring is:
\begin{equation}
p_\mathrm{c}(0,m) = 1-\exp(-(2\thinspace m+1) \frac{t_\mathrm{f}}{t_\mathrm{c}}).
\label{eq:p0}
\end{equation}
With the nonviable outcome $\tilde{n}_\mathrm{g}=1$ already tallied in $p_\mathrm{c}(0,m)$, we also have $p_\mathrm{c}(1,m)=0$.

\begin{figure}
\resizebox{\hsize}{!}{\includegraphics{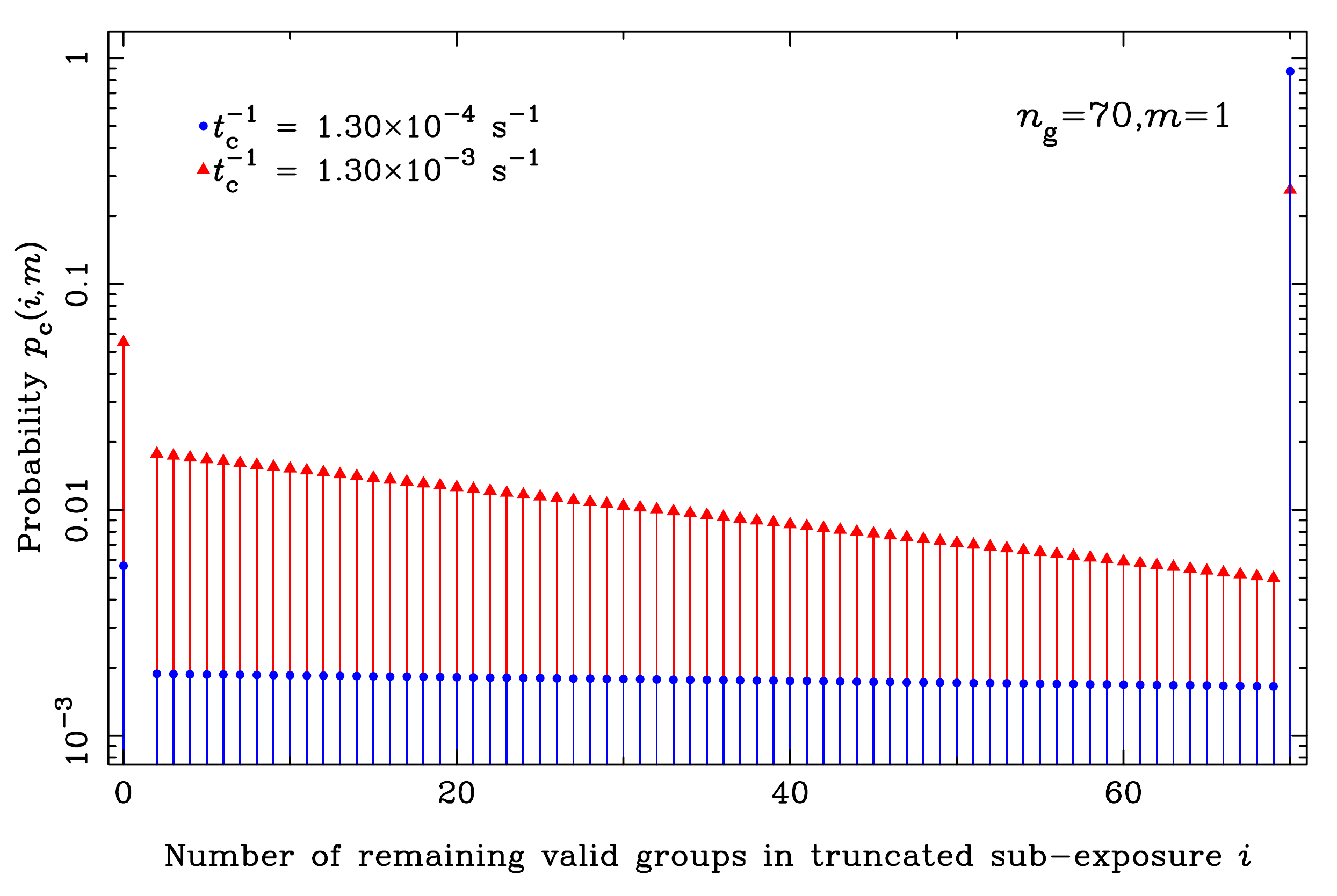}}
\caption{Probability that an exposure ramp extending over $n_\mathrm{g}=70$, $m=1$ groups runs to $\tilde{n}_\mathrm{g}=i$ valid reads before the resolution element spanning 8 detector pixels is hit by a cosmic ray. The blue circles are for the nominal assumed cosmic ray incidence rate. The red triangles are for a ten times higher event rate. Note the logarithmic ordinate scale.}
\label{fig:cr_probs}
\end{figure}

A ramp extending to $\tilde{n}_\mathrm{g}=i \ge 2$ groups results when the first cosmic ray strikes during the accumulation of group $(i+1) \le n_\mathrm{g}$. The probability of this occurring is:
\begin{equation}
p_\mathrm{c}(i,m) =  \exp(-(i \thinspace m+1) \frac{t_\mathrm{f}}{t_\mathrm{c}}) - \exp(-((i+1)\thinspace m+1) \frac{t_\mathrm{f}}{t_\mathrm{c}}).
\label{eq:pi}
\end{equation}
Lastly, the ramp will remain unaffected and extend to $\tilde{n}_\mathrm{g}=n_\mathrm{g}$ if the first hit occurs after accumulation of the last group $n_\mathrm{g}$. This occurs with probability:
\begin{equation}
p_\mathrm{c}(n_\mathrm{g},m) =  \exp(-(n_\mathrm{g}\thinspace m+1) \frac{t_\mathrm{f}}{t_\mathrm{c}}).
\label{eq:pn}
\end{equation}
Equations~(\ref{eq:p0}) -- (\ref{eq:pn}) provide a complete and accurate statistical description of how cosmic ray hits affect a NIRSpec up-the-ramp exposure for any cosmic ray incidence rate $t_\mathrm{c}^{-1}$.
It is easy to see that the above probabilities  $p_\mathrm{c}(i,m)$ for the ($n_\mathrm{g}+1$) possible outcomes sum to unity as they should. 

Figure~\ref{fig:cr_probs} plots the probability distribution $p_\mathrm{c}(i,m)$ for the $n_\mathrm{g}=70$, $m\!=\!1$  case at hand for the nominal assumed event rate of $t_\mathrm{c}^{-1} = 1.30\times10^{-4}$\!~s$^{-1}$ per resolution element and a ten times higher event rate. In the former case, where $t_\mathrm{c}\gg t_\mathrm{f}$ the probability that a given ramp is unaffected by particle hits is reasonably high at $p_\mathrm{c}(n_\mathrm{g},m)$=0.874; the probability that it becomes wholly unusable is low at $p_\mathrm{c}(0,m)$=0.0057; and the duration of a partially truncated ramp is nearly uniformly distributed over $2 \le i <  n_\mathrm{g}$. At the ten times higher event rate, the intact ramps become notably fewer at $p_\mathrm{c}(n_\mathrm{g},m)$=0.261; the ramp  rejection rate rises to $p_\mathrm{c}(0,m)$=0.055; and the durations of the intermediate affected ramps become tilted toward low values of $i$.

Two informative statistics are the mean truncated ramp duration:
\begin{equation}
<\!\tilde{n}_\mathrm{g}\!> = \sum_{i=0}^{n_\mathrm{g}} i\thinspace p_\mathrm{c}(i,m),
\label{eq:avng}
\end{equation}
and its standard deviation:
\begin{equation}
\sigma\tilde{n}_\mathrm{g}=\sqrt{<\!\tilde{n}_\mathrm{g}^2\!>\!-\!<\!\tilde{n}_\mathrm{g}\!>^2};
\label{eq:signg}
\end{equation}
where:
\begin{equation}
<\!\tilde{n}_\mathrm{g}^2\!> = \sum_{i=0}^{n_\mathrm{g}} i^2\thinspace p_\mathrm{c}(i,m).
\label{eq:avng2}
\end{equation}

For the two distributions shown in Fig.~\ref{fig:cr_probs}, the  average truncated ramp duration for the nominal assumed cosmic ray event rate is  $<\!\tilde{n}_\mathrm{g}\!>=65.4$, and the standard deviation $\sigma\tilde{n}_\mathrm{g}=14.2$.  In the enhanced cosmic ray event case, the ramp duration is on the average shortened to $<\!\tilde{n}_\mathrm{g}\!>=37.7$, but with even larger statistical scatter $\sigma\tilde{n}_\mathrm{g}=25.6$.

For low cosmic ray hit levels, the corresponding mean reduced signal-to-noise ratio in a given exposure can be calculated by repeating the calculations of Sect.~\ref{sec:sn} with $<\!\tilde{n}_\mathrm{g}\!>$ substituted for $n_\mathrm{g}$. However, this approach quickly breaks down for increasing cosmic ray rates. The expression for the mean signl-to-noise ratio achievable in a single exposure, valid for an arbitrary cosmic ray flux is:
\begin{equation}
<\!{S\!N}_\mathrm{sub}\!> = \sum_{i=0}^{n_\mathrm{g}} p_\mathrm{c}(i,m)\thinspace {S\!N}_\mathrm{sub}(i,m);  
\label{eq:avsn}
\end{equation}
where ${S\!N}_\mathrm{sub}(i,m)$ is the nominal cosmic ray free signal-to-noise ratio calculated from eqs. (\ref{eq:signalS}) through (\ref{eq:snsub}) for an exposure spanning $i$ groups (cf. Fig.~\ref{fig:snvsnf}).

However, the large standard deviations $\sigma\tilde{n}_\mathrm{g}$ cited above are signaling that cosmic ray events are injecting another important stochastic element into the process. Whenever a long duration NIRSpec exposure is broken up into $n_\mathrm{sub}$ shorter duration subexposures -- as in the two reference cases discussed here -- the resulting subexposures will be truncated by cosmic rays to varying degrees, and therefore display very different signal-to-noise ratios. It follows that when $n_\mathrm{sub}$ such randomly cosmic ray-shortened  subexposures are to be combined, it is vital that they be coadded in a statistically optimal inverse variance-weighted fashion in order that the best possible combined signal-to-noise ratio may be achieved. In this case the signal-to-noise ratio of the combined data set is:
\begin{equation}
{S\!N}_\mathrm{tot} = (\sum_{j=1}^{n_\mathrm{sub}}\  {S\!N}^2_{\mathrm{sub} j})^\frac{1}{2};
\label{eq:snwtot}
\end{equation}
where ${S\!N}_{\mathrm{sub} j}$ is the signal-to-noise ratio achieved in subexposure $j$.
In the mean, (\ref{eq:snwtot}) becomes approximately:
\begin{equation}
<\!{S\!N}_\mathrm{tot}\!> \ \simeq \sqrt{n_\mathrm{sub}}  <\!{S\!N}^2_\mathrm{sub}\!>^{\frac{1}{2}};
\label{eq:snwtot2}
\end{equation}
where the average squared signal-to-noise ratio per subexposure follows from:
\begin{equation}
<\!{S\!N}^2_\mathrm{sub}\!> = \sum_{i=0}^{n_\mathrm{g}} p_\mathrm{c}(i,m)\thinspace {S\!N}_\mathrm{sub}^2(i,m), 
\label{eq:snwtot3}
\end{equation}
and ${S\!N}_\mathrm{sub}(i,m)$ again is the nominal signal-to-noise ratio calculated from eqs. (\ref{eq:signalS}) through (\ref{eq:snsub}) for an exposure spanning $i$ groups.

The astute reader will appreciate that the above model implicitly assumes that only the primary resolution element containing the signal is affected by cosmic rays, and that the background subtraction remains unaffected. Equations (\ref{eq:snwtot2}) and (\ref{eq:snwtot3}) are therefore strictly speaking only applicable in the master background-subtracted case with $1/n_\mathrm{B}=0$ in eq. (\ref{eq:snsub}). The validity of the above model can then readily be verified through detailed pixel-level Monte Carlo simulation of the slope linear fitting process, by assigning suitable Poisson deviates to the incremental increases in the accumulated up-the-ramp charges due to the target, background and dark current, Normal deviates to the read noise per eq. (\ref{eq:readnoise}), and Exponential deviates to the cosmic ray hit times.

\begin{figure}
  \resizebox{\hsize}{!}{\includegraphics{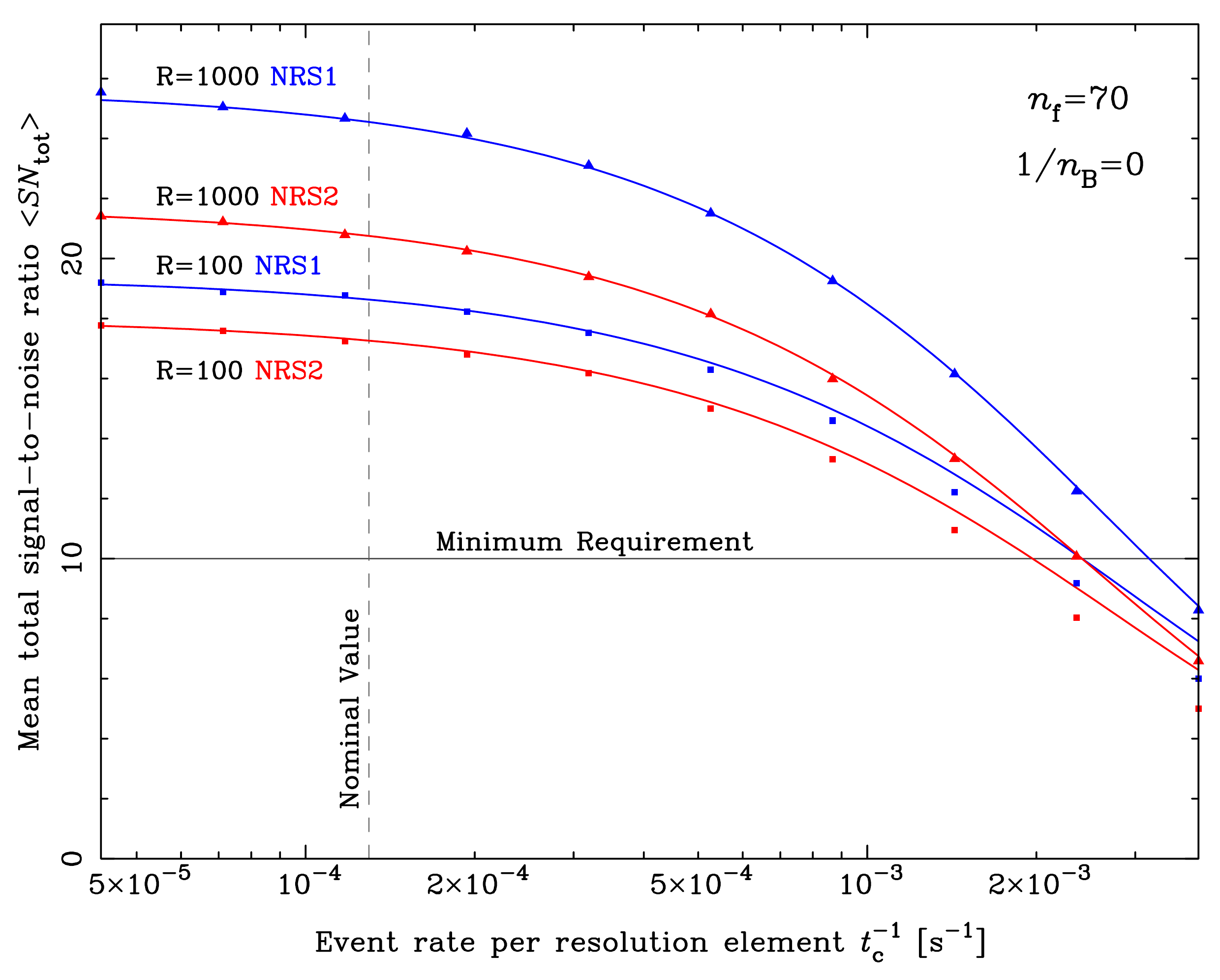}}
  \caption{Mean total signal-to-noise ratio achievable in the two benchmark observations when employing NRSIRS2RAPID subimages spanning $n_\mathrm{g}\!=\!70$, $m=1$ groups plotted against the incident cosmic ray event rate in a resolution element. The triangles and squares mark the outcomes of matching Monte Carlo simulations}
\label{fig:snvscr}
\end{figure}

Figure~\ref{fig:snvscr} plots how the signal-to-noise ratios in the two baseline $1/n_\mathrm{B}=0$ calculations of Tables~\ref{tab:snr100} and \ref{tab:snr1000} are predicted to degrade with increasing cosmic ray event rate according to the model. The outcome of the matching Monte Carlo simulations are overlaid as  squares and triangles for the $R\!=\!100$ and $R\!=\!1000$ cases respectively. It is seen that the agreement is excellent in the two $R\!=\!1000$ cases over the entire range probed, but that eq. (\ref{eq:snwtot2}) predicts slightly too high $R\!=\!100$ signal-to-noise ratios for the more extreme cosmic ray rates where the majority of resolution elements suffer a particle hit. This has to do with the most noisy slope determinations from the severely truncated subexposures being less efficiently averaged down compared to eq.~(\ref{eq:snwtot2}) when only $n_\mathrm{sub}=10$ subexposures are combined, thereby adding to the sample variance in the Monte Carlo simulations. 

For the anticipated nominal cosmic ray rate of $t_\mathrm{c}^{-1} = 1.30\times10^{-4}$\!~s$^{-1}$ per resolution element, the predicted loss in $\left<{S\!N}_\mathrm{tot}\right>$ due to particle hits shortening the net exposure time is {$\simeq\!4\!-\!5$}\% in all cases. Furthermore, Fig. \ref{fig:snvscr} reassuringly suggests that a particle event rate more than an order-of-magnitude higher than the assumed nominal value is required to bring NIRSpec in danger of not being able to meet its benchmark Level~2 requirements. Such high event rates are, however, expected to be encountered during periods of high Solar activity whenever a Solar Coronal Mass Ejection strikes the JWST spacecraft \citep{barth2000}. 

\begin{figure}
  \resizebox{\hsize}{!}{\includegraphics{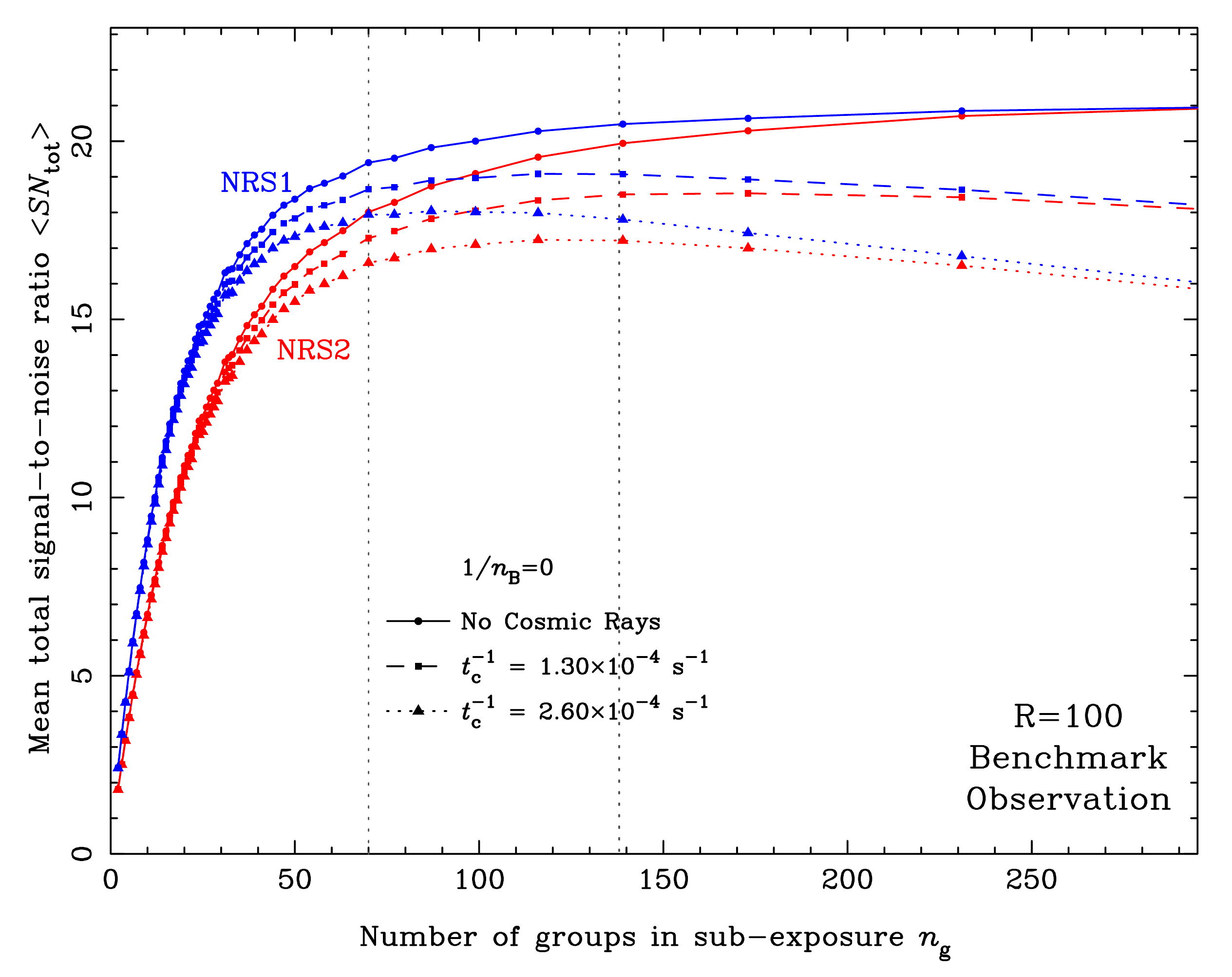}}
  \caption{Mean total signal-to-noise ratio achievable in the benchmark $R\!=\!100$ continuum detection case for a fixed total exposure time of $t_\mathrm{int} = 10^4$~s as a function of the number of $n_\mathrm{g}$, $m\!=\!1$ groups in each subexposure.}
\label{fig:sn100}
\end{figure}

\begin{figure}
  \resizebox{\hsize}{!}{\includegraphics{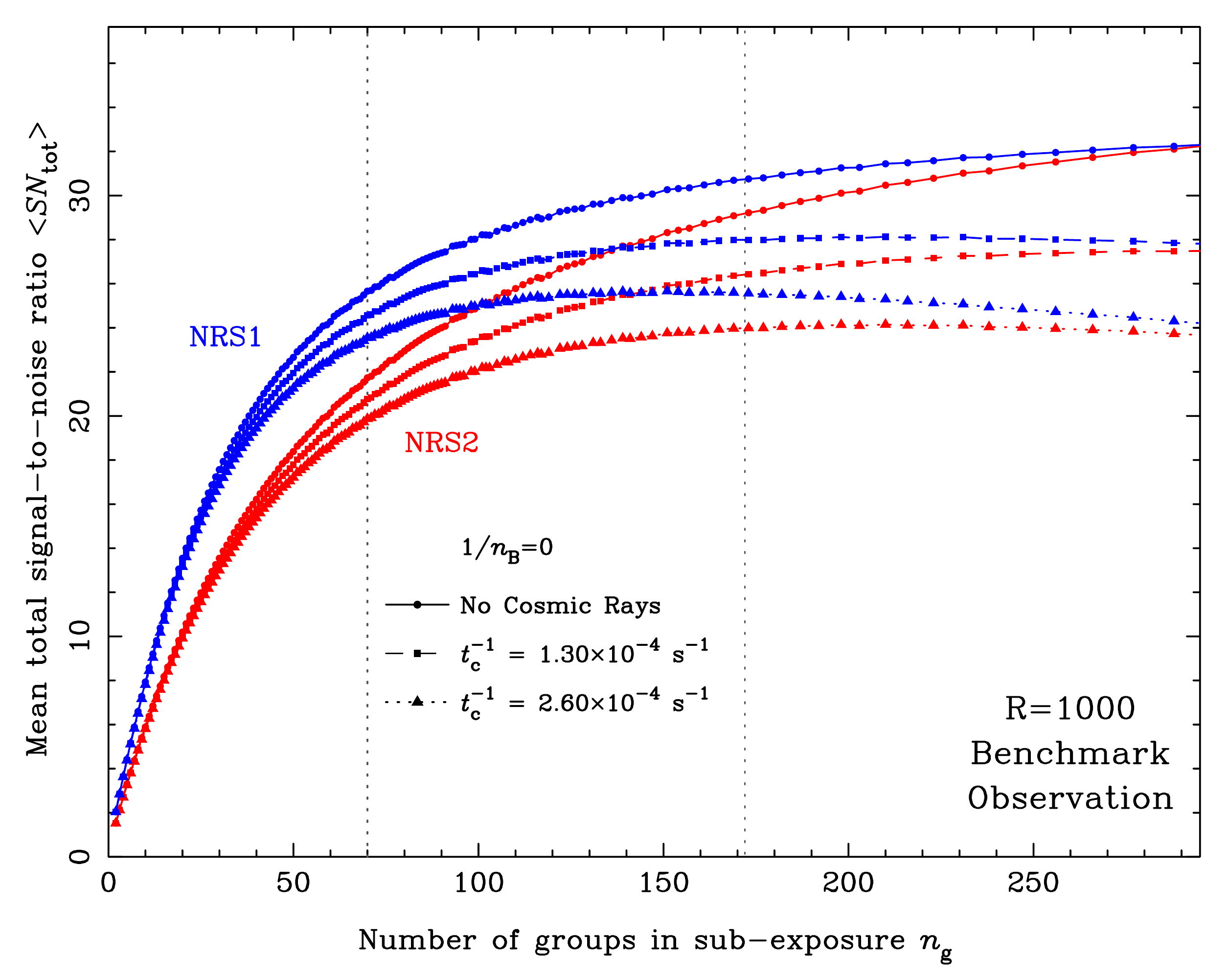}}
  \caption{Mean total signal-to-noise ratio achievable in the benchmark $R\!=\!1000$ emission line detection case for a fixed total exposure time of $t_\mathrm{int} = 10^5$~s as a function of the number of $n_\mathrm{g}$, $m\!=\!1$ groups in each subexposure.}
\label{fig:sn1000}
\end{figure}

\subsection{The potential for employing longer integrations}
\label{sec:int}

Considering that the dominant total detector read noise according to eqs. (\ref{eq:readfac}) and (\ref{eq:readnoise}) stays roughly constant  when  $n_\mathrm{g}$ is increased, there is a potential gain in net signal-to-noise to be had provided the cosmic ray event rate allows splitting the benchmark  observations up into fewer, but longer subexposures. Such a change would also release a savings in observatory overhead.

With the aim of quantifying this  gain, Figs.~\ref{fig:sn100} and \ref{fig:sn1000} show the results of repeating the two benchmark signal-to-noise calculations for all combinations of $n_\mathrm{g}$ and $n_\mathrm{sub}$ that provide a total exposure time $t_\mathrm{int}$ per eq. (\ref{eq:exptime}) for $m=1$ that falls just short of the specified total exposure times by no more than 3\%. The filled circles connected by full lines show the results ignoring cosmic ray hits calculated from eq. (\ref{eq:sntot0}). The filled squares connected by dashed lines are calculated from eqs.~(\ref{eq:snwtot2}) and (\ref{eq:snwtot3}) for the nominal cosmic ray event rate of $t_\mathrm{c}^{-1} = 1.30\times10^{-4}$\!~s$^{-1}$, and the triangles connected by dotted lines for twice this rate. The slight raggedness of the locations of the points in these plots reflects the small 3\% differences in total integration time between the different combinations of $n_\mathrm{g}$ and $n_\mathrm{sub}$. Depending on how successful the up-the-ramp processing is in recovering the ramp slopes following a particle hit \citep{giar19}, the actual NIRSpec performance should fall somewhere between the upper and lower curves in Figs.~\ref{fig:sn100} and \ref{fig:sn1000}.

Not surprisingly, the topmost cosmic ray-free curves demonstrate that consideration of the detector performance alone indeed pushes toward the longest possible subexposures, so that the photon noise can overwhelm the nearly constant total read noise. Since the total exposure time is held constant for all shown combinations of $n_\mathrm{g}$ and $n_\mathrm{sub}$, the topmost uncorrected curves flatten out at large values of $n_\mathrm{g}$ once the Poisson statistics dominate, and the two sets of curves for NRS1 and NRS2 converge in this limit since the two devices are assumed to have the same quantum efficiency.

Fig.~\ref{fig:sn100} shows that in the $R\!=\!100$ case, the predicted modest $\simeq4$\% average reduction in $\left<{S\!N}_\mathrm{tot}\right>$ caused by cosmic ray events occurring at the assumed nominal rate for $n_\mathrm{g}\!=\!70$ can be  gained back -- at least for NRS2 -- by doubling the subexposure duration and instead splitting the observation into  $n_\mathrm{sub}\!=\!5$ subexposures each of $n_\mathrm{f}\!=\!140$ frames duration. Nonetheless, both sets of $R\!=\!100$ curves are reassuringly flat in the vicinity of $n_\mathrm{g}\!=\!70$ suggesting that the assumed maximum subexposure length of $t_\mathrm{sub}\simeq 1000$~s is not too far from optimal. 

The gain obtained by lengthening the subexposure duration in the more severely detector noise-limited $R\!=\!1000$ benchmark case is more substantial. Figure~\ref{fig:sn1000} suggests that breaking up the $10^5$\!~s total integration into, say, $n_\mathrm{sub}\!=\!40$ subexposures extending up to $n_\mathrm{g}\!=\!172$ groups improves the nominal cosmic ray reduced net signal-to-noise ratio by $\simeq14$\% in the case of NRS1 and $\simeq27$\% in the case of NRS2.  While too much stock should probably not be placed in these inferences given the uncertainty on the actual range of $t_\mathrm{c}^{-1}$ that NIRSpec will operate under, and how well the data processing software will be able to detect and deal with particle hits, Fig.~\ref{fig:sn1000} does justify that effort be spent on orbit determining the optimal compromise common maximum subexposure duration for the two slightly different flight detector arrays of NIRSpec.

\end{appendix}

\end{document}